\documentclass[twocolumn,tighten,times]{aastex631}

\usepackage{graphicx}
\usepackage[T1]{fontenc}
\usepackage{gensymb}
\usepackage{natbib}
\usepackage[utf8]{inputenc}
\usepackage{multirow}
\usepackage[caption=false]{subfig}
\usepackage[percent]{overpic}

\let\oldput\put
\def\put(#1,#2)#3{%
  \oldput(#1,#2){\sffamily #3}%
}
\definecolor{navyblue}{RGB}{0,50,250}
\definecolor{purple}{RGB}{100,0,250}
\hypersetup{linkcolor=purple,citecolor=navyblue,filecolor=cyan,urlcolor=magenta}

\def\hi{{\ion{H}{1}}}

\newcommand{\co}{CO}

\def\mum154{{154\,$\mu$m}}
\def\s{{$\sim$}}
\def\n{{\n}{$-$}}
\def\hawc{{HAWC\nolinebreak[4]\hspace{-.02em}\raisebox{.17ex}{+}}}
\def\sofiahawc{{SOFIA/HAWC\nolinebreak[4]\hspace{-.02em}\raisebox{.17ex}{+}}}
\def\m51{{M\nolinebreak[4]\hspace{0.08em}51}}

\newcommand{\mPsi}{\ensuremath{\overline{\Psi}}}

\newcommand{\lineco}{$^{12}$\co(1--0)}
\def\columndensity{\ensuremath{N_{\mathrm{HI+2H_{2}}}}}

\newcommand{\APsiFIR}{\ensuremath{\mPsi_{\mathrm{FIR}}^{\mathrm{Arms}}}}

\newcommand{\IAPsiFIR}{\ensuremath{\mPsi_{\mathrm{FIR}}^{\mathrm{IA}}}}
\newcommand{\FDPsiFIR}{\mPsi_\ensuremath{\mathrm{FIR}}^{\mathrm{FD}}}
\newcommand{\MPsiFIR}{\ensuremath{\mPsi_{\mathrm{FIR}}^{\mathrm{Morph}}}}

\newcommand{\APsithreecm}{\ensuremath{\mPsi_{\mathrm{3\,cm}}^{\mathrm{Arms}}}}

\newcommand{\IAPsithreecm}{\ensuremath{\mPsi_{\mathrm{3\,cm}}^{\mathrm{IA}}}}
\newcommand{\FDPsithreecm}{\ensuremath{\mPsi_{\mathrm{3\,cm}}^{\mathrm{FD}}}}
\newcommand{\MPsithreecm}{\ensuremath{\mPsi_{\mathrm{3\,cm}}^{\mathrm{Morph}}}}

\newcommand{\APsisixcm}{\ensuremath{\mPsi_{\mathrm{6\,cm}}^{\mathrm{Arms}}}}

\newcommand{\IAPsisixcm}{\ensuremath{\mPsi_{\mathrm{6\,cm}}^{\mathrm{IA}}}}
\newcommand{\FDPsisixcm}{\ensuremath{\mPsi_{\mathrm{6\,cm}}^{\mathrm{FD}}}}
\newcommand{\MPsisixcm}{\ensuremath{\mPsi_{\mathrm{6\,cm}}^{\mathrm{Morph}}}}

\newcommand*{\logten}{\mathop{\log_{10}}}
\newcommand{\um}{$\mu$m}
\renewcommand*{\thefootnote}{\fnsymbol{footnote}}

\begin{document}
\title{Extragalactic Magnetism with SOFIA (Legacy Program)\\ I: The magnetic field in the multi-phase interstellar medium of \m51 \protect\footnote{The SOFIA Legacy Group for Magnetic Fields in Galaxies software repository is available in https://github.com/galmagfields/hawc, via the official project website: http://galmagfields.com/, and Zenodo/GitHub: https://doi.org/10.5281/zenodo.5116134}}
\shorttitle{The magnetic field in the multi-phase interstellar medium of \m51}
\shortauthors{Borlaff et al.}

\correspondingauthor{Borlaff, A. S.}
\email{a.s.borlaff@nasa.gov}

\author{Alejandro S. Borlaff}
\affil{NASA Ames Research Center, Moffett Field, CA 94035, USA}

\author{Enrique Lopez-Rodriguez}
\affil{Kavli Institute for Particle Astrophysics \& Cosmology (KIPAC), Stanford University, Stanford, CA 94305, USA}

\author{Rainer Beck}
\affil{Max-Planck-Institut f\"ur Radioastronomie, Auf dem H\"ugel 69, 53121 Bonn, Germany}

\author{Rodion Stepanov}
\affil{Institute of Continuous Media Mechanics, Korolyov str.\,1, 614013 Perm, Russia}

\author{Eva Ntormousi}
\affil{Scuola Normale Superiore, Piazza dei Cavalieri 7, 56126 Pisa, Italy}

\author{Annie Hughes}
\affil{CNRS, IRAP, 9 Av. du Colonel Roche, BP 44346, 31028 Toulouse cedex 4, France}
\affil{Universite de Toulouse, UPS-OMP, IRAP, 31028 Toulouse cedex 4, France}

\author{Konstantinos Tassis}
\affil{Institute of Astrophysics, Foundation for Research and Technology-Hellas, 71110 Heraklion, Greece}
\affil{Department of Physics, and Institute for Theoretical and Computational Physics, University of Crete, 70013 Heraklion, Greece}

\author{Pamela M. Marcum}
\affil{NASA Ames Research Center, Moffett Field, CA 94035, USA}

\author{Lucas Grosset}
\affil{Kavli Institute for Particle Astrophysics \& Cosmology (KIPAC), Stanford University, Stanford, CA 94305, USA}

\author{John E. Beckman}
\affil{Instituto de Astrof\'isica de Canarias, C/ Via L\'actea s/n, 38200 La Laguna, Tenerife, Spain}
\affil{Departamento de Astrof\'isica, Universidad de La Laguna, Avda. Astrofísico Fco. S\'anchez s/n, 38200 La Laguna, Tenerife, Spain}

\author{Leslie Proudfit}
\affil{SOFIA Science Center, NASA Ames Research Center, Moffett Field, CA 94035, USA}

\author{Susan E. Clark}
\affil{Institute for Advanced Study, 1 Einstein Drive, Princeton, NJ 08540, USA}

\author{Tanio D\'iaz-Santos}
\affil{Institute of Astrophysics, Foundation for Research and Technology-Hellas (FORTH), Heraklion, 70013, Greece}

\author{Sui Ann Mao}
\affil{Max-Planck-Institut f\"ur Radioastronomie, Auf dem H\"ugel 69, 53121 Bonn, Germany}

\author{William T. Reach}
\affil{SOFIA Science Center, NASA Ames Research Center, Moffett Field, CA 94035, USA}

\author{Julia Roman-Duval}
\affil{Space Telescope Science Institute, 3700 San Martin Drive, Baltimore, MD 21218}

\author{Kandaswamy Subramanian}
\affil{Inter-University Centre for Astronomy and AstrophysicsPost Bag 4, Ganeshkhind, Pune 411007}

\author{Le Ngoc Tram}
\affil{SOFIA Science Center, NASA Ames Research Center, Moffett Field, CA 94035, USA}

\author{Ellen G. Zweibel}
\affil{Department of Astronomy, U. Wisconsin-Madison, 475 N Charter Street, Madison, WI 53706, USA}
\affil{Department of Physics, U. Wisconsin-Madison, 1150 University Avenue, Madison, WI 53706 USA}

\author{Daniel Dale}
\affil{Department of Physics \& Astronomy, University of Wyoming, Laramie, WY 82070}

\author{Legacy Team}



\begin{abstract} 
The recent availability of high-resolution far-infrared (FIR) polarization observations of galaxies using HAWC+/SOFIA has facilitated studies of extragalactic magnetic fields in the cold and dense molecular disks.
We investigate if any significant structural differences are detectable in the kpc-scale magnetic field of the grand design face-on spiral galaxy \m51\ when traced within the diffuse (radio) and the dense and cold (FIR) interstellar medium (ISM). 
Our analysis reveals a complex scenario where radio and FIR polarization observations do not necessarily trace the same magnetic field structure. We find that the magnetic field in the arms is wrapped tighter at \mum154\ than at 3 and 6\,cm; statistically significant lower values for the magnetic pitch angle are measured at FIR in the outskirts ($R\ge7$ kpc) of the galaxy. This difference is not detected in the interarm region. We find strong correlations of the polarization fraction and total intensity at FIR and radio with the gas column density and \lineco\ velocity dispersion. We conclude that the arms show a relative increase of small-scale turbulent B-fields at regions with increasing column density and dispersion velocities of the molecular gas. No correlations are found with \hi\ neutral gas. The star formation rate shows a clear correlation with the radio polarized intensity, which is not found in FIR, pointing to a small-scale dynamo-driven B-field amplification scenario.
This work shows that multi-wavelength polarization observations are key to disentangling the interlocked relation between star formation, magnetic fields, and gas kinematics in the multi-phase ISM.
\end{abstract} 



\section{Introduction}
\label{sec:introduction}

\begin{figure*}[ht!]
\begin{center}
\includegraphics[trim={0 0 0 0}, clip, width=0.49\textwidth]{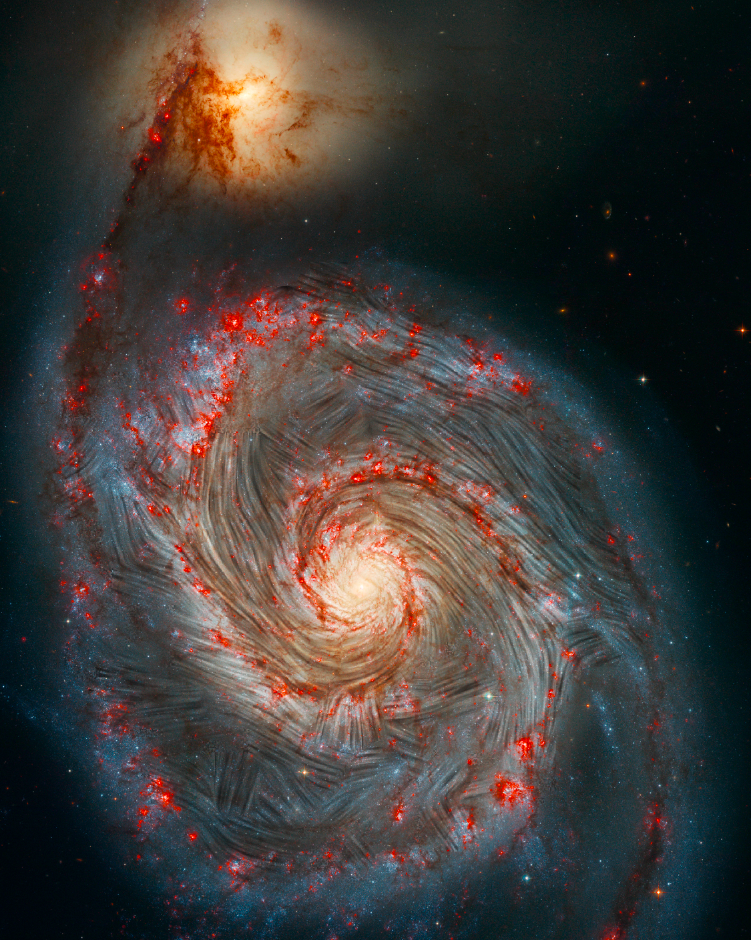}
\includegraphics[trim={0 0 0 0}, clip, width=0.49\textwidth]{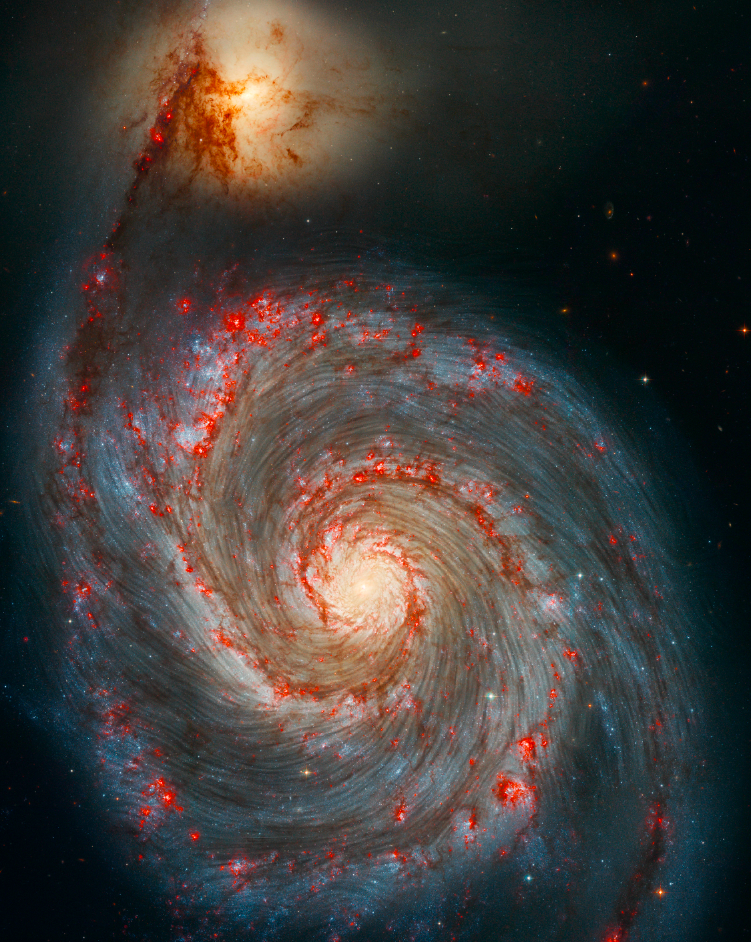}
\caption{Far-infrared (\mum154\ from this work, \emph{left}) and radio polarization \citep[$6$ cm from][\emph{right}]{Fletcher2011} magnetic field orientation in the plane of the sky represented over the optical morphology of \m51. \emph{RGB background:} \textit{Hubble Space Telescope} observations of \m51\ with the F658N (H$\alpha$) and F814W (red), F555W (green), and F435W (blue) bands using the Advanced Camera for Surveys (ACS). \emph{Overlayed stripped texture:} The Line Integral Convolution \citep[LIC,][]{LIC} technique was used to show the orientations of the B-field at FIR and radio, where only polarization measurements with $P/\sigma_{P} \ge 3$, a resample scale of 5, and a contrast of 2 was used.}
\label{fig:m51_apod}
\end{center}
\end{figure*}

Pioneering optical polarimetric observations in galaxies detected the interstellar polarization due to aligned dust grains, which motivated the discussion of magnetic fields (B-fields) in galaxies \citep[i.e.][]{Elvius1951,Aller1958, Elvius1964,Piddington1964, Segalovitz1976,Scarrott1987}. The formation and sustainability of B-fields in the galactic disks, as well as their possible role in the evolution of their hosts, are still outstanding questions of modern astrophysics. Primordial magnetic fields are not strong enough \citep{Rees1987, Gnedin2000, Subramanian2016} to explain the observations in spiral galaxies by simple gravitational collapse \citep{Beck1996}. Extragalactic B-fields are thought to be generated by galactic dynamos, which rely on small-scale turbulent velocity fields and differential rotation of the galactic disk to amplify and order the B-fields \citep[i.e.][]{Beck1996, Gressel2008b, Gressel2008a, Gent2012, Bendre2015}. Current dynamo theories can be divided into large-scale dynamos, which produce regular B-fields on scales larger than the flow scale; and into small-scale dynamos, generated at scales smaller than the energy-carrying eddies \citep{Rees1987, Gnedin2000,BS2005, Gressel2008b, Gressel2008a, Subramanian2016}. The  coherence length scale of supernova-driven turbulence is 50--100 pc \citep[i.e.][]{Haverkorn2008}. The most prominent theory for large-scale dynamos is given by the mean-field approach, where the velocity and B-fields are decomposed into averaged components and fluctuating components, whose average can either be an ensemble average or some kind of spatial average \citep{BAS2012}. Recently, more attention has been given to small-scale dynamos as they are more generic in terms of flow requirements and exhibit much faster B-field growth. The amplification timescale of small-scale B-fields are of the order of the smallest turbulent eddy turnover time scale. This is important because the small-scale dynamos allow amplification of the B-fields even in galaxy clusters or elliptical galaxies \citep{BS2005,SS2021}. The small-scale B-fields may also explain strong B-fields in high redshift galaxies when the universe was much younger and large-scale dynamo amplification times were not sufficient \citep{AB2009}.

The dynamical role of magnetic fields on galactic scales is strongly debated. Magnetic fields in galaxies are strong enough to turn a significant amount of kinetic energy into magnetic energy, driving gas mass inflows into the galactic core \citep{Kim2012}. Magnetic fields have even been considered as a hidden contributor to flattening rotation curves \citep{Battaner2007,RuizGranados2010,Tsiklauri2011, RuizGranados2012, Jalocha2012a, Jalocha2012b}. However, various studies have posited that the local conditions of magnetic fields might be too turbulent to add a significant kinematic support to the gas disk or to create a systematic stellar migration \citep{SanchezSalcedo2013,Elstner2014}. In spite of these arguments, recent magneto-hydrodynamic simulations of Milky Way mass objects with magnetic fields have shown that the resulting galaxies present more extended disks, showing more gas and more atomic hydrogen in their halos than those models without them \citep{vandeVoort2020}. \citet{Tabatabaei2016} found a correlation between the large-scale magnetic field strength and the rotation speed of galaxies showing the effect of the gas dynamics in ordering the magnetic fields in galaxies. Different authors consider that these B-fields are able to significantly influence disk galaxies, dominating the fragmentation pattern \citep{Kortgen2019} and affecting the global rotation of the gas \citep{Martin-Alvarez2020}. 

Most of our knowledge about extra-galactic magnetism comes from radio polarimetric observations \citep[i.e.][]{Mathewson1972, Vollmer2013, Beck2015b, Krause2020} by means of synchrotron polarized emission from energetic particles in the diffuse interstellar medium (ISM) and intergalactic medium (IGM). Using a sample of $13$ galaxies from the CHANG-ES radio continuum survey, \citet{Krause2018} found that the 4--30 cm radio observations are sensitive to average scale heights of 1--2 kpc. Synchrotron emission measures the total magnetic field strength and the magnetic field component in the plane of the sky (POS), while the magnetic field component along the line-of-sight (LOS) is inferred using the effect of Faraday rotation. Synchrotron polarization provides a measurement of the degree of order of the B-field, where the ordered field can be a regular (dynamo-generated) and/or an anisotropic turbulent one. The fractional polarization can decrease due to beam depolarization, bandwidth depolarization, and/or wavelength-dependent depolarization. Beam depolarization occurs due to tangled B-fields within the beam size of the observations. Bandwidth depolarization arises from the rotation of the plane of polarization at different frequencies within the frequency range of the observations. Wavelength-dependent depolarization is caused by Faraday rotation along the LOS or within the source. Major efforts have been performed to estimate the B-fields orientation of galaxies using optical \citep{Elvius1951,Elvius1964, Scarrott1987, Fendt1998} and near-IR \citep[NIR,][]{Jones1997,Jones2000,Pavel2012} polarization techniques via dichroic absorption. However, dust/electron scattering seems to be the dominant polarization mechanism in some of these observations, where after careful subtraction, the B-field can be inferred \citep[i.e. M\,82,][]{Jones1997,Jones2000}.

Magnetic fields in galaxies have also been measured using thermal emission from magnetically aligned dust grains at far-IR (FIR) \citep{LopezRodriguez2018, LopezRodriguez2020,LopezRodriguez2021,ELR2021b, Jones2019, Jones2020} and sub-mm ($850$ \um) wavelengths \citep[i.e.][]{Greaves2000,Matthews2009}. 
These studies have shown that the FIR wavelength range (50--220 \um) can characterize the strength and structure of B-fields in galaxies. FIR polarimetric observations of the edge-on galaxies Centaurus\,A \citep{ELR2021b}, M\,82 \citep{Jones2019,LopezRodriguez2021}, NGC\,253 \citep{Jones2019}, and NGC\,891 \citep{Jones2020} show scale heights $<500$ pc for the galactic disks. At these wavelengths, the spectral energy distribution of galaxies is dominated by the thermal emission from interstellar dust at temperatures of 10 -- 100 K, which traces deeper regions of the molecular disk than those from optical, NIR, and radio. Dust grains have their long axes aligned perpendicularly to the local B-field, as described by the radiative torque alignment theories \citep[RATs, i.e.][]{Hoang2014,Andersson2015}. Thus, thermal polarized emission measures the B-field orientation in the POS, after the polarization angles are rotated by $90^{\circ}$. As in radio wavelengths, thermal polarization provides a measurement of the degree of order of B-fields. The thermal polarization fraction is affected by beam depolarization, turbulence at scales smaller than the observational beam, and physical properties of the dust grains including temperature, column density, and alignment efficiency.

The ISM of the spiral galaxies is highly heterogeneous. The cold and dense clouds and the diffuse ISM \citep{Field1969} dominate different regions of the galactic disk. Molecular gas is closer to the galactic plane, while the scale-height of the diffuse ISM can be one order of magnitude larger \citep{Ferriere2001}. Molecular gas also is more rotationally supported than the diffuse ionized gas component, which has higher dispersion in velocity \citep{Davis2013, Levy2018}. As most of the studies on kpc-scale magnetic fields in galactic disks are based on radio-polarimetric observations, our knowledge is mainly focused on the B-field tracing the diffuse ISM rather than of the cold dense molecular clouds and filaments. However, it is inside the molecular clouds where star formation takes place and where turbulence and magnetic fields can be dominant forces \citep{Santos2016,Pillai2017, Pillai2020}. The geometry of the magnetic field in observations of Galactic polarized dust emission suggests that the magnetic field structure may influence the formation of molecular clouds. The magnetic field is aligned preferentially parallel to molecular cloud structures at low densities, and preferentially perpendicular at higher densities and back to parallel at even higher densities \citep{Planck2016, Soler2017, Fissel2019}.

Using the High-resolution Airborne Wideband Camera-plus \citep[\hawc,][]{Vaillancourt2007, Dowell2010, Harper2018} installed on the 2.7-m Stratospheric Observatory for Infrared Astronomy (SOFIA) FIR polarization observations, \citet{Pillai2020} found evidence for a multi-phase processing scenario where gas filaments merge into a central region in the molecular clouds, reorienting the magnetic field in dense gas flows compared to the orientation of the surrounding ISM. These transitions in the orientation of the magnetic fields may be related to small-scale gas accretion kinematics and the subsequent magnetic field line dragging, as reported by magneto-hydrodynamic simulations \citep{Gomez2018}. The morphological and kinematic differences between the diffuse ISM and the molecular clouds elicit a basic yet unresolved question: How does the multi-phase ISM in galaxies affect the B-field? Motivated by the potentially important role of magnetic fields in the dense ISM, we quantify the morphology and degree of order of the B-field in the multi-phase ISM traced by FIR and radio polarimetric observations.

Given that polarization studies are strongly limited by the signal-to-noise ratio (SNR), local bright galaxies are the most extensively studied objects. One of these objects is the grand design face-on spiral \m51. Although there have been attempts to measure the B-field in \m51 using optical \citep{Scarrott1987} and NIR \citep{Pavel2012} wavelengths, these observations have been found to be dominated by dust/electron scattering. The kpc-scale B-field of \m51 has been traced using radio polarimetric observations \citep{Mathewson1972, Beck1987, Neininger1992, Horellou1992, Patrikeev2006, Fletcher2011, Kierdorf2020}. These studies have shown an ordered kpc-scale B-field where turbulent B-fields dominate in the arms, while a regular B-field dominates in the inter-arm. In addition, \citet{Kierdorf2020} measured that the turbulent B-field strength and/or the thermal electron density decrease toward larger radii. In a recent study, \citet{Jones2020} presented the inferred B-field orientation of \m51 traced by $154$ \um\ thermal emission of magnetically aligned dust grains using HAWC+/SOFIA. The authors show the general B-field structure of the disk and compared it with results from previous radio-polarization observations at $6$ cm \citep{Fletcher2011}. The authors concluded that the magnetic fields traced in radio and FIR have a similar general structure showing no obvious differences on inspection by-eye. 

Detecting systematic differences in the magnetic field between radio and FIR  wavelengths requires precise and quantitative statistics to be estimated using both data sets. Since the star formation rate (SFR) is not homogeneous across the galactic disks of spiral galaxies, variances between the polarization maps at radio and FIR would be expected to likewise have an inhomogeneous spatial distribution as the multi-phase ISM affect the galactic B-field. Thus, our investigation is particularly focused on the radial variation and differences between disk regions (arms vs. interarm). In the particular case of \m51, we also look for a possible variation of the magnetic field orientation between the northern region (closer to the interacting companion \m51b) and the southern section. In this paper we revisit the magnetic field structure of \m51, using deeper observations than those presented by \citet{Jones2020}, to investigate quantitatively how the properties of \m51's magnetic field structure correlate with wavelength, morphological region, and the ISM phase. 

The paper is organized as follows: We describe the different data sets used to study the multi-phase ISM, the morphology of the galaxy for different tracers, and the magnetic structure of \m51 in Sec.\,\ref{Sec:ArchivalData}. We present the statistical methods used to parameterize them in Sec.\,\ref{Sec:Methods}. Sec.\,\ref{Sec:magnetic_results} is dedicated to the analysis of the magnetic and morphological spiral structure of \m51. In Sec.\,\ref{subsec:ISM_results} we analyze the properties of the ISM of \m51 as a function of the column density, FIR and radio polarization, and gas kinematics for multiple phases of the galactic gaseous disk. Finally, Sec.\,\ref{Sec:Discussion} and \,\ref{Sec:Conclusions} contain the discussion and conclusions respectively. In this paper, we assume a distance to \m51 of $8.58 \pm 0.28$ Mpc (1\arcsec $\sim$ 41.6 pc), based on the results from \citet{McQuinn2017} from the analysis of the tip of the red giant branch. 

\section{Archival Data}
\label{Sec:ArchivalData}

\subsection{Far-infrared polarimetry} 
\label{subsec:ArchivalData_fir}

Publicly available \sofiahawc\ observations of \m51 obtained under proposals with IDs 70\_0509 (Guaranteed Time Observations by the \hawc\ Team), 76\_0003 (Discretionary Director Time), and 08\_0260 (PI: Dowell, D.) from 2017 to 2020 (see Fig.\,\ref{fig:m51_apod}) were used. Table \ref{tab:table_HAWC} summarizes the observations combined in this work. Polarimetric observations with \hawc\ simultaneously measure two orthogonal components of linear polarization in two arrays of $32\,\times\,40$ pixels each. Observations were performed using Band~D with a characteristic central wavelength of $154$ \um, bandwidth of $34$ \um, pixel scale of $6\farcs90$, and beam size (FWHM) of $13\farcs6$ \citep{Harper2018}. For \m51, FWHM$_{\mathrm{\hawc}} = 0.565$ kpc. Observations were performed in a four-position dither square pattern with a distance of several detector pixels in the equatorial sky coordinates system (ERF) as shown in Table \ref{tab:table_HAWC} (column 8). The ERF for these observations was used, so a positive increase of angles is in the counterclockwise direction. In each dither position, four half-wave plate (HWP) position angles (PA) were taken in the standard sequence 5$^{\circ}$, 27.5$^{\circ}$, 50$^{\circ}$, and 72.5$^{\circ}$. These dither sequences of four HWP PA will be referred to as \emph{sets} hereafter. A chop-frequency of 10.2\,Hz was used, with the chop-angle, chop-throw, and nod time as listed in Table \ref{tab:table_HAWC}. The chop-angle is defined as the angle in the east of north direction along which the telescope chops with a given chop-throw. 

The total observation time (on-source time + overheads) is 7.21\,h, of which 2.78\,h is the time on-source. Low-quality exposures due to bad tracking, vignetting by the observatory's door in flight F547, or other technical issues at the time of observations are listed within the parenthesis in the sets column. The observations require time on the off-position due to the chop-nod technique as well as time to take internal calibrators right before and after each set of four HWP PA, which translates to an overhead of approximately $\times2.6$. Note that the previously published results of \m51 by \citet{Jones2020} used only a subset of the data presented here. Specifically, \citet{Jones2020} used observations from 70\_0509 and 76\_0003, with a total time of $4.6\,$h, where our observations encompass a total observing time of $7.21\,$h. We present here observations with larger integration time and better sensitivity, which allow us to perform a quantitative analysis of the inner and outer arms of \m51. In addition, our data reduction pipeline, supported by the SOFIA Science Center, is the most updated version (v2.3.2) in comparison with that used by \citet{Jones2020}, v1.3.0beta3. The new pipeline version corrects for background subtraction and propagation of errors from the timestreams, so no inflated errors using a $\chi^{2}$ analysis is required, and smoothing techniques have been implemented to account for correlated pixels. A direct comparison between both datasets is beyond the scope of this manuscript and we refer the reader to the update of the pipeline by the SOFIA Science Center for further details. 

\begin{deluxetable*}{cccccccccc}[ht!]
\tablecaption{Summary of \hawc\ polarimetric observations. \emph{Columns, from left to right:} a) Observation plan identifier. b) Observation date. c) Flight ID. d) Sea-level altitude during the observations (ft). e) Chop-angle (degrees) f) Chop-throw (arcsec). g) Time between nodding iterations (s). h) Amplitude of the dithering pattern (arcsec). i) Number of observation sets obtained (and rejected). j) Total observation time (on source + overheads) (s).\label{tab:table_HAWC}}
	\tablewidth{0pt}
\tablehead{\colhead{PlanID}	&	\colhead{Date} 	&	\colhead{Flight ID}	 &	\colhead{Altitude} &	\colhead{Chop-Angle} & \colhead{Chop-Throw} &	\colhead{Nod Time} &  \colhead{Dith. scale} & \colhead{\# Sets (bad)} &	\colhead{t$_{\mathrm{obs\_time}}$} \\
&	\colhead{(YYYYMMDD)}	&	&	\colhead{(ft)}	&	\colhead{($^{\circ}$)} & \colhead{(\arcsec)} & \colhead{(s)}	& \colhead{\arcsec} &  & \colhead{(s)} \\ 
(a) & (b) & (c) & (d) & (e) & (f) & (g) & (h) & (i) & (j)}
		\startdata
		 70\_0509   &	20171109	&	F450	&	43000	&	105	&	400	&	40	&	20	&	6   &	1263\\
		 		    &			    &		    &			&		&		&	50	&	33	&	4   &	1003\\
				    &			    &		    &			&		&		&	35	&	33	&	3   &	573\\
				    &	20101115	&	F452	&	43000	&	105	&	400	&	40	&	33	&	7(2)&	1495\\
				    &			    &		    &			&		&		&		&	35	&	8	&	1696\\
				    &			    &		    &			&		&		&	35	&	35	&	3	&	575\\
				    &	20171117	&	F454	&	43000	&	105	&	400	&	40	&	33	&	10	&	2122\\
		76\_0003    &	20190212	&	F545	&	42000	&	90	&	450	&	45	&	20	&	8	&	1852\\
			        &	20190220	&	F547	&	43000	&	90	&	450	&	50	&	20	&	8	&	2135\\
				    &			    &		    &			&		&		&		&	27	&	4(4)&	1338\\
		08\_0260	&	20200118	&	F651	&	43000	&	105	&	450	&	50	&	20	&	17	&	4200\\
				    &			    &		    &			&		&		&		&	28	&	8	&	1993\\
				    &	20200125	&	F653	&	43000	&	105	&	450	&	50	&	20	&	8(1)&	1993\\	
				    &			    &		    &			&		&		&		&	28	&	15	&	3715\\
		 \enddata
\end{deluxetable*}

The observations were reduced using the \textsc{hawc\_drp\_pipeline v2.3.2}. The pipeline procedure described by \citet{Harper2018} was used to background-subtract and flux-calibrate the data and compute Stokes parameters and their uncertainties. The final degree and PA of polarization are corrected for instrumental polarization, bias, and polarization efficiency. Typical standard deviations of the degree of polarization after subtraction of $\sim0.8$\% are estimated. We generated final reduced images with a pixel scale equal to half beam size, which corresponds to $6\farcs8$. Further analysis and high-level displays were performed with custom \textsc{python} routines, described in Sec.\,\ref{subsec:Methods_magnetic_pitch_angle}. We discard all those measurements with a signal-to-noise ratio (SNR) lower than 2 in polarized intensity ($p_{\mathrm{lim}}=0.05$, probability higher than 95\% of having signal higher than the noise level) in order to avoid regions dominated by noise. We also discard those pixels with a SNR in total intensity lower than $\sqrt{2}/p_{\mathrm{lim}} \sim 28.28$. We refer the reader to Sec.\,4 in \citet[][]{Gordon2018} for more details on \sofiahawc\ quality cuts. The inferred B-field orientation at 154 \um~is shown as streamlines using the Line Integral Convolution \citep[LIC,][]{LIC} technique in Fig. \ref{fig:m51_apod} (left panel), where only polarization measurements with $P/\sigma_{P} \ge 3$ were used, with $\sigma_P$ is the uncertainty in the polarization fraction. A resample scale of 5 and a contrast of 2 were used to compute the LIC image. The total intensity and polarization map is shown in Fig.\,\ref{fig:mag_pitch_full_hawc}. Inferred B-field orientations, and the \sofiahawc\ footprint at \mum154\ are shown in Figure \ref{fig:m51_arms}. In all figures, the observed PAs of polarization have been rotated by $90^{\circ}$. These observations are used to trace the magnetic fields in the cold and dense ISM regions of \m51. 

\subsection{Radio polarimetry}
\label{subsec:ArchivalData_radio}

We make use of the 3\,cm and 6\,cm radio polarimetric maps at a resolution of 8\arcsec from \citet{Fletcher2011}. These datasets were obtained using a combination of observations from the Karl G. Jansky Very Large Array (VLA) and the Effelsberg 100\,m single-dish radio-telescopes. We refer to the original paper for a complete description of the observations and data reductions of the datasets used in our work. Longer wavelength (18, 20\,cm) observations from \citet{Fletcher2011} can be strongly affected by Faraday rotation \citep{Beck2013}, and are thus not considered in this work. For our analysis, Stokes $IQU$ were convolved with a Gaussian kernel to match a FWHM$_{\mathrm{\hawc}} = 13\farcs6$ and reprojected to the \hawc\ observations. Then, the degree and PA of polarization and polarized flux were computed, accounting for the level of polarization bias as a function of the SNR \citep{Wardle1974}. We show the magnetic field streamlines of the 6\,cm dataset in Fig.\,\ref{fig:m51_apod} (right panel), compared to those of the \mum154/\hawc\ observations (left panel). A resample scale of 5 and a contrast of 2 were used to compute the LIC image. Final inferred B-field orientations at 3\,cm and 6\,cm are shown in Fig.\,\ref{fig:mag_pitch_full_radio} middle and bottom panels respectively, where the observed PAs of polarization have the same length. To avoid biased results due to the number of measurements across the galaxy, we only use radio polarization measurements that are spatially coincident with the \hawc\ observations. The radio polarization maps are used to spatially correlate the polarization arising from synchrotron emission with that arising from thermal emission by means of magnetically aligned dust grains observed with \hawc, as detailed in Sec.\,\ref{Sec:magnetic_results}.

\subsection{\co\ and \hi\ observations}
\label{subsec:ArchivalData_COHI}

\lineco\ observations were obtained from the Plateau de Bure interferometer (PdBI) and Arcsecond Whirpool Survey (PAWS\footnote{PAWS data at \url{https://www2.mpia-hd.mpg.de/PAWS/PAWS/Home.html}}), which uses the PdBI and IRAM-30\,m data to image at high angular resolution the emission from the molecular gas disk in \m51. Data are described in \citet{Pety2013} and \citet{Colombo2014}. Specifically, we used moments 0 (integrated emission line) and 2 (intensity weighted dispersion, velocity dispersion) of the \lineco\ emission line at angular resolutions of $6\arcsec$. For our analysis, moments 0 and 2 were convolved using a Gaussian kernel to match the \hawc\ beam size of 13.6\arcsec and then reprojected to the grid of the \hawc\ observations. \hi\ data were obtained from The \hi\ Nearby Galaxy Survey (THINGS\footnote{THINGS project: \url{https://www2.mpia-hd.mpg.de/THINGS/Data.html}}) described in \citet{Walter2008}. Moments 0 and 2 were used to trace the neutral gas in the disk of \m51. For our analysis, these observations were processed using the same method as \lineco\ observations. 

Both \lineco\ and \hi\ datasets are used to trace the velocity dispersion as a proxy of the turbulence in the molecular and neutral gas. We note that the \lineco\ integrated emission-line images of IRAM-30\,m at a resolution of $23\arcsec$ cover the full FOV of the HAWC+ observations. However, due to the low angular resolution of the IRAM-30\,m  observations, any comparison between structures of the galaxy (arms, interarms) and polarization observations are not physically meaningful as structures are hardly distinguished. In addition, we used THINGS \hi\ $21\,$cm datasets to generate the morphological mask that separates the arm and interarm regions (see Sec.\,\ref{subsec:Methods_mask}). 

\subsection{Column density map}
\label{subsec:ArchivalData_TNh}

Column density map, \columndensity, was estimated using the integrated emission-line (moment 0) neutral, HI, and molecular, \lineco, gas of M51. The IRAM-30 m \lineco\ integrated emission-line observations with a resolution of $23\arcsec$ were used for this analysis. These observations cover the full FOV of the HAWC+ observations, while the $6\arcsec$ observations used in Section \ref{subsec:ArchivalData_COHI} only cover the central $\sim3\arcmin$ of M51. Specifically, we used the following HI, and \lineco, conversions to $N_{\mathrm{HI}}$, and $N_{\mathrm{2H_{2}}}$:

\begin{equation}
N_{\mathrm{HI}} = 1.105 \times 10^{21} \frac{I_{\mathrm{HI}}}{\mathrm{FWHM}_{\mathrm{HAWC+}}^{2}}~(\mathrm{cm}^{-2})
\end{equation}
\noindent
by \citet{hunter2012}, where I$_{\mathrm{HI}}$ is the integrated emission line (moment 0) of \hi\ in units of Jy beam$^{-1}$ m s$^{-1}$, and FWHM$_{\mathrm{HAWC+}}$ is the beamsize of HAWC+ at 154 \um~in units of arcsec.

\begin{equation}
N_{\mathrm{2H_{2}}} = X_{\mathrm{CO}} I_{\mathrm{CO}}~(\mathrm{cm}^{-2})
\end{equation}
\noindent
by \citet{Bolatto2013}, where I$_{\mathrm{CO}}$ is the integrated emission line (moment 0) of \lineco\ in units of K km s$^{-1}$, and $X_{\mathrm{CO}}$ is the conversion factor of value $2 \times 10^{20}$ cm$^{-2}$ (K km s$^{-1}$)$^{-1}$. 

Final column density is estimated such as $\columndensity = N_{\mathrm{HI}} + N_{\mathrm{2H_{2}}}$. Column density values range from $\logten(\columndensity\ [$cm$^{-2}]) = [20.4-22.11]$, in agreement with \citet[][]{MC2012}. The computed column density is used for the analysis of the multi-phase ISM as well as the estimation of the star formation rate.


\section{Methods}
\label{Sec:Methods}

\subsection{Magnetic and morphological pitch angle analysis}\label{subsec:Methods_magnetic_pitch_angle}
\begin{figure}[]
 \begin{center}
\centering
\includegraphics[trim={5 0 0 5}, clip, width=0.5\textwidth]{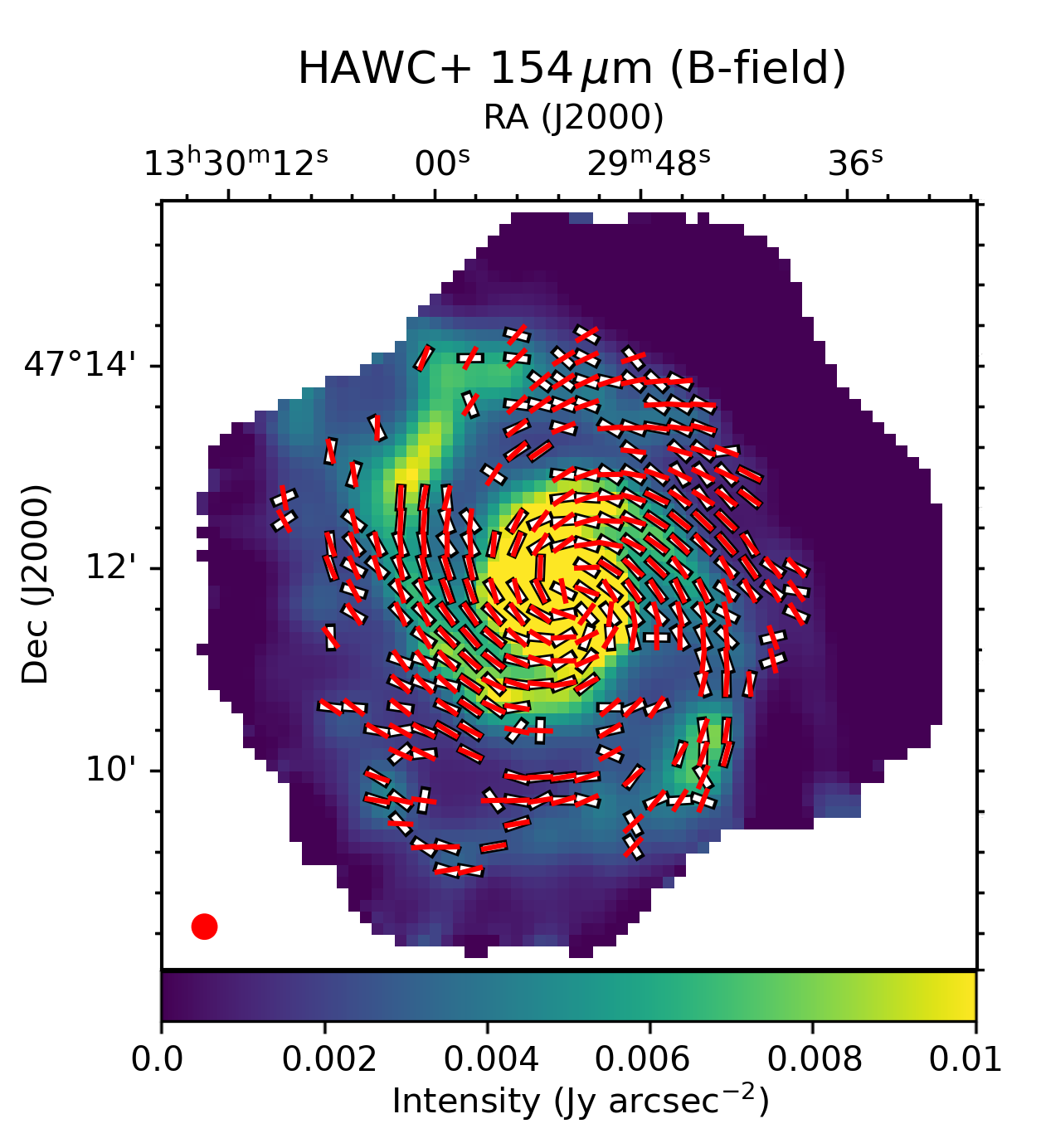}
\caption{\emph{Top to bottom:} B-field orientation maps of \m51 from 154 $\mu$m\,/~\hawc~(this work) polarimetric observations. The white lines represent the B-field orientations, where the lengths have been normalized to unity. Red lines are the average polarization orientation estimated from the magnetic pitch angle profile. The background color map represents the total surface brightness intensity in their respective wavelengths.} 
\label{fig:mag_pitch_full_hawc}
\end{center}
\end{figure}

\begin{figure*}[]
 \begin{center}
\centering
\includegraphics[trim={5 0 0 5}, clip, width=0.5268\textwidth]{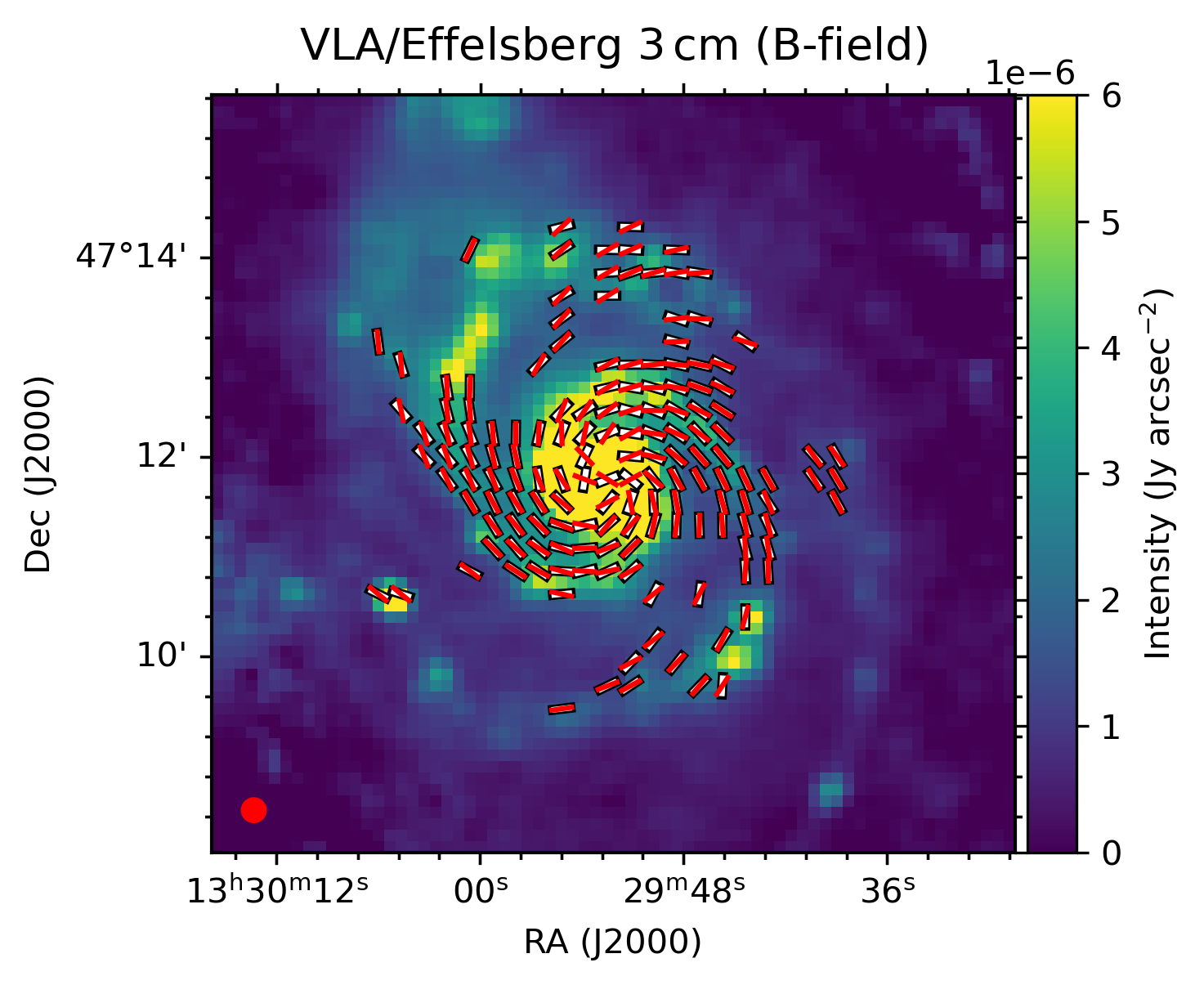}
\includegraphics[trim={56 0 0 5}, clip, width=0.4632\textwidth]{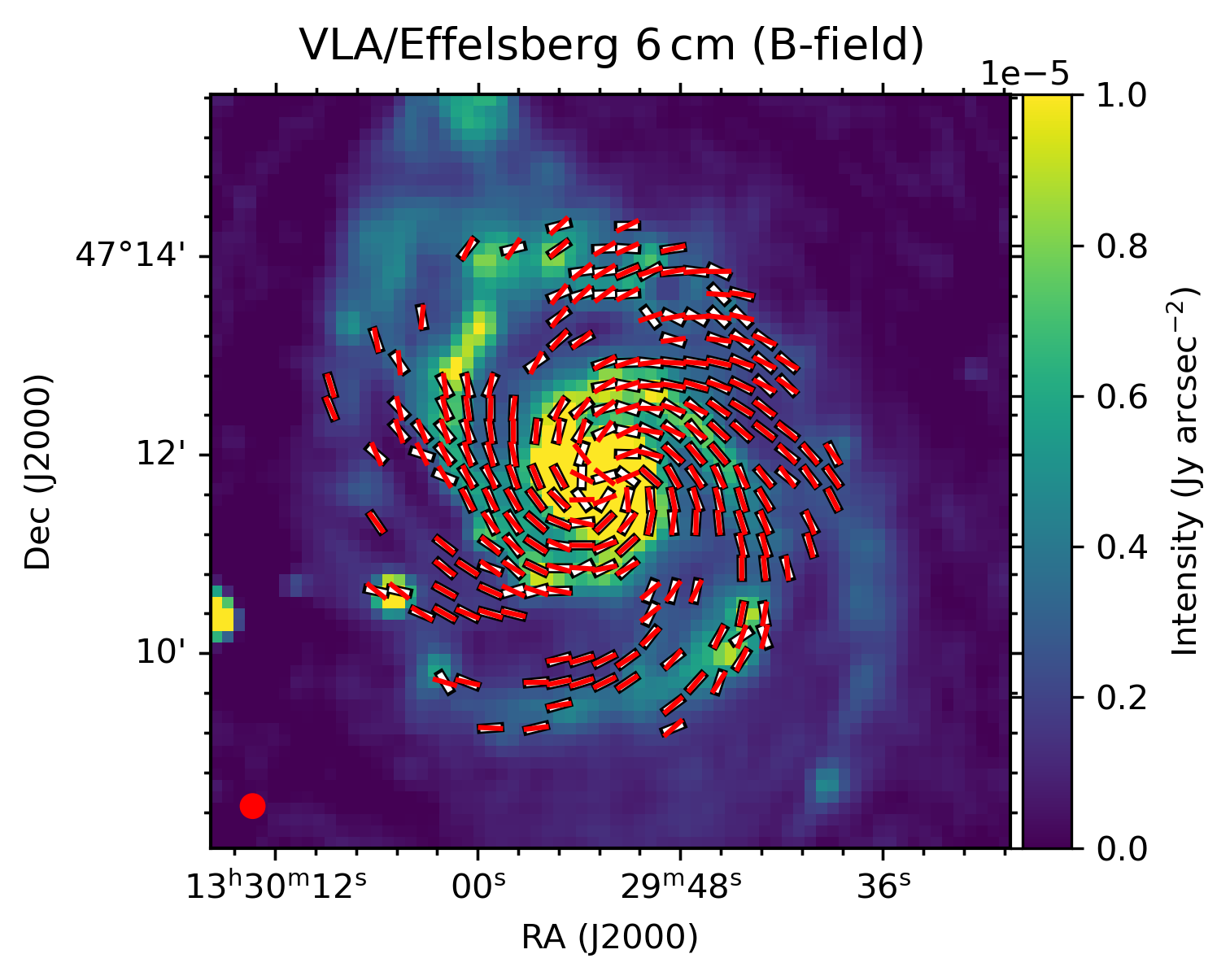}
\caption{B-field orientation maps of \m51 from radio polarimetric observations at 3\,cm (left) and 6\,cm (right) \citep{Fletcher2011}. The white lines represent the B-field orientations, where the lengths have been normalized to unity. Red lines are the average polarization orientation estimated from the magnetic pitch angle profile. The background color map represents the total surface brightness intensity in their respective wavelengths.} 
\label{fig:mag_pitch_full_radio}
\end{center}
\end{figure*}

\begin{figure}[t!]
\begin{center}
\includegraphics[trim={0 40 0 40}, clip, width=0.5\textwidth]{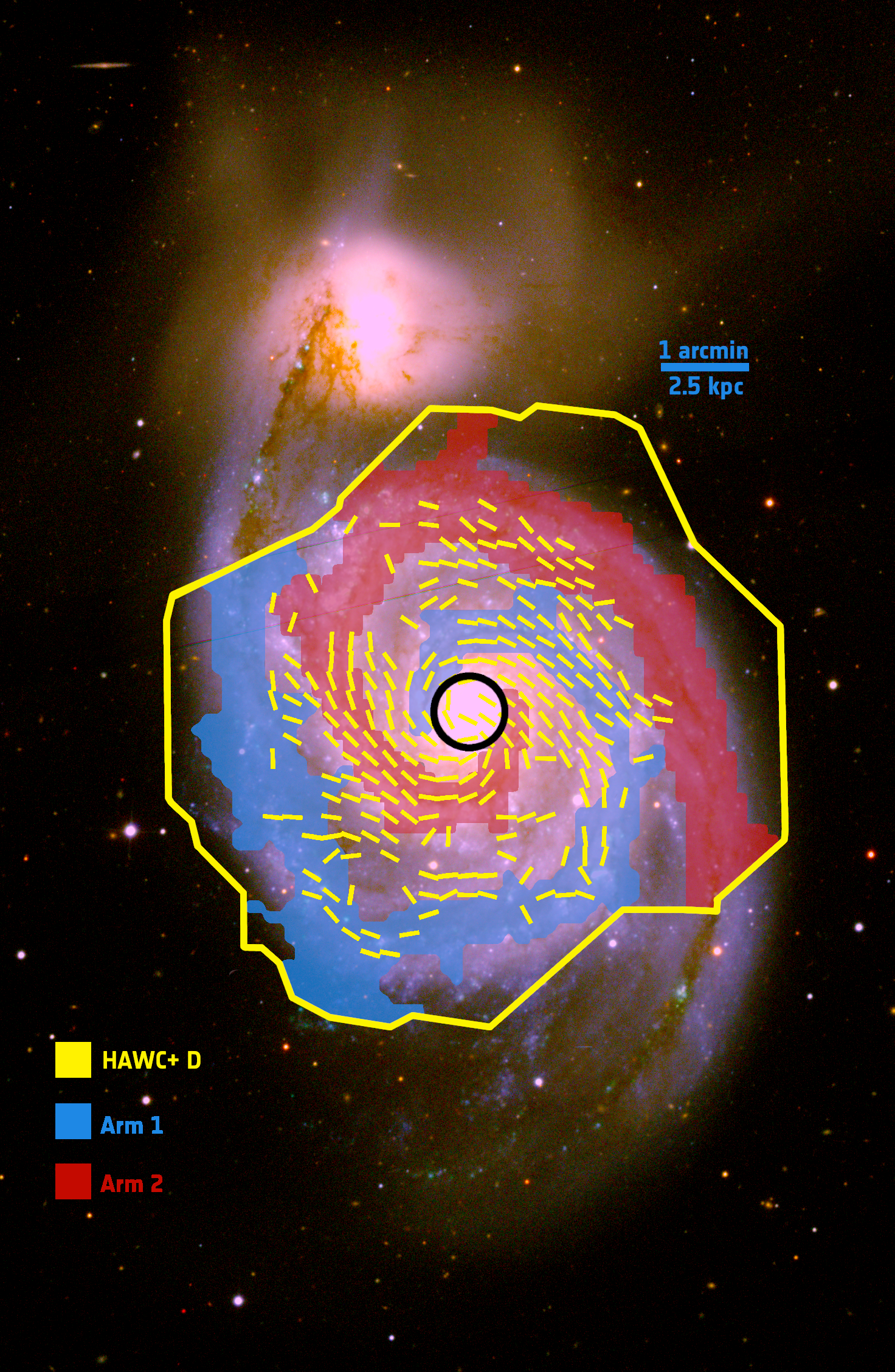}
\caption{General description of the \m51 regions for this analysis. \emph{Background:} RGB image based on SDSS $gri$ imaging \citep{Gunn2006}. \emph{In yellow:} Footprint and B-field orientations with their lengths normalized to unity from \hawc\ observations. \emph{Red and blue shaded regions:} Mask and arms definitions, see Sec.\,\ref{subsec:Methods_mask}.  \emph{Black circle:} Limiting radius of the M51 core region. See the legend for labeling and physical scaling.} 
\label{fig:m51_arms}
\end{center}
\end{figure}

\begin{figure*}[ht!]
 \begin{center}
 \includegraphics[trim={0 0 0 0}, clip, width=0.99\textwidth]{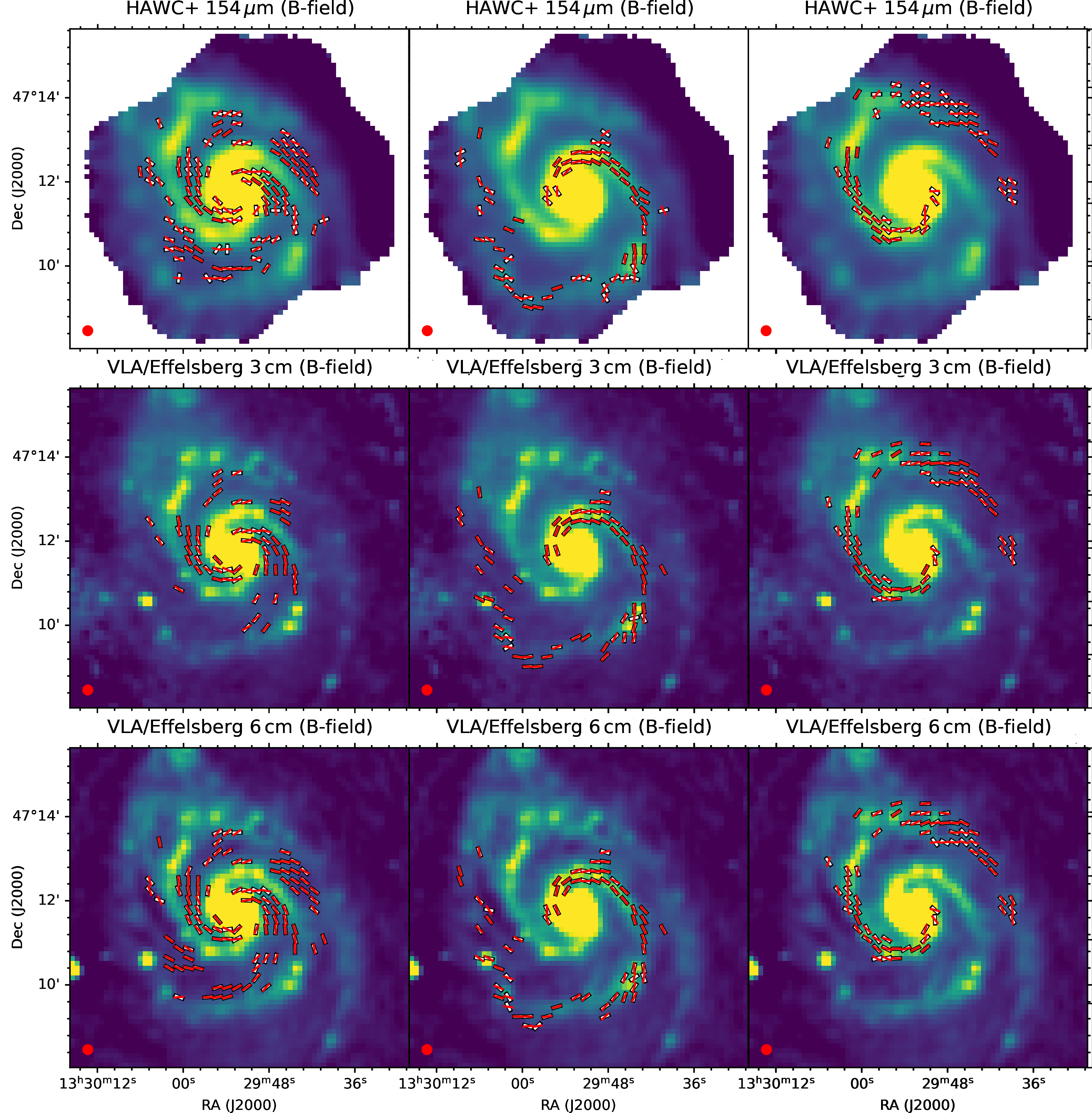}
\caption{B-field orientation maps of \m51 of 154$\,\mu$m\,/\,\hawc (top row), radio polarimetric observations at 3\,cm (middle row) and 6\,cm (bottom row) \citep{Fletcher2011}, for the interarm (left column), Arm 1 (middle column), and Arm 2 (right column) morphological regions defined in Sec.\,\ref{subsec:Methods_mask} (see Fig.\,\ref{fig:m51_arms}). The white lines represent the measured B-field orientations, for which the lengths have been normalized to unity. Red lines show the average polarization orientation estimated from the magnetic pitch angle profile. Total intensity is displayed in the background. See the colorbar in Figs.\,\ref{fig:mag_pitch_full_hawc} and \ref{fig:mag_pitch_full_radio} for reference.}
\label{fig:mag_pitch_arms_interarms}
\end{center}
\end{figure*}

In this section, we describe the methodology used to estimate the magnetic and morphological pitch angles of \m51. The algorithm described here is used to analyze the FIR, radio polarimetric observations, and the velocity fields that result from the wavelet analysis of their total intensity maps (see Sec.\,\ref{subsec:Methods_wavelet_analysis}).

The magnetic pitch angle profile is estimated as follows: 
\begin{enumerate}
    \item The debiased polarization level and its associated uncertainty are computed using the Stokes $IQU$ parameters and their uncertainties $\delta I, \delta Q, \delta U$:
\begin{equation}
\label{eq:polarization_level}
P_{\rm debias} = \sqrt{P^2 - \delta P^2}
\end{equation}

where:

\begin{equation}
\label{eq:polarization_level2}
P = \sqrt{\left(\frac{Q}{I}\right)^2 + \left(\frac{U}{I}\right)^2}
\end{equation}

and:

\begin{equation}
\label{eq:polarization_level_uncert}
    \delta P = \frac{1}{I} \sqrt{\frac{(Q \cdot \delta Q)^2 + (U \cdot \delta U)^2}{Q^2 + U^2} + \delta I^2\frac{Q^2 + U^2}{I^2}}
\end{equation}

\item To reproject the observations, our method requires the coordinates of the galactic center ($\alpha$, $\delta$), the galactic disk inclination, $i$, and tilt angle, $\theta$. Morphological parameters were adopted from \citet{Colombo2014} and have the following values: $\alpha=202.4699^{\circ}$, $\delta=+47.1952^{\circ}$, $i=22\pm5^{\circ}$, $\theta=-7.0\pm3.0^{\circ}$, where $\alpha$ and $\delta$ are the equatorial coordinates of the center of M51, $i$ is the apparent inclination of the disk with the line of sight (where face-on corresponds to $i =0^{\circ}$), and $\theta$ is the apparent tilt angle of the major axis with positive values in the east of north direction (where north corresponds to $\theta =0^{\circ}$).

\item We compute the radius ($R$) and azimuthal position ($\phi$) of every pixel in galactocentric coordinates, where all pixels are assumed to be located in the galactic plane ($z=0$). Radial and angular masks are generated with the same tilt angle and inclination as \m51.

\item An azimuthal angular mask is created. This is generated such that the deprojected vector at each pixel location is perpendicular to the radial direction. We will refer to this idealized field as the \emph{zero pitch angle} field or $\Xi$. The observed debiased polarization measurements ($P_{\mathrm{debias}}$) are deprojected to the galactic plane frame ($P^{\prime}$) using two chained rotation matrices, one to account for the inclination of the galaxy, $R_{x}[i]$, and a second for the tilt angle, $R_{z}[\theta]$. The projected matrix is estimated to be $P^{\prime}= R_{x}[i]R_{z}[\theta] P$.

\item A method was devised to account for the $180^{\circ}$ degeneracy in the direction of \hawc's PAs. An effective averaging of the directions of several pixels requires resolution of the degeneracies. The $\Xi$ zero pitch angle frame from the previous step is used to correct the PAs, setting them arbitrarily to a common outward-pointing direction. This is performed by measuring the relative angle difference with $\Xi$, and adding or subtracting 180$^{\circ}$ as required. Note that the result is independent of the reference angle of choice, and it is only used for averaging purposes. As a consequence of this correction, the magnetic pitch angle profile also suffers a 180$^{\circ}$ degeneracy.

\item We project the measured B-field orientations to a new reference frame in which the galaxy is observed face-on. We used the morphological parameters of inclination and tilt angles ($i$, $\theta$), and the measured PAs of the B-field orientation corrected for 180$^{\circ}$-degeneracy from the previous step. 

\item The pitch angle $\Psi(x,y)$ is calculated as the difference between the measured PAs of the B-field orientation and the $\Xi$ vector field. 

\item $\Psi(x,y)$ is then averaged at each radius from the core. The radial bins are linearly spaced, and the number of them is optimized as a compromise between SNR and spatial resolution. The angular average is performed as follows: 
\begin{equation}
\label{eq:vector_averaging}
    \mPsi(R) = {\rm atan2} \left( \frac{<\cos\,\Psi(x,y)>}{<\sin\, \Psi(x,y) >} \right)
\end{equation}

where the $<>$ operator indicates a robust median value (based on Monte Carlo simulations) and \mPsi$(R)$\ is the averaged magnetic pitch angle value for a certain radial bin. For each map, the process detailed below is repeated 10\,000 times, using Monte Carlo simulations to include the uncertainties of the tilt angle, inclination, and the Stokes parameters. An independent Gaussian probability distribution for each parameter is assumed, with a standard deviation $\sigma$ equal to their uncertainties. Each of these Monte Carlo simulations produces a magnetic pitch angle array. The results of the Monte Carlo simulations are stored in a data cube, which are later used to calculate the pitch angle profiles (\mPsi$(R)$). Finally, for each radial bin, the median \mPsi$(r_i)$ value and the 68\% and 95\% (equivalent to the 1$\sigma$, 2$\sigma$) uncertainty intervals are computed. For all the analyses, we will consider a critical level of at least $p=0.05$ (95\%) to declare statistical significance. 
\end{enumerate}

This method was implemented in \texttt{Python} and is available on the project website\footnote{SOFIA Legacy Project for Magnetic Fields in Galaxies: http://galmagfields.com/} \citep{borlaff_2021_5116134}. In Appendix \ref{appendix:mock_pitch} we test this method over a set of 8 mock \hawc\ polarization observations using different tilt angles, inclinations, SNR, and magnetic pitch angles. Our method allows us to estimate the magnetic pitch angle profile without strong dominating systematic errors at an uncertainty level of $p>0.05$. Using mock polarization observations with a $P/\sigma_{p} \ge 2$, an accuracy $\le5^{\circ}$ is expected in the \mPsi$(R)$. 

Our magnetic pitch angle estimation method entails processing the data on a pixel-by-pixel basis, allowing the user to separate different regions of the galaxy by using masks. Section \ref{Sec:magnetic_results} describes how this masking technique was leveraged to produce measures of the magnetic pitch angle orientation for different regions in \m51: a) Full-disk. b) Arm vs. Interarm. c) Arm\,1 vs. Arm\,2.

\subsection{Morphological wavelet analysis}
\label{subsec:Methods_wavelet_analysis}

To compare the magnetic spiral structure with the morphology of the total intensity using several tracers, a measure of the pitch angle of the spiral arms is required. To identify the orientation of the spiral arms in the \mum154, 3\,cm, and 6\,cm observations, we take advantage of the technique applied in \citet{Patrikeev2006, Frick2016} -- the two-dimensional anisotropic wavelet transform -- for the identification of elongated structures.
Wavelet transforms allow recovery of the position angle of the maximum amplitude wavelet at each pixel where the signal is significant, returning a map of wavelet orientations representing the local pitch angle of the image. The wavelet scale used is 13.8\arcsec, twice the size of the pixel scale. We refer to the original articles \citep{Patrikeev2006, Frick2016} and the references therein for a complete explanation of the method and its mathematical description.


\begin{figure}[]
\begin{center}
\includegraphics[trim={0 0 10 0}, clip, width=0.5\textwidth]{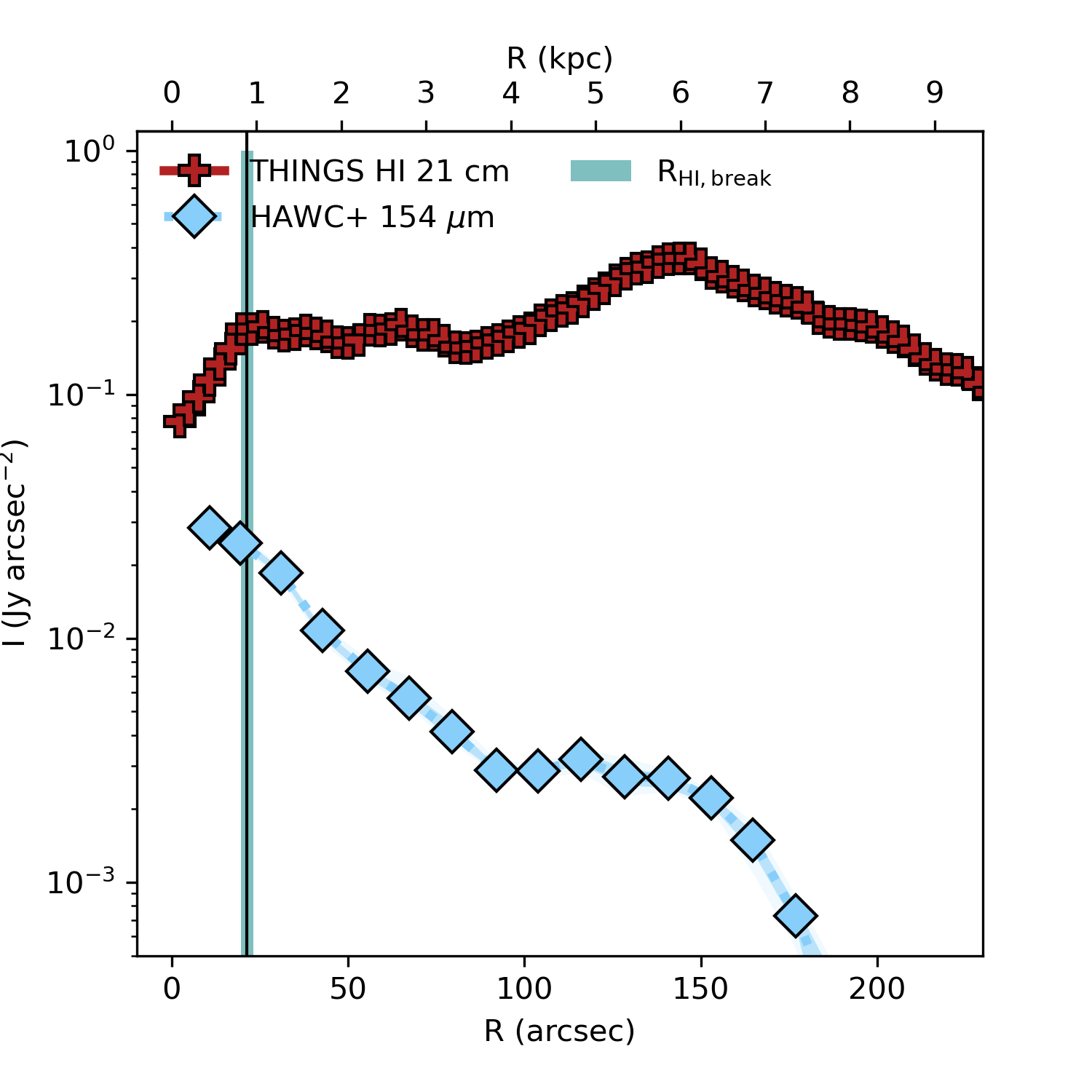}
\caption{Surface brightness profile analysis of the \hi\ (\emph{red crosses}) and FIR components (\emph{blue diamonds}) of M51. The vertical black solid line and the teal region represent the location of the \hi\ surface brightness break ($R_{\rm{break}}=21.2^{+1.8}_{-1.6}\arcsec$, $0.88^{+0.08}_{-0.07}$ kpc), as estimated using {\tt{Elbow}} \citep{Borlaff2017}.} \label{fig:HI_profile}
\end{center}
\end{figure}

In Sec.\,\ref{subsubsec:results_morphological_pitch_angle} we present the wavelet transform maps for the $154$ \um\ FIR, 3 and 6\,cm radio intensity images, \lineco, and 21\,cm \hi\ observations. The lines inside the spiral arms closely follow the local structure of the spiral arms for each tracer. Conveniently, the orientation of the wavelet transform can be decomposed into its corresponding Stokes $Q$ and $U$, allowing analysis of their structure using the same pitch angle method and software described in Sect.\,\ref{subsec:Methods_magnetic_pitch_angle}.

\subsection{Morphological masks}
\label{subsec:Methods_mask}

The THINGS 21\,cm observations of the \hi\ gas disk (Sec.\,\ref{subsec:ArchivalData_COHI}) and the morphological wavelet analysis from Sec\,\ref{subsec:Methods_wavelet_analysis} are used to separate the different morphological regions of \m51 (spiral arms, interarms, and core). The resulting masks are shown in Fig.\,\ref{fig:m51_arms}, and the polarization fields separated by the morphological masks for the different wavelengths are shown in Fig.\,\ref{fig:mag_pitch_arms_interarms}. We choose the \hi\ gas to define the arm-interarm mask based on two factors: 1) we have high-resolution, deep observations of \m51, and more importantly 2) it allow us to trace the spiral arms closer to the inner core of the galaxy, something that is not possible with lower resolution data such as those of our FIR observations.

As a first step, the core region is defined by studying the surface brightness profile of the \hi\ disk (see Fig.\,\ref{fig:HI_profile}). The inner region of the profile ($R<100\arcsec$, $<4.16$ kpc) shows a nearly constant surface brightness, with a notable decrease of the 21 cm emission at $R<22\arcsec$ ($<0.9$ kpc), corresponding to the core region. We fit the location of the break-in surface brightness profile using the software {\tt{Elbow}}\footnote[2]{{\tt{Elbow}}: a statistically robust method to fit and classify the surface brightness profiles.  The code is publicly available at GitHub (\url{https://github.com/Borlaff/Elbow})} \citep{Borlaff2017}, obtaining a break radius of $R_{\rm{break}}=21.2^{+1.8}_{-1.6}\arcsec$ ($0.88^{+0.08}_{-0.07}$ kpc), statistically significant at a level of $p<10^{-5}$. We define this region as the radial limit for the core region in the morphological mask. 

In a second step, the intensity image of the \hi\ observations is analyzed using the wavelet transformation method (see Sec.\,\ref{subsec:Methods_wavelet_analysis}). The amplitude of the wavelet transformed image provides us with a probability map of the spatial distribution of elongated structures, like spiral arms. We define as statistically significant (and thus, part of a spiral arm) every pixel whose associated wavelet amplitude is higher than twice the standard deviation ($2\sigma$) of the background noise in the wavelet transformed image. By doing this, we only select regions that have at least a $\sim95\%$ probability to be part of an elongated \hi\ structure. Finally, we separate the two spiral arms using a visually defined polygon over the resulting mask, taking into account the morphology of the galaxy in the FIR, \lineco, 3 and 6\,cm, and \hi\ datasets (see Fig.\,\ref{fig:m51_arms}). 


\section{Magnetic pitch angle results}
\label{Sec:magnetic_results}


This section describes the results of the magnetic pitch angle profile for different wavelengths (\mum154, 3\,cm, and 6\,cm) and morphological regions (full disk, arms, and interarms). In order to avoid systematic effects in the results caused by the different spatial resolutions from different datasets, we convolve and rebin the radio observations to the \sofiahawc\ \mum154\ resolution (FWHM$_{\mathrm{\hawc}}=13.6\arcsec$). In addition, we use the same location of the polarization measurements in FIR and in radio observations, which allows us to study the same LOS at both wavelengths regimes. As the FIR observations have lower SNR than the radio observations, we select statistically significant polarization measurements, $P/\sigma_P\ge2$. The common resolution scale enables the comparison of maps at the same positions, a particularly critical requirement for the analysis of the arms and interarms regions (Secs.\,\ref{subsubsec:results_pitch_angle_full_arms} and \ref{subsubsec:results_pitch_angle_full_interarms}).

\subsection{Radial axisymmetric profile of the magnetic pitch angle: Full Disk}\label{subsubsec:results_pitch_angle_full}

\begin{figure*}[t]
 \begin{center}
\includegraphics[trim={60 25 82 0}, clip, width=\textwidth]{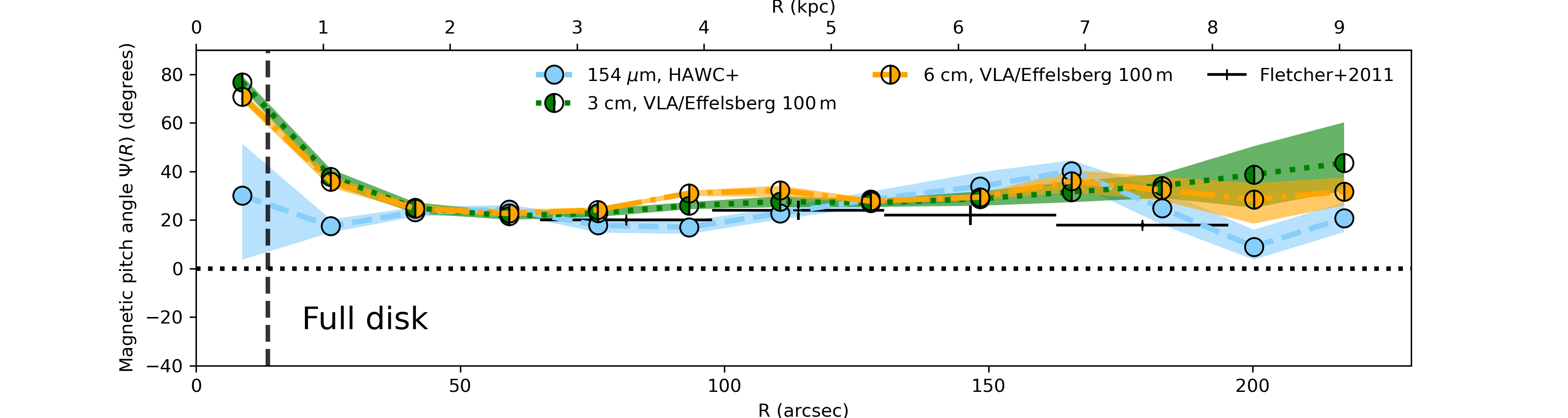}

\includegraphics[trim={60 25 82 20}, clip, width=\textwidth]{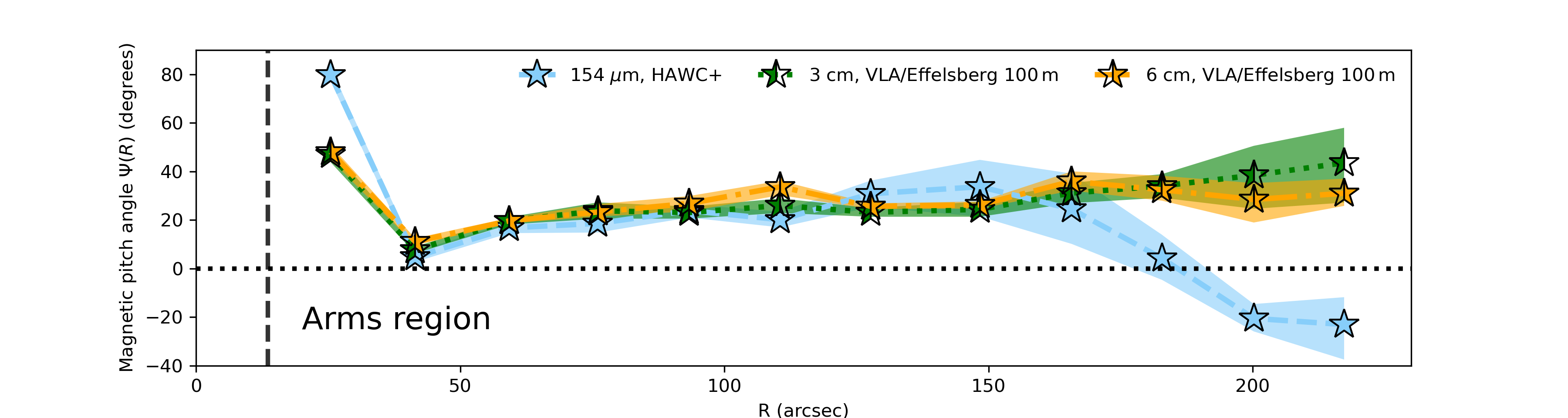}

\includegraphics[trim={60 0 82 20}, clip, width=\textwidth]{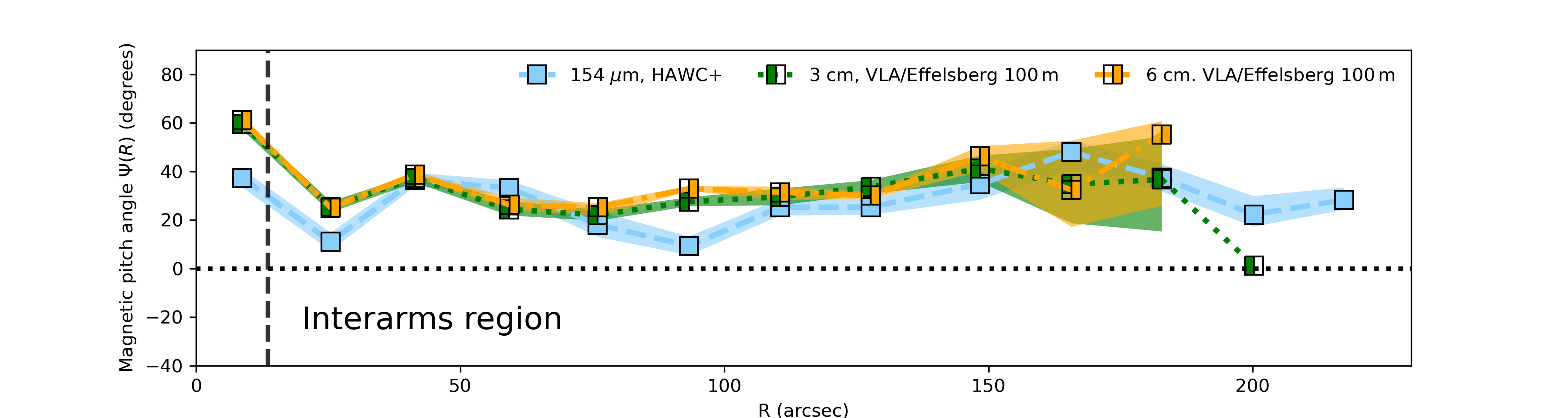}

\caption{Magnetic pitch angle profiles for the FIR \mum154 \hawc\ observations (this work) and the 3\,cm and 6\,cm radio polarimetric observations \citep{Fletcher2011} of \m51. On the vertical axis, we represent the average magnetic pitch angle profile $\Psi(R)$ per radial bin, as a function of radius. \emph{Top panel:} Profile for the full disk region, assuming axisymmetry and homogeneity. \emph{Central panel:} Arms region profile. \emph{Bottom panel}: Interarm region profile. See the legend for the color and linetype. The central beam of the observations is shown as a black vertical dashed line in each figure.} 
\label{fig:mag_pitch_full_radialprofile}
\end{center}
\end{figure*}


The properties of the magnetic pitch angle across the \m51\ galactic disk are first analyzed across the full disk mask, with no partition into arm and interarm regions (see Figs.\,\ref{fig:mag_pitch_full_hawc} and \,\ref{fig:mag_pitch_full_radio}). The top panel of Fig.\,\ref{fig:mag_pitch_full_radialprofile} shows the radial profiles of the magnetic pitch angles for the full disk after applying the methodology presented in Section \ref{subsec:Methods_magnetic_pitch_angle}. For the radio polarization observations, we find that the magnetic pitch angle profile is mostly flat up to a radius of $220$\arcsec\ ($9.15$ kpc). Similarly, for our FIR observations, the magnetic pitch angle is mostly flat up to a radius of $160$\arcsec\ ($6.66$ kpc), for galactocentric radii larger than $R>160\arcsec$ ($>6.66$ kpc) we find signs of a drop in the magnetic pitch angle profile. The central beam of the observations is shown as a black vertical dashed line in each figure. The pitch angle increases at the center due to resolution effects produced by the small number of polarization measurements available at the core.

For the full disk (Figs.\,\ref{fig:mag_pitch_full_hawc}, \,\ref{fig:mag_pitch_full_radio} and \ref{fig:mag_pitch_full_radialprofile}), we estimate an average magnetic pitch angle of $\FDPsiFIR=+23.9^{+1.2\circ}_{-1.2}$ for the \mum154/\,\hawc\ dataset. For the 3\,cm and 6\,cm observations we obtain $\FDPsithreecm=+26.0^{+0.9\circ}_{-0.8}$ and $\FDPsisixcm=+28.0^{+0.8\circ}_{-0.6}$, which are compatible at some of the bins with the results from \citet[][see their Table A1]{Fletcher2011}. The 3 and 6\,cm magnetic pitch angle profiles presented in this work are slightly higher on average than those presented in \citet{Fletcher2011} but compatible on the low end in some regions. Specifically, comparing Fig.\,\ref{fig:mag_pitch_full_radialprofile} with line 3 of Table A1 in \citet{Fletcher2011}, there is reasonable agreement within the error bars in the first three radial ranges. Only in the outer range ($6.0$--$7.2$ kpc), the absolute value of the pitch angle from \citeauthor{Fletcher2011} decreases, while it increases in Fig.\,\ref{fig:mag_pitch_full_radialprofile}. Nevertheless, there is a substantial difference between the two analyses that we must consider: First, their profiles combine polarization observations from 3, 6, 18, 20\,cm datasets, while we are analyzing the 3 and 6\,cm wavelengths independently. Second, their pitch angle ($p_0$) represents the average pitch angle for the dominant of two different large-scale modes of the regular magnetic field, while our profiles represent a non-parametric measurement of the magnetic pitch angle, including variations on smaller scales. For these reasons, we should consider a direct comparison between both profiles with care.



\subsection{Radial magnetic pitch angle profile - spiral arms}\label{subsubsec:results_pitch_angle_full_arms}

\begin{figure*}[t]
 \begin{center}
\includegraphics[trim={60 25 82 0}, clip, width=\textwidth]{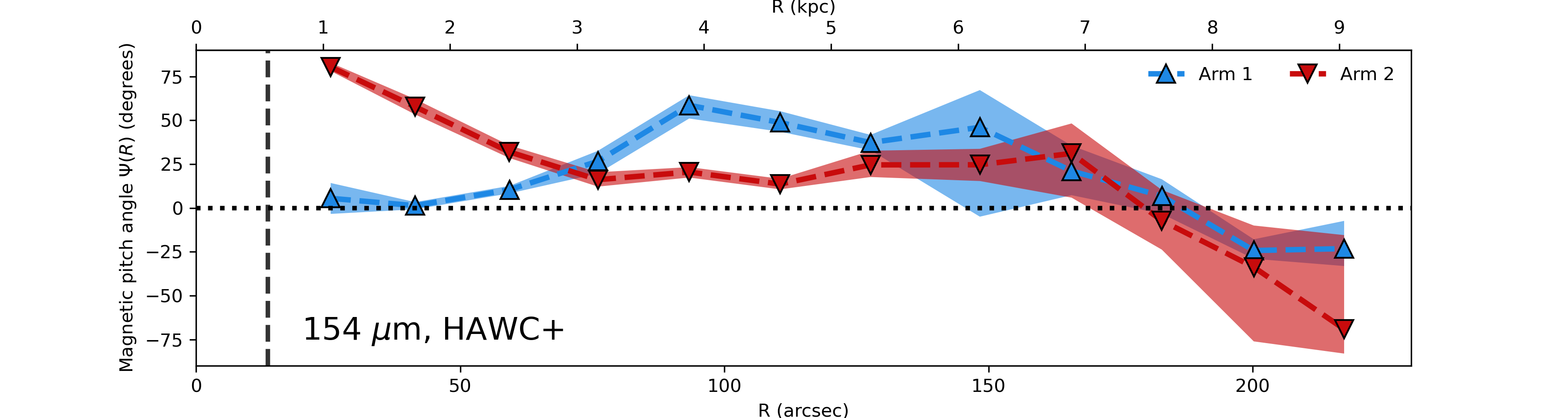}

\includegraphics[trim={60 25 82 20}, clip, width=\textwidth]{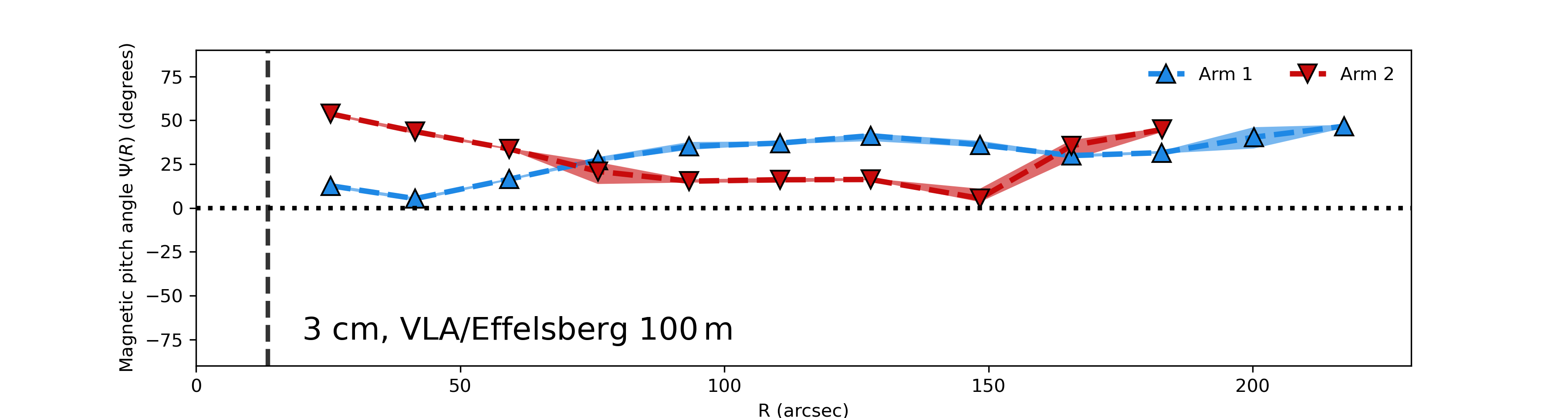}

\includegraphics[trim={60 0 82 20}, clip, width=\textwidth]{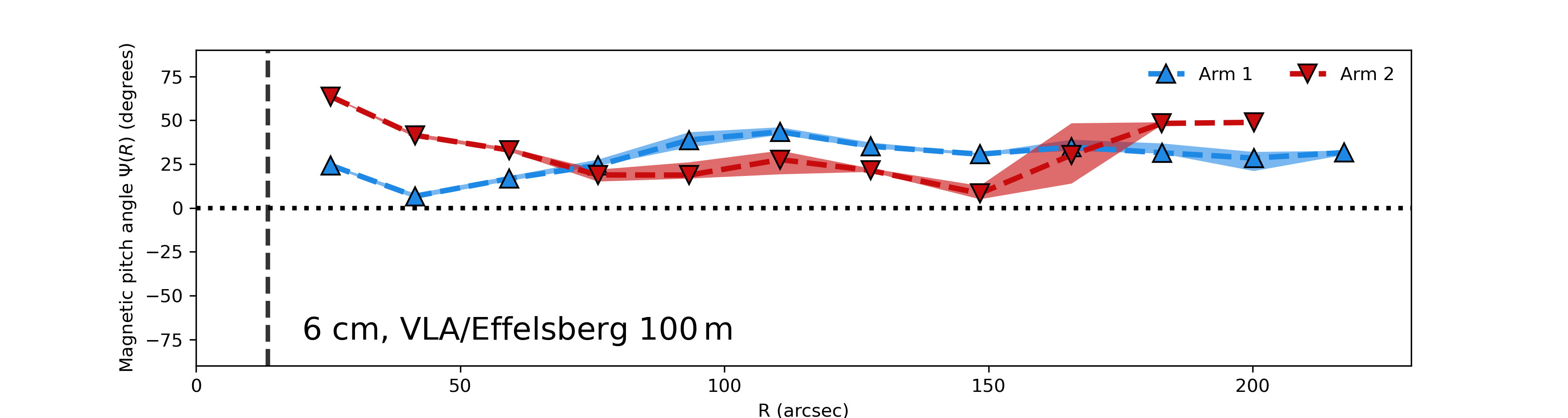}

\caption{Magnetic pitch angle profiles for the spiral Arm 1 (blue) and Arm 2 (red) of M51 as a function of wavelength. In each panel we present the average magnetic pitch angle profile $\Psi(R)$ per radial bin, as a function of radius. \emph{Top panel:} Profile for the \mum154/\hawc\ observations. \emph{Central panel:} Magnetic pitch angle profile for 3\,cm. \emph{Bottom panel}: Magnetic pitch angle profile for 6\,cm. See the legend for the color and linetype.} 
\label{fig:mag_pitch_arm}
\end{center}
\end{figure*}


Given the angular resolution of the FIR and radio observations, the interarm and arms regions can be separated and analyzed independently. Using the mask described in Sec.\,\ref{subsec:Methods_mask}, we generate three radial profiles of the magnetic pitch angle: Arm\,1, Arm\,2, and both spiral arms combined (`Arms region' in Fig. \ref{fig:mag_pitch_full_radialprofile}). We adopt the same notation for the spiral arms of \m51 as in \citet[][see their Fig.\,3]{Patrikeev2006}. From the outskirts of the galaxy, Arm 2 is the most northern arm close to M51b, while Arm 1 is the most southern arm (Fig.\,\ref{fig:m51_arms}). We show the polarization measurements used for each region in  Fig.\,\ref{fig:mag_pitch_arms_interarms}. The results of the magnetic pitch angle profile for both arms combined are shown in the central panel of Fig.\,\ref{fig:mag_pitch_full_radialprofile}, labeled as `Arms region'. Figure \,\ref{fig:mag_pitch_arm} shows the magnetic pitch angle profiles for Arm 1 and Arm 2 separately at \mum154, 3\,cm, and 6\,cm. 

\begin{figure*}[t]
 \begin{center}
 
\begin{overpic}[trim={0 32 0 0}, clip, width=0.32\textwidth]{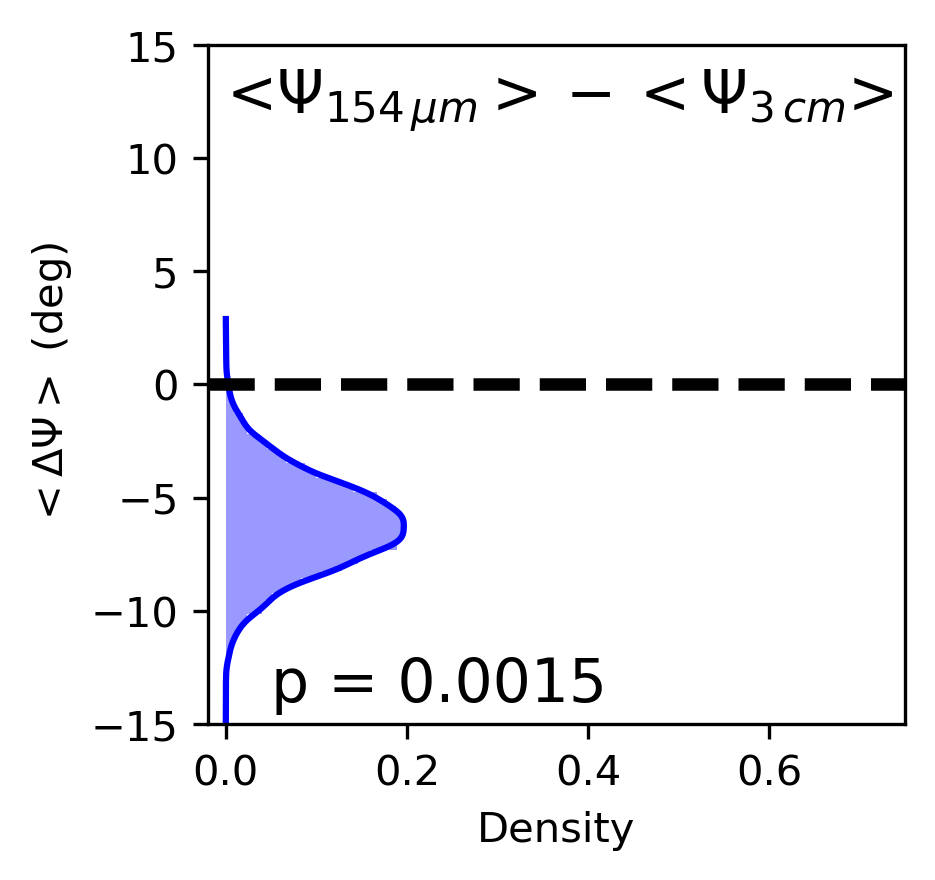} 
\put(25,60){\color{black} \colorbox{white}{\textbf{ARMS}}}
\end{overpic}
\begin{overpic}[trim={0 32 0 0}, clip, width=0.32\textwidth]{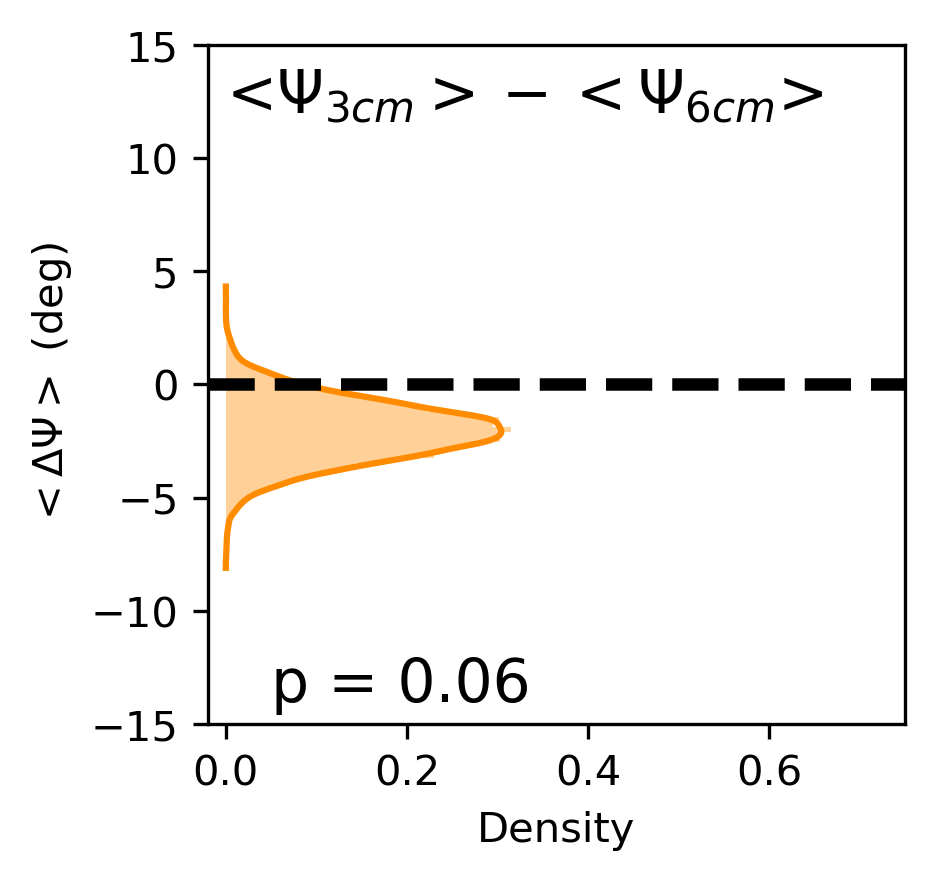} 
\put(25,60){\color{black} \colorbox{white}{\textbf{ARMS}}}
\end{overpic}
\begin{overpic}[trim={0 32 0 0}, clip, width=0.32\textwidth]{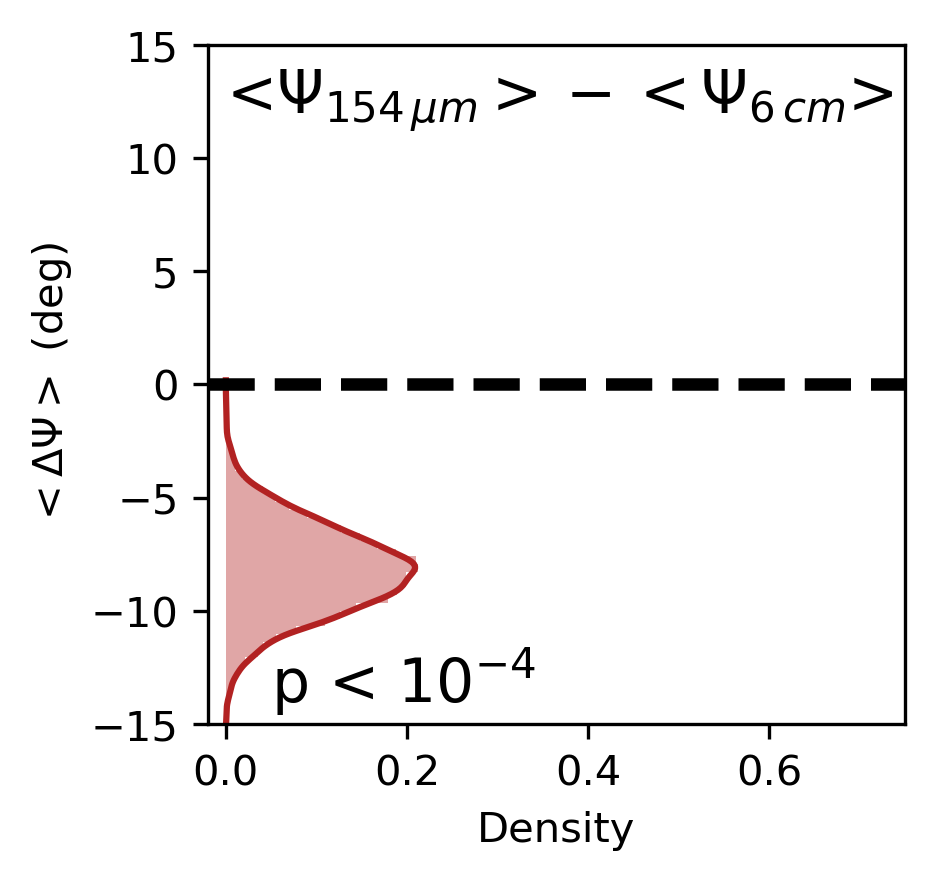} 
\put(25,60){\color{black} \colorbox{white}{\textbf{ARMS}}}
\end{overpic}

\begin{overpic}[trim={0 0 0 2}, clip, width=0.32\textwidth]{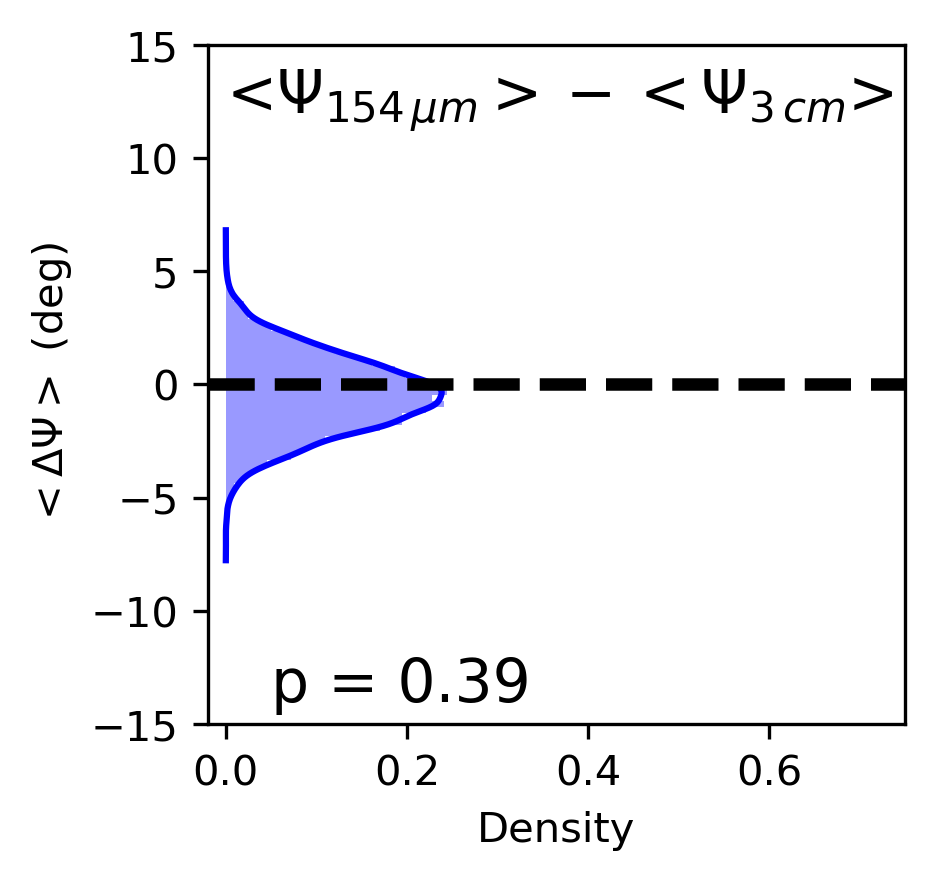} 
\put(25,74){\color{black} \colorbox{white}{\textbf{INTERARM}}}
\end{overpic}
\begin{overpic}[trim={0 0 0 2}, clip, width=0.32\textwidth]{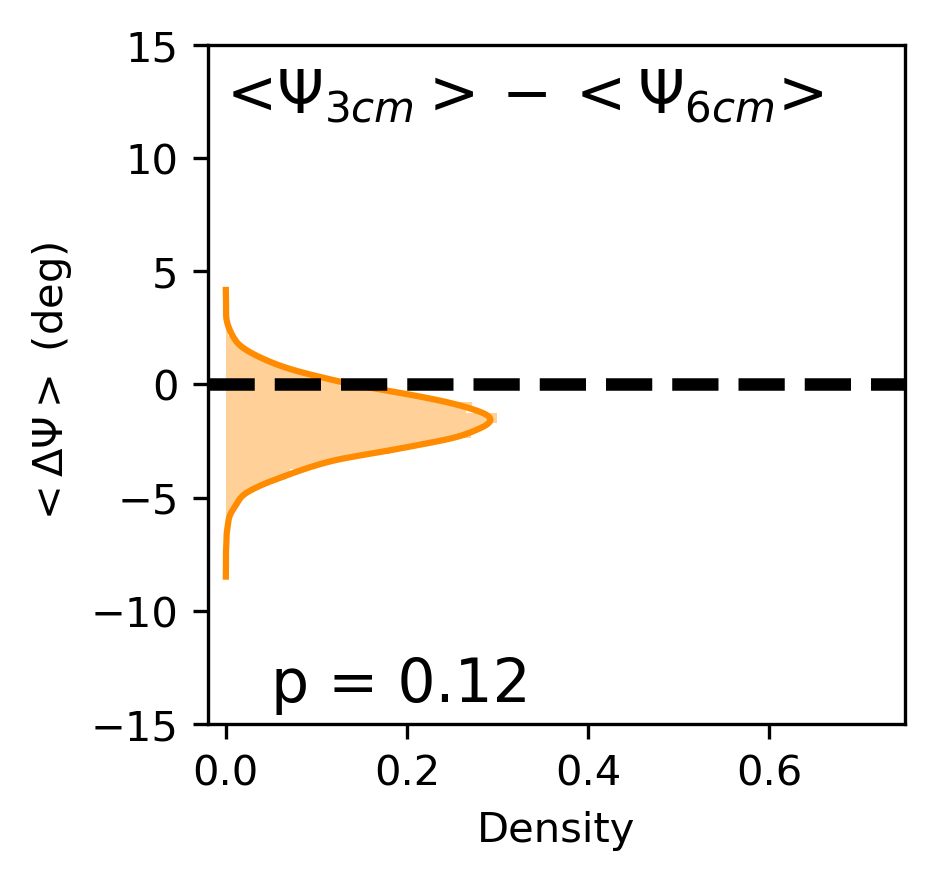} 
\put(25,74){\color{black} \colorbox{white}{\textbf{INTERARM}}}
\end{overpic}
\begin{overpic}[trim={0 0 0 2}, clip, width=0.32\textwidth]{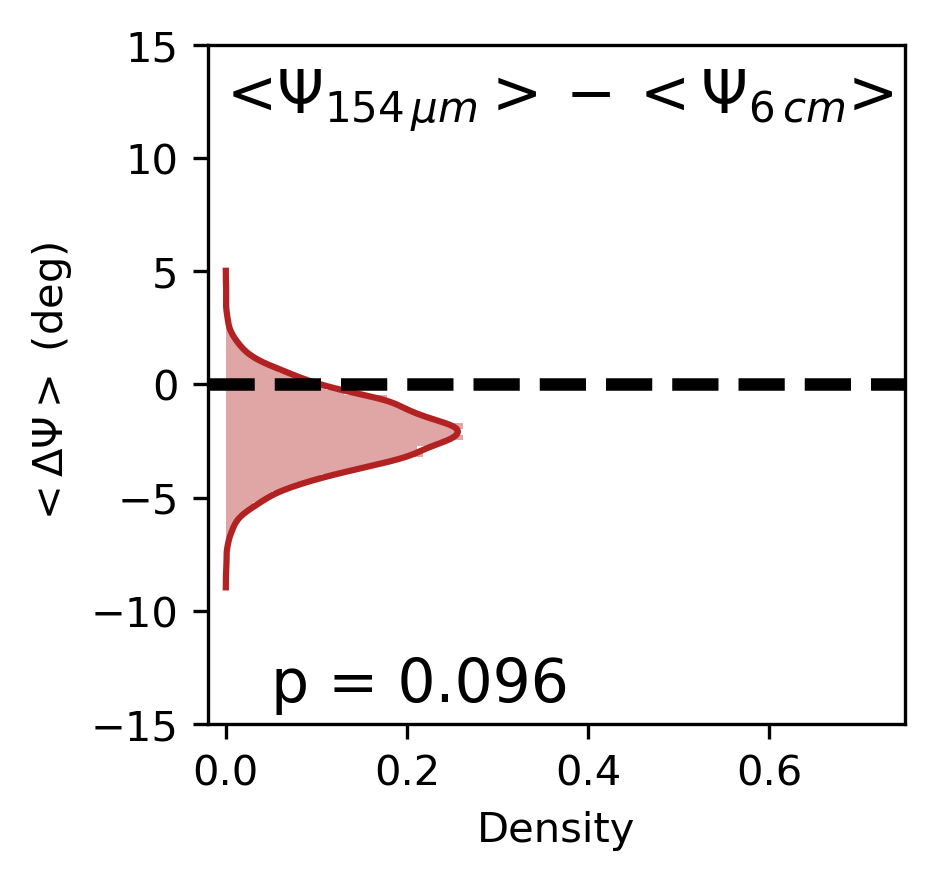} 
\put(25,74){\color{black} \colorbox{white}{\textbf{INTERARM}}}
\end{overpic}

\caption{Probability density distributions of difference in median magnetic pitch angle ($<\Delta\Psi>$, vertical histograms). \emph{Columns from left to right:} a) \mum154\ vs. 3\,cm. b) 3\,cm vs. 6\,cm. c) \mum154\ vs. 6\,cm. \emph{Rows from top to bottom:} a) Arms region (Arm 1 + Arm 2). b) Interarm region. The horizontal black dashed line represents the zero level (no difference). The $p$-value on each panel represents the probability that the distribution is compatible with zero (no difference).} 
\label{fig:median_pitch_angle}
\end{center}
\end{figure*}

For the arms region, we estimate an average magnetic pitch angle of \APsiFIR$=+16.9^{+1.8\circ}_{-1.7}$ for the \mum154 observations, \APsithreecm$=+23.1^{+1.1\circ}_{-1.0}$ and \APsisixcm$=+25.1^{+0.8\circ}_{-0.8}$ for the 3\,cm and 6\,cm observations, respectively. The magnetic pitch angle profiles of the spiral arms reveal an interesting scenario. The radio polarization maps at 3\,cm and 6\,cm trace a relatively flat pitch angle up to a radius of $220$\arcsec~($9.15$ kpc) -- showing some steady increase with radius.  The FIR magnetic pitch angle suffers a strong break at a radius $\sim150\arcsec$ ($\sim6.24$ kpc) decreasing suddenly towards negative values. Statistical analysis of the probability distributions obtained with the Monte Carlo simulations of each bin beyond the $\sim150\arcsec$ break reveals that the difference is significant ($p<0.05$) and consistent up to the limiting radius of observation on \m51. 

The observed break in the magnetic pitch angle profile of the arms region has a significant impact on the average value. In Fig.\,\ref{fig:median_pitch_angle} we compare the global differences in the magnetic pitch angle between FIR and radio wavelengths. We measure the difference in average magnetic pitch angle for each pair of datasets (\mum154, 3\,cm, and 6\,cm) and arms regions of \m51. The vertical histograms on Fig.\,\ref{fig:median_pitch_angle} represent the probability distribution for the difference in the median pitch angle as a function of the wavelengths and regions compared. These probability distributions are generated based on the 10\,000 Monte Carlo simulations obtained for the magnetic pitch angle analysis. The distributions take into account the uncertainties in position angle, inclination, and the Stokes $IQU$ from the different sets of polarization maps. Using these simulations, we are able to reconstruct the realistic probability density distribution of the average difference between the magnetic pitch angle profiles. We find a statistically significant difference in the magnetic pitch angle between FIR and radio wavelengths in the arms. Averaged across the complete extension of both arms, the FIR magnetic pitch angle is $-6.2^{+2.1\circ}_{-2.0}$ and $-8.3^{+2.0\circ}_{-1.9}$ lower than that measured in 3 and 6\,cm, a result significant with $p$-values of 0.002 and $<10^{-4}$, respectively. 

We now analyze the two arms separately in Fig.\,\ref{fig:mag_pitch_arm}. Results show that the two arms have different radial profiles of the magnetic pitch angles across the galactocentric radius. At small radii ($R<75\arcsec$, $<3.12$ kpc), Arm\,1 shows a lower magnetic pitch angle than Arm\,2, $\mPsi^{A1} < \mPsi^{A2}$. The magnetic pitch angle profile is inverted at $R>75\arcsec$ ($>3.12$ kpc), where $\mPsi^{A1} > \mPsi^{A2}$. This inversion is observed at all wavelengths up to $R\sim160\arcsec$ ($6.66$ kpc). At $R>160\arcsec$ ($>6.66$ kpc), the magnetic pitch angle of both arms shows a sharp decrease towards zero and negative values in FIR, but not in the 3 and 6\,cm radio polarization observations. For the 3\,cm and 6\,cm radial profiles, the magnetic pitch angle of Arm\,1 is mostly flat beyond $R>75\arcsec$ ($>3.12$ kpc), while Arm 2 presents an upturn at $R>150\arcsec$ ($>6.24$ kpc). A high pitch angle dispersion region is found in Arm 1 at $R\sim150\arcsec$ ($\sim6.24$ kpc) on the \mum154\,/HAWC+ magnetic pitch angle profile.

\begin{figure*}[]
 \begin{center}
\centering
\includegraphics[trim={0 0 35 30}, clip, width=0.378\textwidth]{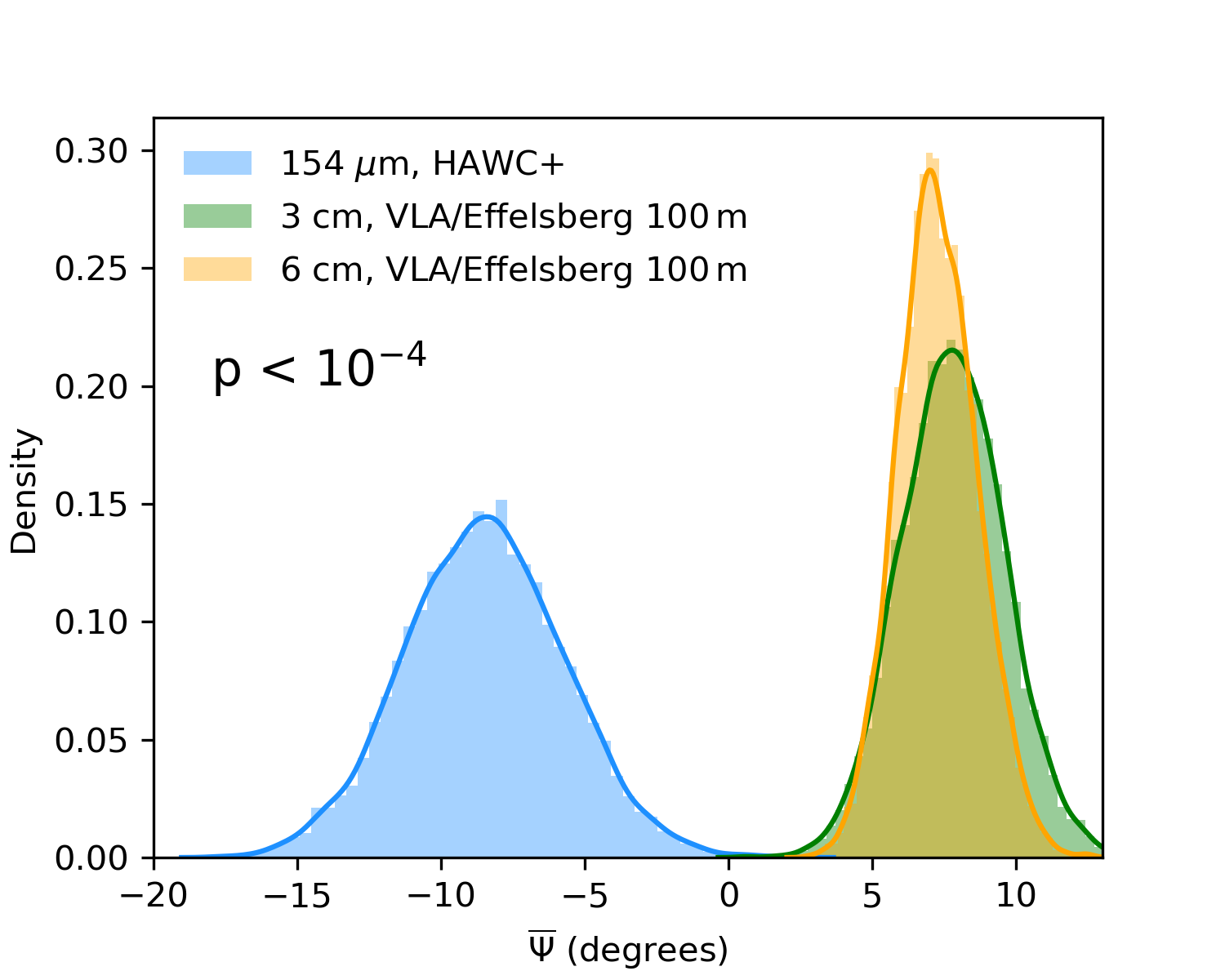}
\includegraphics[trim={10 0 32 10}, clip, width=0.613\textwidth]{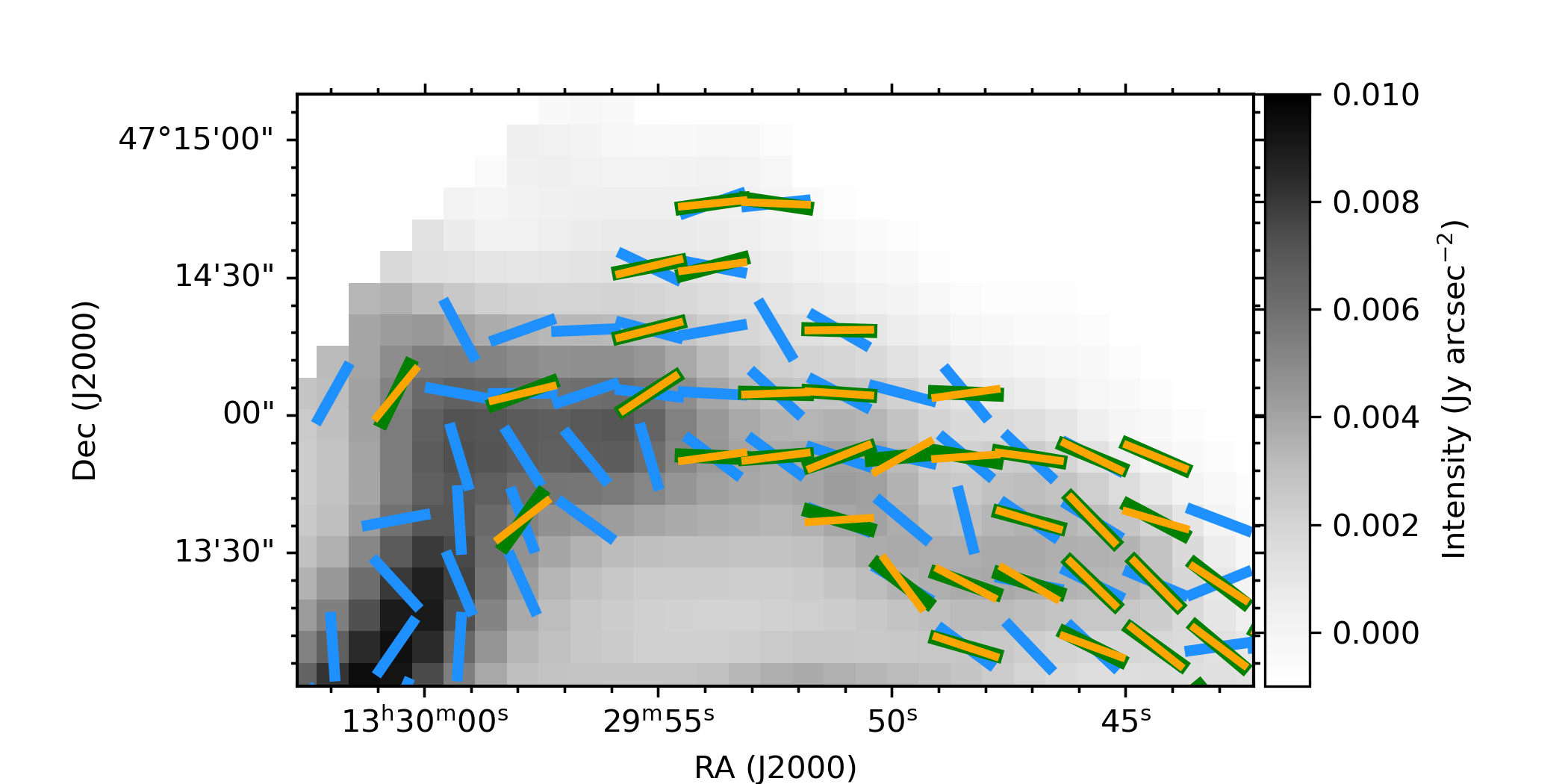}
\includegraphics[trim={0 0 35 30}, clip, width=0.378\textwidth]{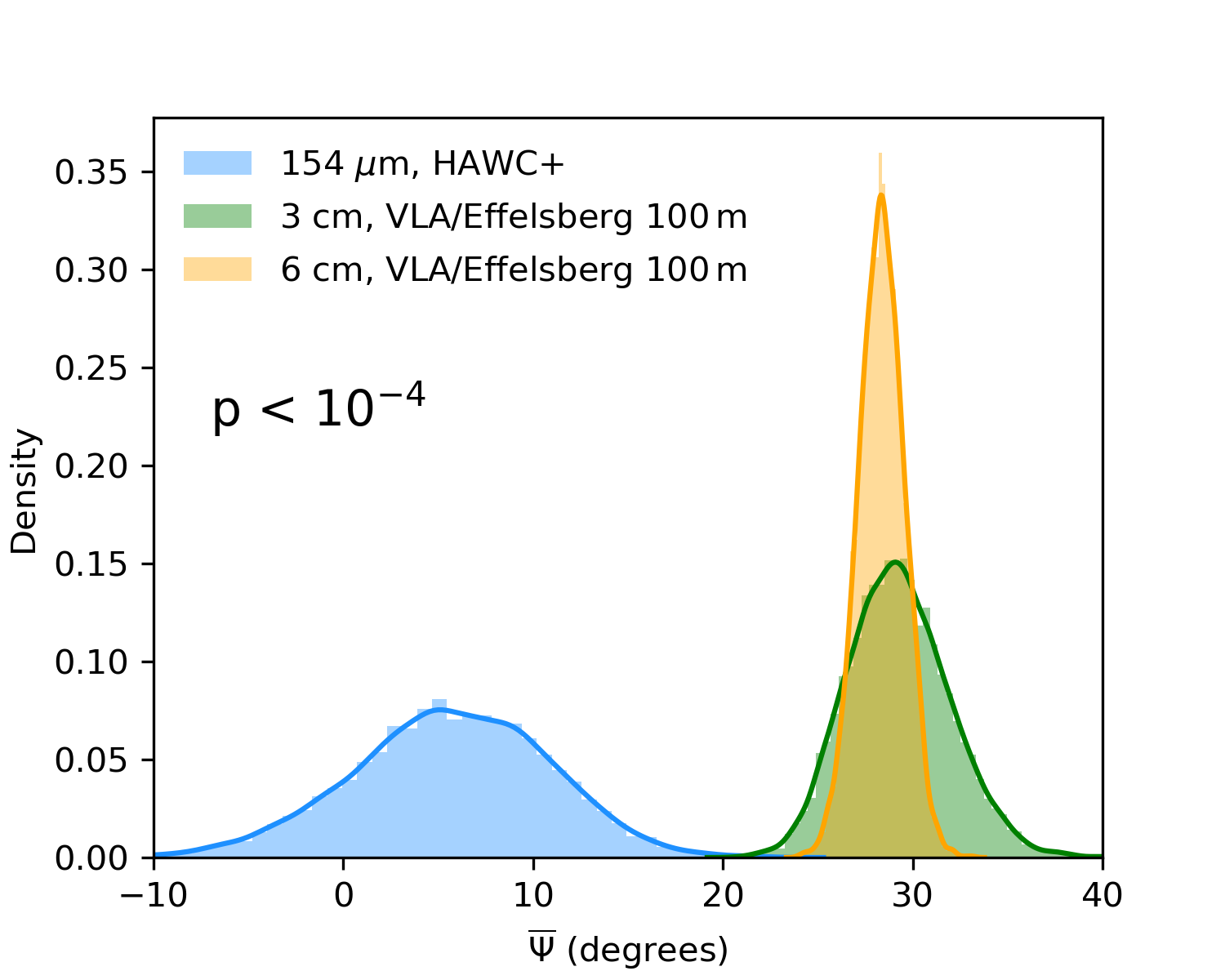}
\includegraphics[trim={10 0 32 10}, clip, width=0.613\textwidth]{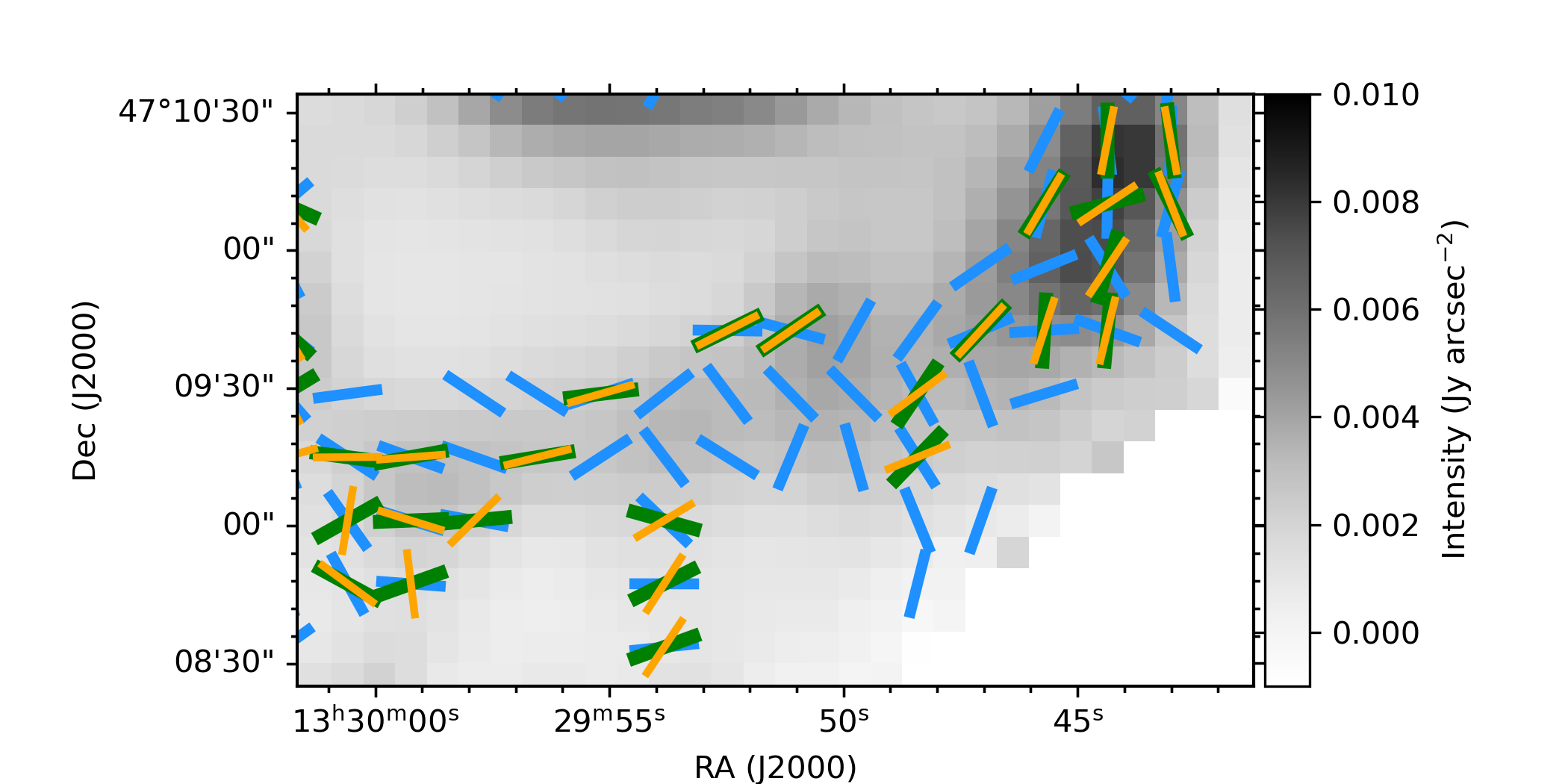}
\caption{Analysis of the magnetic pitch angle difference in the northern (top row) and southern (bottom row) region of the \m51\ spiral arms. \emph{Left panel:} Probability distribution of the median magnetic pitch angle for the \mum154, 3\,cm, and 6\,cm observations. \emph{Right panel:} B-field orientations for \mum154, 3\,cm, and 6\,cm. The grey-scaled background image shows the FIR total intensity from Fig.\,\ref{fig:mag_pitch_full_hawc}. For better visualization, only one in every two polarization measurements is represented. See the color legend in the left panel for reference.}
\label{fig:north_local_test}
\end{center}
\end{figure*}

We further explore the pitch angle difference for FIR and radio polarization observations in the northern section of Arm 2, one of the closest -- but not physically connected -- spiral arm regions to M51b. We study the distribution of magnetic pitch angles in a rectangular aperture of $3.45\times2.07$ arcmin$^2$. ($8.6\times5.2$ kpc$^{2}$) centered at $\alpha=202.47^{\circ}$, $\delta=47.23^{\circ}$. Fig.\,\ref{fig:north_local_test} shows the B-field orientations for the 154 \um, 3\,cm, and 6\,cm observations. Visual inspection of the three B-fields shows that on average the magnetic field at 154 \um~shows a different orientation with a smaller pitch angle than those from radio polarimetric observations.  In the left panel, we show the probability distributions for the average value of the pitch angle in that aperture. The results show a systematic difference ($p<10^{-4}$) between the FIR and the two radio observations. The magnetic pitch angles of the 3\,cm and 6\,cm  are compatible with each other. The average magnetic pitch angles in this region are \mPsi$_{\rm{FIR}}= -8.5^{+2.8\circ}_{-2.7}$, \mPsi$_{\rm{3\,cm}}= +7.8^{+1.8\circ}_{-1.8}$, and \mPsi$_{\rm{6\,cm}}= +7.2^{+1.4\circ}_{-1.3}$.


We repeat the analysis on an equivalent aperture located in the southern region of Arm 1, symmetrically separated from the core ($\alpha=202.46^{\circ}$, $\delta=47.16^{\circ}$, also with an area of $8.6\times5.2$ kpc$^{2}$). The results show that the average magnetic pitch angle in this region is $\mPsi_{\mathrm{FIR}}=+5.8^{+5.2\circ}_{-5.3}$, which is significantly ($p<10^{-4}$) lower than those measured in 3\,cm ($\mPsi_{\mathrm{3\,cm}}=+29.1^{+2.8\circ}_{-2.6}$) and 6\,cm ($\mPsi_{\mathrm{6\,cm}}=+28.3^{+1.2\circ}_{-1.5}$). These results  -- including the magnetic pitch angle profiles -- confirm that the magnetic field in the outskirts of \m51\ traced by radio and FIR polarization observations are different. 

Our results show that the structure of the magnetic field is not isotropic or homogeneous across the galactic disk. Interestingly, the independent trends of the two spiral arms in the inner region of the disk ($R<150\arcsec$, $<6.24$ kpc) are detected in the three wavelengths independently, ensuring that the quality of the observations and the analysis is high enough to confirm that the radial changes in magnetic pitch angle are not caused by statistical uncertainty. In addition, we found that this feature is systematically present in both spiral arms at FIR wavelengths, confirming that the change in magnetic pitch angle are a detectable feature of the magnetic spiral structure of \m51. 





\subsection{Radial magnetic pitch angle profile -  interarms}
\label{subsubsec:results_pitch_angle_full_interarms}

We analyze the interarm region in Fig.\,\ref{fig:mag_pitch_full_radialprofile}, whose polarization measurements and models are shown in Fig. \ref{fig:mag_pitch_arms_interarms}. At all wavelengths, the interarm magnetic pitch angle shows a fairly constant structure up to $220\arcsec$ ($9.15$ kpc). We estimate the average magnetic pitch angles to be \IAPsiFIR$=+28.6^{+1.3\circ}_{-1.3}$, \IAPsithreecm$=+29.1^{+1.0\circ}_{-1.0}$ and \IAPsisixcm$=+30.6^{+1.0\circ}_{-0.8}$ for the \mum154, 3\,cm and 6\,cm observations, respectively. The magnetic pitch profiles and their average values show that the interarm magnetic field structure of \m51\ is the same at FIR and radio wavelengths. However, we find that the interarm magnetic pitch angles are higher than the corresponding values for the arm regions (Sec.\,\ref{subsubsec:results_pitch_angle_full_arms}). This is significant at a $p$-value $<10^{-4}$ for \mum154, 3\,cm, and 6\,cm.  

The most striking result from the comparison of the interarm magnetic pitch angle profiles is that the \mum154\ observations do not show signs of the same distortions or radial variations as those detected in the spiral arms (Sec.\,\ref{subsubsec:results_pitch_angle_full_arms} and Fig.\,\ref{fig:mag_pitch_arms_interarms}). The interarm radial profile appears to be relatively smooth and constant across the galaxy disk up to the observed outer radius of $220\arcsec$ ($9.15$ kpc). In Fig.\,\ref{fig:median_pitch_angle} (bottom row) we compare the global differences in the magnetic pitch angle between FIR and radio wavelengths, this time for the interarm region. We do not find any significant difference between the average magnetic pitch angle value of the FIR and radio-polarization dataset in the interarm region, confirming the results from the previous profiles. 

\begin{deluxetable}{cccccl}[th!]
\tablecaption{Magnetic field pitch angles in the radial range of $21.2$--$220$\arcsec~($0.88$--$9.15$ kpc) from Fig.\,\ref{fig:mag_pitch_full_radialprofile}. \label{tab:Bpitchangles}}
	\tablewidth{0pt}
		\tablehead{\colhead{Wavelength}	&	\colhead{Full disk} 	&	\colhead{Arms region} & \colhead{Interarms region} \\
			&	\colhead{($\Psi^{\mathrm{FD}}$,$^{\circ}$)} 	&	\colhead{($\Psi^{\mathrm{Arms}}$,$^{\circ}$)} & \colhead{($\Psi^{\mathrm{IA}}$,$^{\circ}$)}}
		\startdata
		$154$ \um   &   $23.9^{+1.2}_{-1.2}$ &   $16.9^{+1.8}_{-1.7}$    &   $28.6^{+1.3}_{-1.3}$ \\
		$3$ cm      &   $26.0^{+0.9}_{-0.8}$  &   $23.1^{+1.1}_{-1.0}$    &   $29.1^{+1.0}_{-1.0}$    \\
		$6$ cm      &   $28.0^{+0.8}_{-0.6}$  &   $25.1^{+0.8}_{-0.8}$   &   $30.6^{+1.0}_{-0.8}$ \\
		 \enddata
\end{deluxetable}

A summary of the average magnetic pitch angles within the radial range of 21.2\arcsec--220\arcsec ($0.88$--$9.15$ kpc) is shown in Table \ref{tab:Bpitchangles}. Based on the results from previous sections, we conclude that: 
\begin{enumerate}
    \item The outer ($R>6.24$ kpc) magnetic spiral structure of the spiral arms in \m51\ is wrapped tighter when measured in FIR than in radio-wavelengths.
    \item The FIR interarm magnetic pitch angle structure is similar to that traced with the radio polarization observations in the diffuse ISM.
\end{enumerate}

These results suggest that the outer field decoupling of the FIR and radio magnetic fields is only associated with the spiral arms. This result is further confirmed with the observations of the magnetic pitch angle profiles and the custom apertures studied in Sec.\,\ref{subsubsec:results_pitch_angle_full_arms}. We note that this difference is significant despite the fact that the radial binning and the combination in azimuthal coordinates may be smoothing the differences found in the histograms from this section. We discuss the implications of these results in Sec.\,\ref{Sec:Discussion}.


\begin{figure}[]
 \begin{center}
\centering
\includegraphics[trim={5 35 0 5}, clip, width=0.5\textwidth]{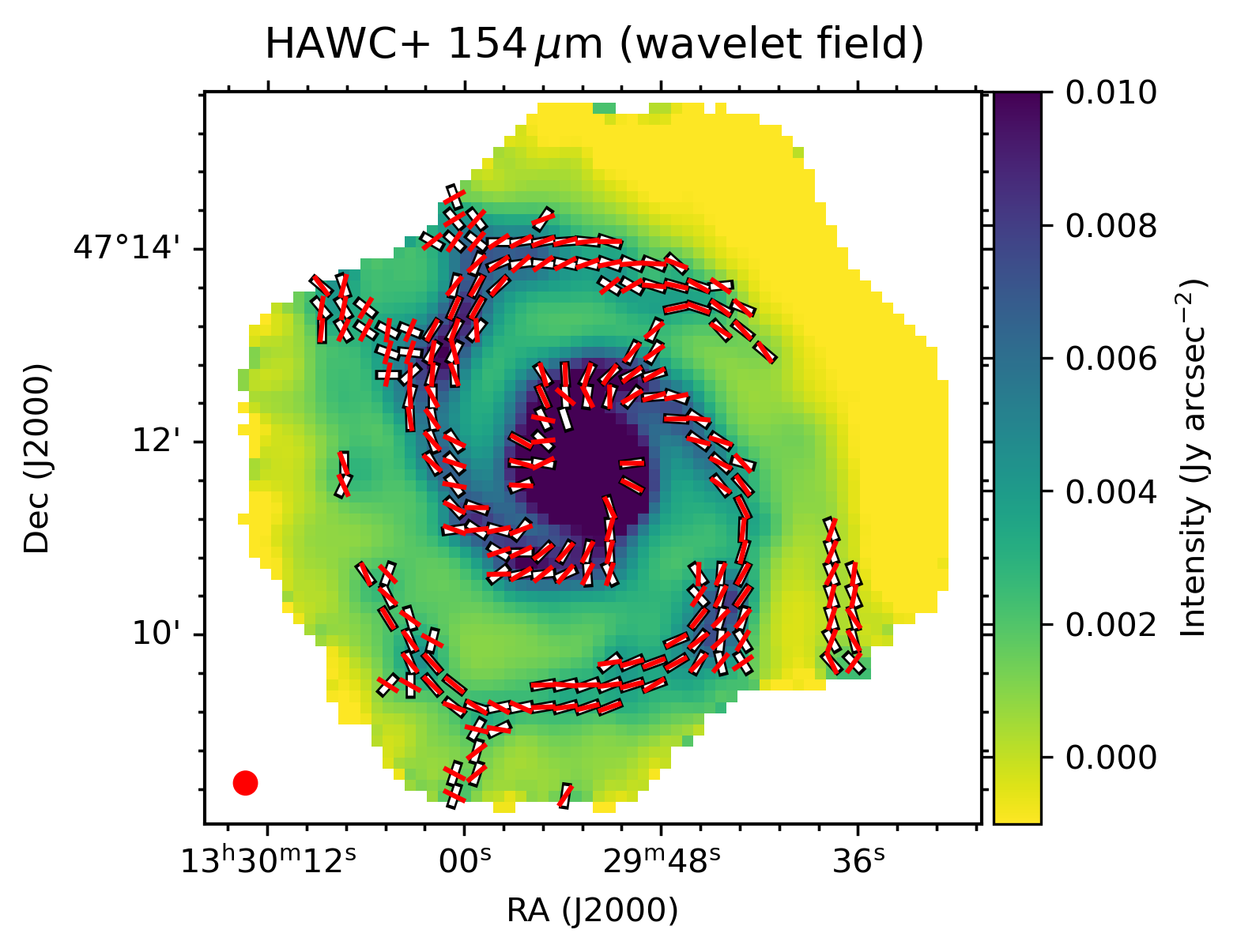}
\includegraphics[trim={5 35 0 5}, clip, width=0.472\textwidth]{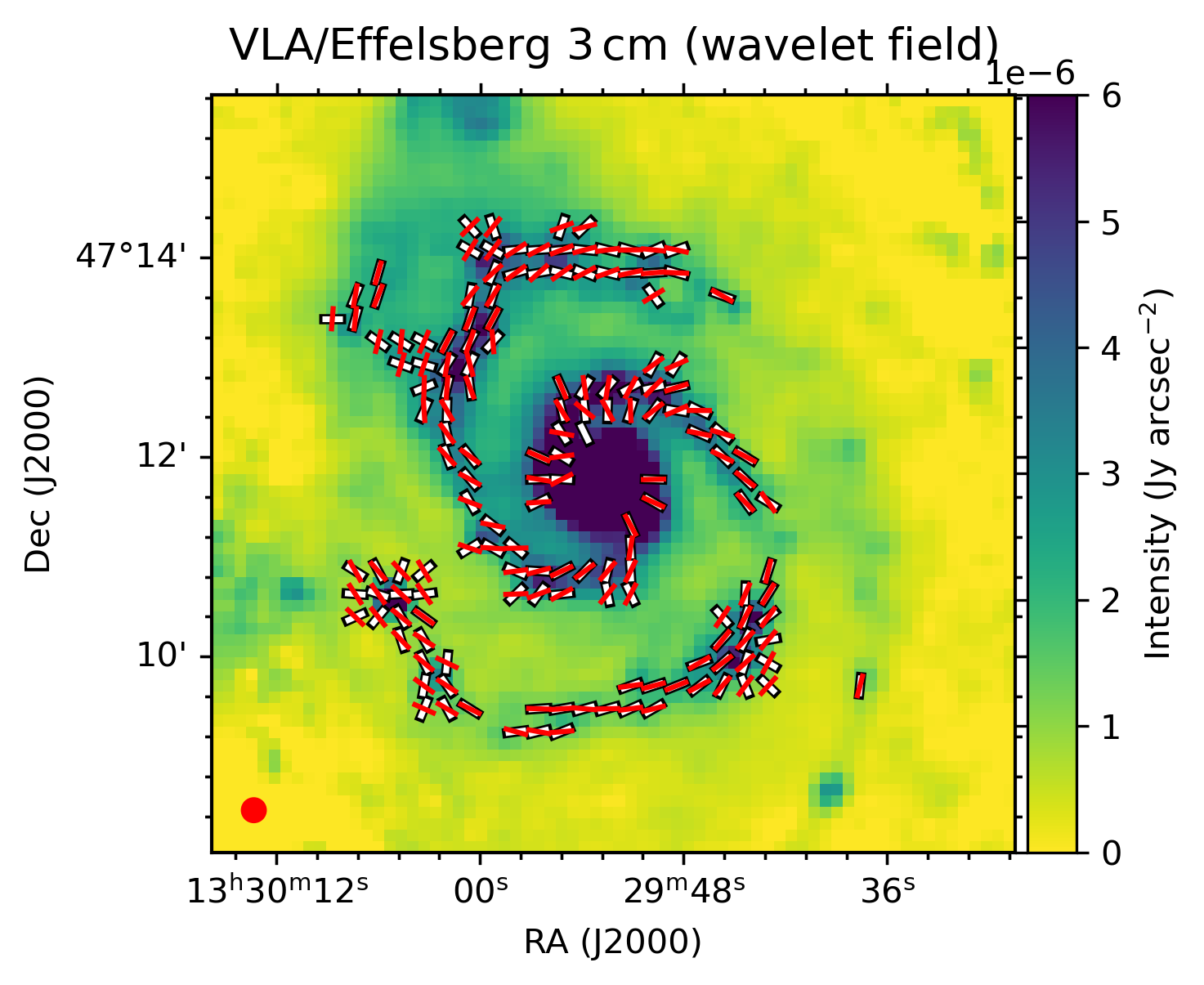}
\includegraphics[trim={5 0 0 5}, clip, width=0.484\textwidth]{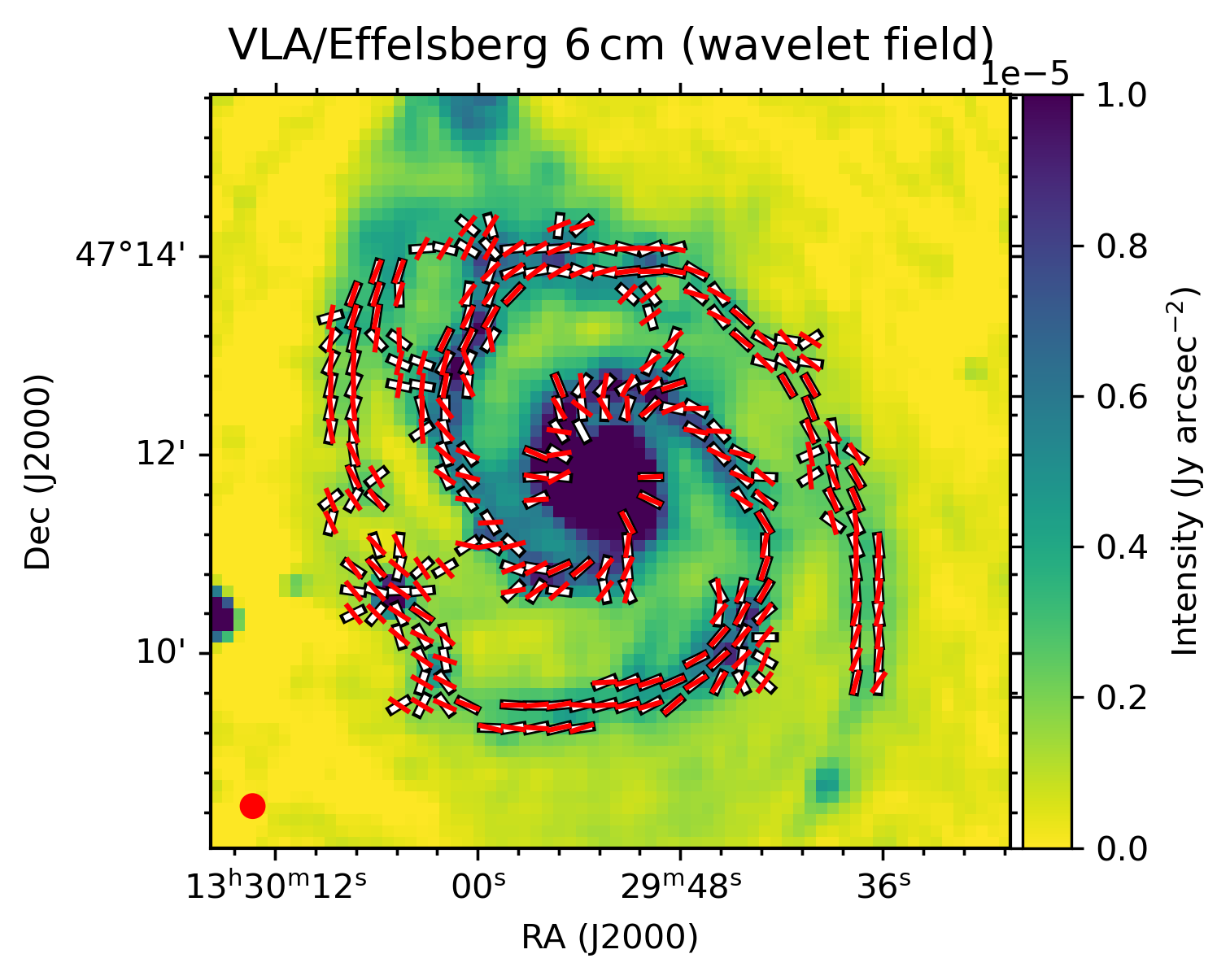}
\caption{\emph{Top to bottom:} Surface brightness distributions for 1) \hawc 154 \um, 2) VLA/Effelsberg 3\,cm and, 3) VLA/Effelsberg 6\,cm with the morphological wavelet line plotted in red. In white, we show the azimuthally averaged morphological pitch angle directions. \emph{Red circle:} Resolution element (beam size) of the analyzed maps.} \label{fig:wavelets_hawc_3cm_6cm}
\end{center}
\end{figure}

\begin{figure}[]
 \begin{center}
\centering
\includegraphics[trim={0 0 0 0}, clip, width=0.5\textwidth]{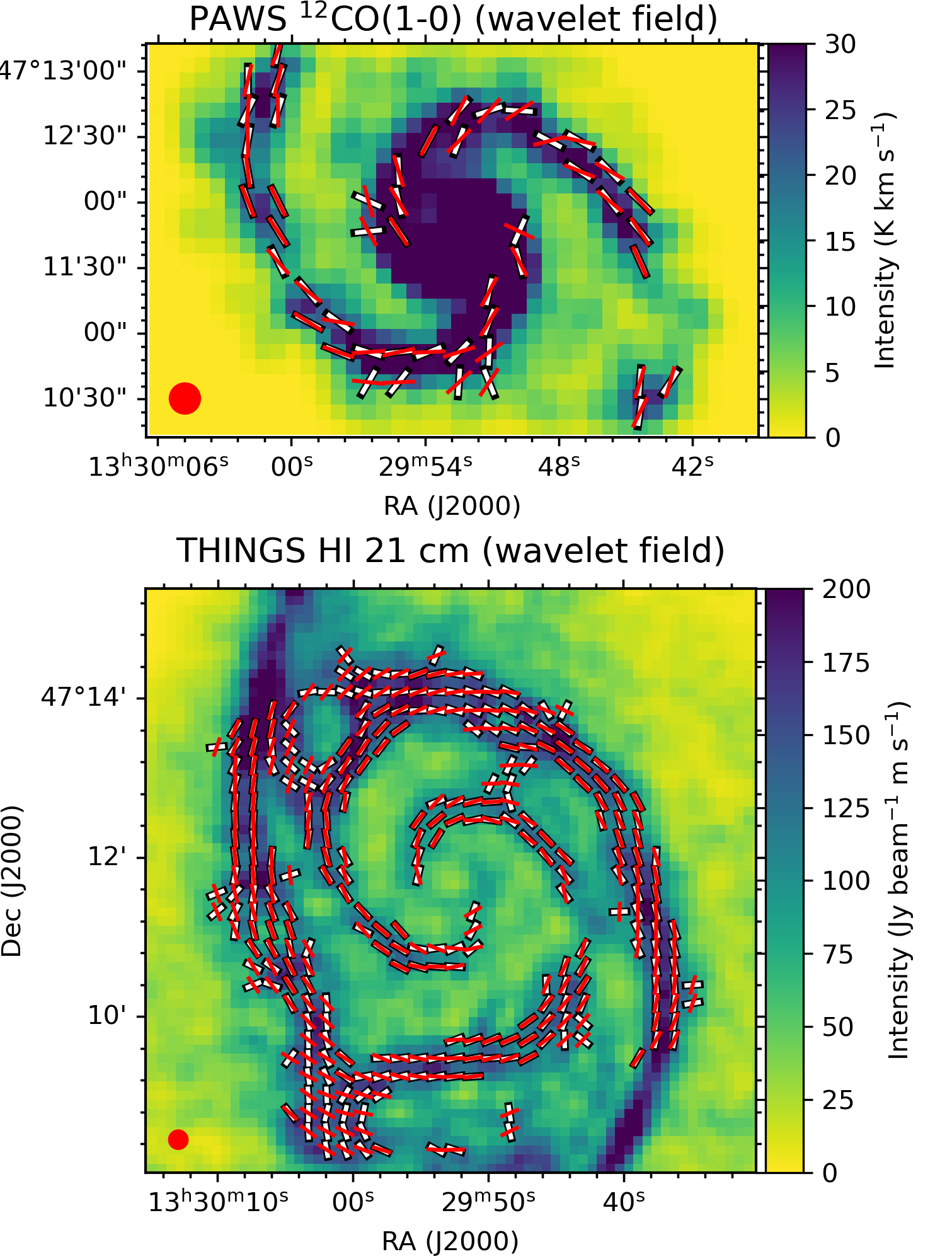}
\caption{ \emph{Top to bottom:} Surface brightness distributions for 1) \lineco\ PAWS and 2) THINGS \hi\ observations with the morphological wavelet line overplotted (red). In white, we show the averaged morphological pitch angles profile. \emph{Red circle:} Resolution element (beam size) of the analyzed maps.} \label{fig:wavelets_co_hi}
\end{center}
\end{figure}

\subsection{Comparison with the morphological pitch angle of the spiral arms}
\label{subsubsec:results_morphological_pitch_angle}

\begin{figure*}[h!]
 \begin{center}
\includegraphics[trim={60 28 85 0}, clip, width=0.95\textwidth]{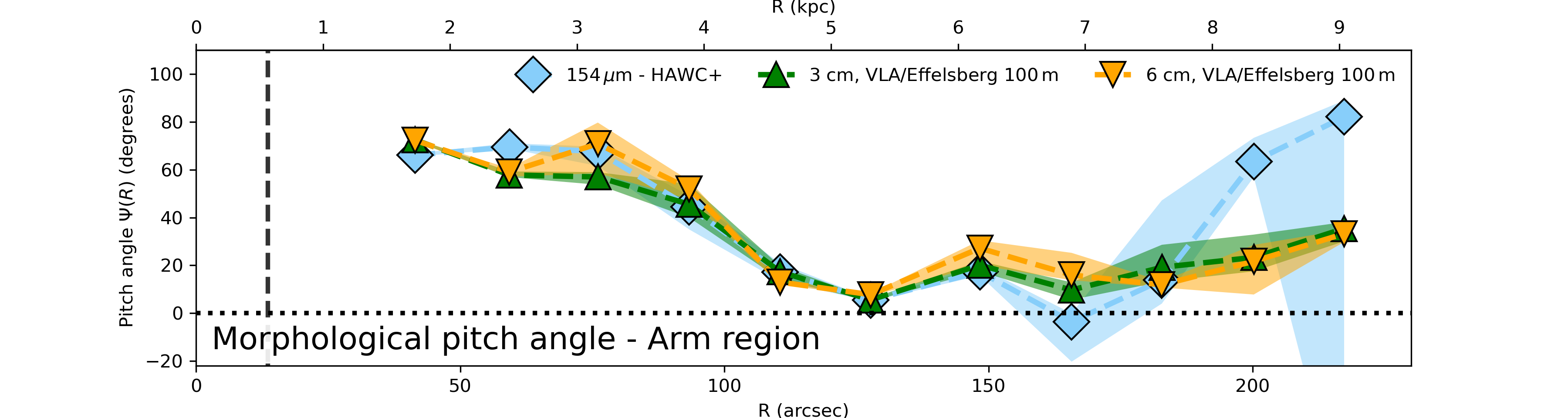}

\includegraphics[trim={60 28 85 26}, clip, width=0.95\textwidth]{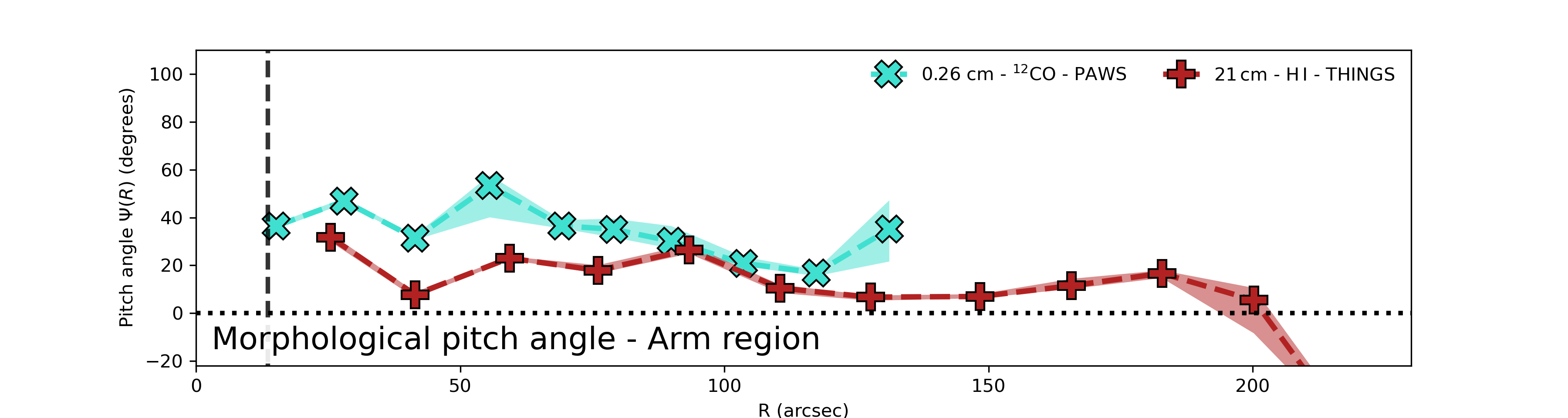}

\includegraphics[trim={60 28 85 26}, clip, width=0.95\textwidth]{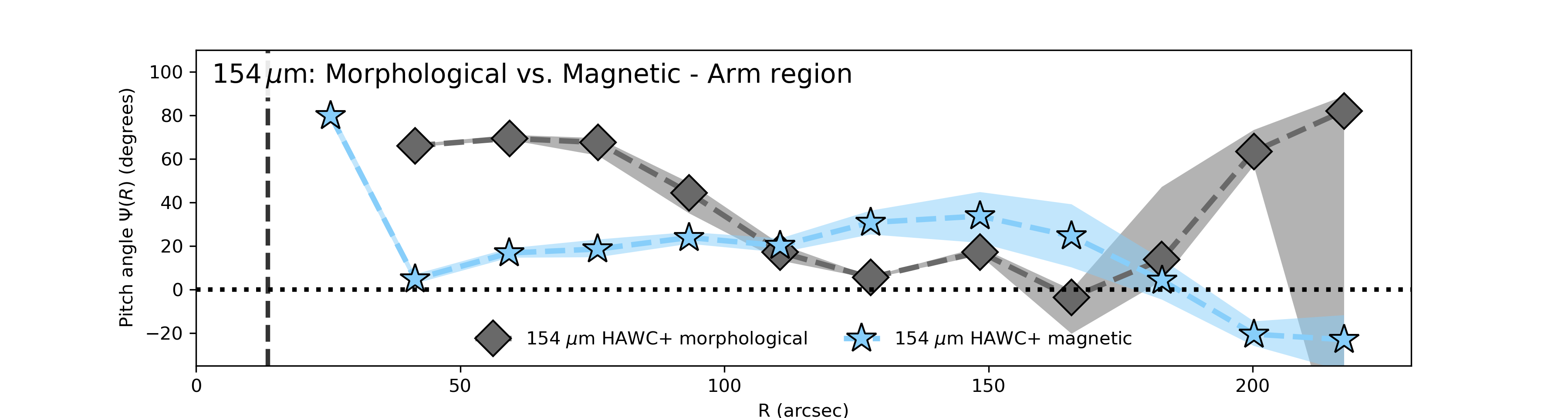}

\includegraphics[trim={60 28 85 26}, clip, width=0.95\textwidth]{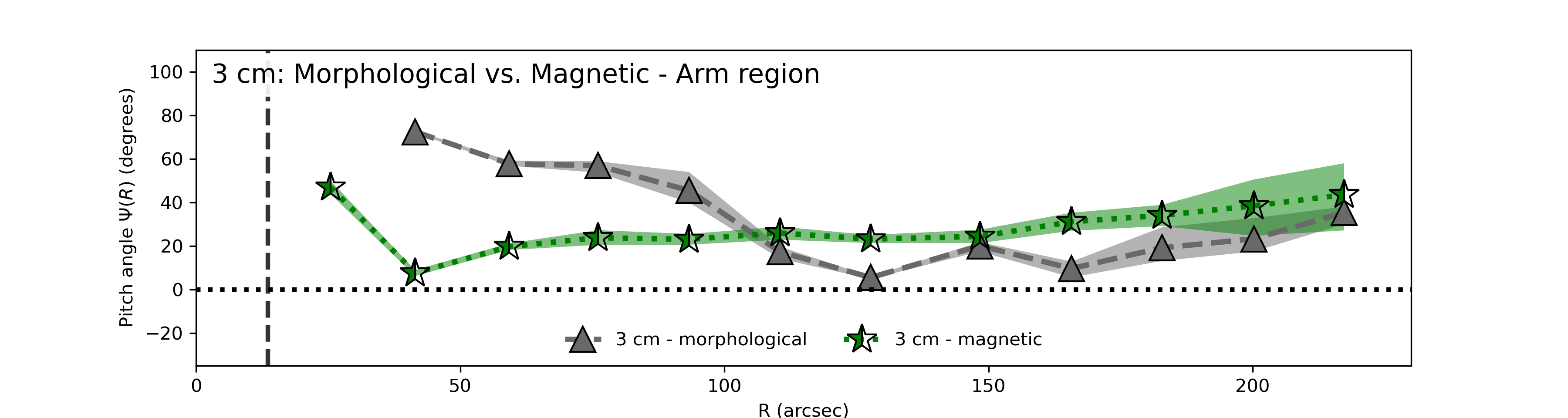}

\includegraphics[trim={60 0 85 26}, clip, width=0.95\textwidth]{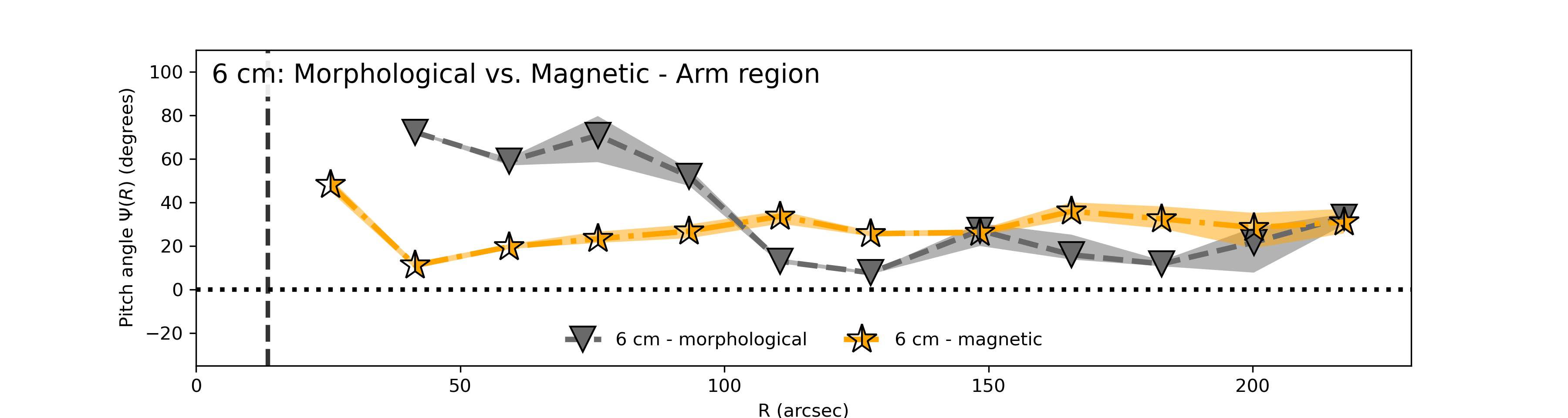}

\caption{Morphological pitch angle profiles (pitch angle $\Psi(R)$ as a function of the galactocentric radius, $R$). \emph{Top to bottom:} 1) \mum154, 3\,cm and, 6\,cm. 2) \lineco\ and 21 cm. 3) \mum154 morphological vs. magnetic profile. 4) 3\,cm morphological vs. magnetic. 5) 6\,cm morphological vs. magnetic. The profiles are calculated on the arms region of \m51. See the legend for the color and linetype.} 
\label{fig:morph_vs_magnetic}
\end{center}
\end{figure*}

Figs.\,\ref{fig:wavelets_hawc_3cm_6cm} and \ref{fig:wavelets_co_hi} show the morphological pitch angle maps of the \mum154, 3\,cm, 6\,cm, \lineco, and 21\,cm \hi\ observations. These maps have been constructed from the total intensity images and the wavelet transform method described in Secs.\,\ref{subsec:Methods_magnetic_pitch_angle}--\ref{subsec:Methods_wavelet_analysis}. To avoid selection effects due to the different resolution of the images, we convolved every dataset to the \mum154~HAWC+ beam size, as we did in the previous section for the VLA/Effelsberg 100\,m observations. In Fig.\,\ref{fig:morph_vs_magnetic} we present the morphological pitch angle profiles of the spiral arms for the five different datasets considered, plus the comparison of the magnetic and morphological pitch angle for \mum154, 3\,cm, and 6\,cm. 

The morphological pitch angles of \mum154, 3\,cm, and 6\,cm have a similar radial profile, i.e. $\MPsiFIR \sim \MPsithreecm \sim \MPsisixcm$. At low radii ($<120\arcsec$, $<5.0$ kpc), the morphological pitch angle is relatively high, starting at $\sim 60$--$70^{\circ}$. At higher radii ($>120\arcsec$, $>5.0$ kpc), the morphological pitch angle decreases to $0$--$10^{\circ}$ with a relatively slow increase showing some scatter in the outskirts ($>200\arcsec$, $8.32$ kpc), especially for \mum154. We also compare the distribution of the morphological pitch angle with the magnetic pitch angle profiles obtained in Secs.\,\ref{subsubsec:results_pitch_angle_full}--\ref{subsubsec:results_pitch_angle_full_interarms} (see lower panels of Fig.\,\ref{fig:morph_vs_magnetic}). The analysis shows that for the three bands analyzed, the magnetic pitch angle is lower than the morphological equivalent up to a radius of $\sim100\arcsec$ ($\sim4.16$ kpc). At larger radii ($>100\arcsec$, $>4.16$ kpc), the magnetic pitch angle is larger than the morphological pitch angle. The exception is in the outermost region ($>175\arcsec$, $>7.28$ kpc) of the \mum154/\hawc\ data, due to the magnetic pitch angle break reported in Sec.\,\ref{subsubsec:results_pitch_angle_full_arms}. 


For \lineco, we find a relatively constant, albeit with large scatter, pitch angle profile of $\mPsi_{\mathrm{CO}}^{\mathrm{Morph}} \sim 40$--$60^{\circ}$ up to the limit of the PAWS observations ($R=120\arcsec$, $\sim5$ kpc), with an average of $\mPsi_{\mathrm{CO}}^{\mathrm{Morph}} = 30.7 ^{+0.5\circ}_{-0.4}$. For the 21 cm \hi\ observations, we find a relatively constant morphological pitch angle of $\mPsi_{\mathrm{21\,cm}}^{\mathrm{Morph}} =9.9^{+0.3\circ}_{-0.5}$ across the whole observable disk. We find that the morphological pitch angle of \hi\ is smaller than at FIR, radio, and \lineco\ within the central $120\arcsec$ ($5$ kpc). But it is approximately similar to that of the outer region ($R>120\arcsec$, $>5$ kpc) when compared with the \mum154\ (\MPsiFIR $=8.4^{+0.5\circ}_{-0.5}$), 3\,cm (\MPsithreecm $=10.6^{+0.7\circ}_{-0.9}$) and 6\,cm (\MPsisixcm$=13.0^{+0.7\circ}_{-0.6}$). For reference, the average magnetic pitch angles in the outer region of the spiral arms are: $\APsiFIR=15.2^{+4.0\circ}_{-4.2}$, $\APsithreecm=25.9^{+1.5\circ}_{-1.5}$, and $\APsisixcm=27.5^{+1.1\circ}_{-1.1}$. 

We find that the magnetic field pitch angles are higher than the morphological pitch angles of the \hi\ in the outskirts of the spiral arms of \m51. The $p$-value for this difference in average values is lower than $10^{-4}$ for the 3 and 6\,cm observations (highly significant) and $p=0.044$ for the \mum154\ observations.  The higher values for the outermost bins of the FIR morphological pitch angle profile are possibly an artifact caused by the boundaries of the \hawc\ footprint with the wavelet algorithm, thus we consider them negligible. In addition, the lower significance at \mum154\ is caused by its observed magnetic pitch angle break in the outskirts, which combined with the outer distortions on the morphological profile, reduce the difference between the morphological and magnetic values.


%

Close inspection of the total intensity distribution on Figs.\,\ref{fig:wavelets_hawc_3cm_6cm} and \ref{fig:wavelets_co_hi} reveal that \mum154, 3\,cm, 6\,cm, and \lineco\ datasets show bright emission in the core of \m51. Contrarily, 21 cm \hi\ observations show no detectable emission at small radii, as previously mentioned in Sec.\,\ref{subsec:Methods_mask}. These different distributions can be responsible for the difference in the morphological pitch angle distributions at inner radii ($<120\arcsec$, $<5$ kpc). The main reason is that the direction of the wavelet field is affected by the presence of a large, bright, radial central gradient from the core. Nevertheless, the fact that 1) we observe this relatively higher pitch angle value up to $R\sim100\arcsec$ ($4.16$ kpc) far away from the main component of the total intensity of the core and, 2) the \lineco\ dataset also shows higher morphological pitch angle than the \hi\ observations, suggests that the morphological differences of the pitch angle for the spiral arms is not caused entirely by systematic effects from the central zone.

In summary, we find that for the spiral arms: 
\begin{enumerate}
    \item The morphological pitch angles change as a function of the multi-phase ISM, such as $\mPsi_{\mathrm{HI}}^{\mathrm{Morph}}< \mPsi_{\mathrm{CO}}^{\mathrm{Morph}} <
    \MPsiFIR \sim \MPsithreecm \sim \MPsisixcm$.
    \item The morphological pitch angles at FIR and radio wavelengths are similar across the full disk of \m51.
    \item At FIR and radio and within the inner $100\arcsec$ ($4.16$ kpc), the magnetic pitch angles are wrapped tighter than the morphological pitch angles. 
    \item  At FIR and radio and at radius $>100\arcsec$ ($>4.16$ kpc),  the magnetic pitch angles of the spiral arms are larger than those from the morphological structure. The exception is the FIR, whose magnetic pitch angle becomes tighter than the morphological pitch angle at radius $>200\arcsec$ ($>8.32$ kpc).
\end{enumerate}


\section{Magnetic fields in the multi-phase ISM}
\label{subsec:ISM_results}

\subsection{The multi-phase ISM}


\begin{figure*}[ht!]
 \begin{center}
\includegraphics[trim={0 0 0 0}, clip, width=0.497\textwidth]{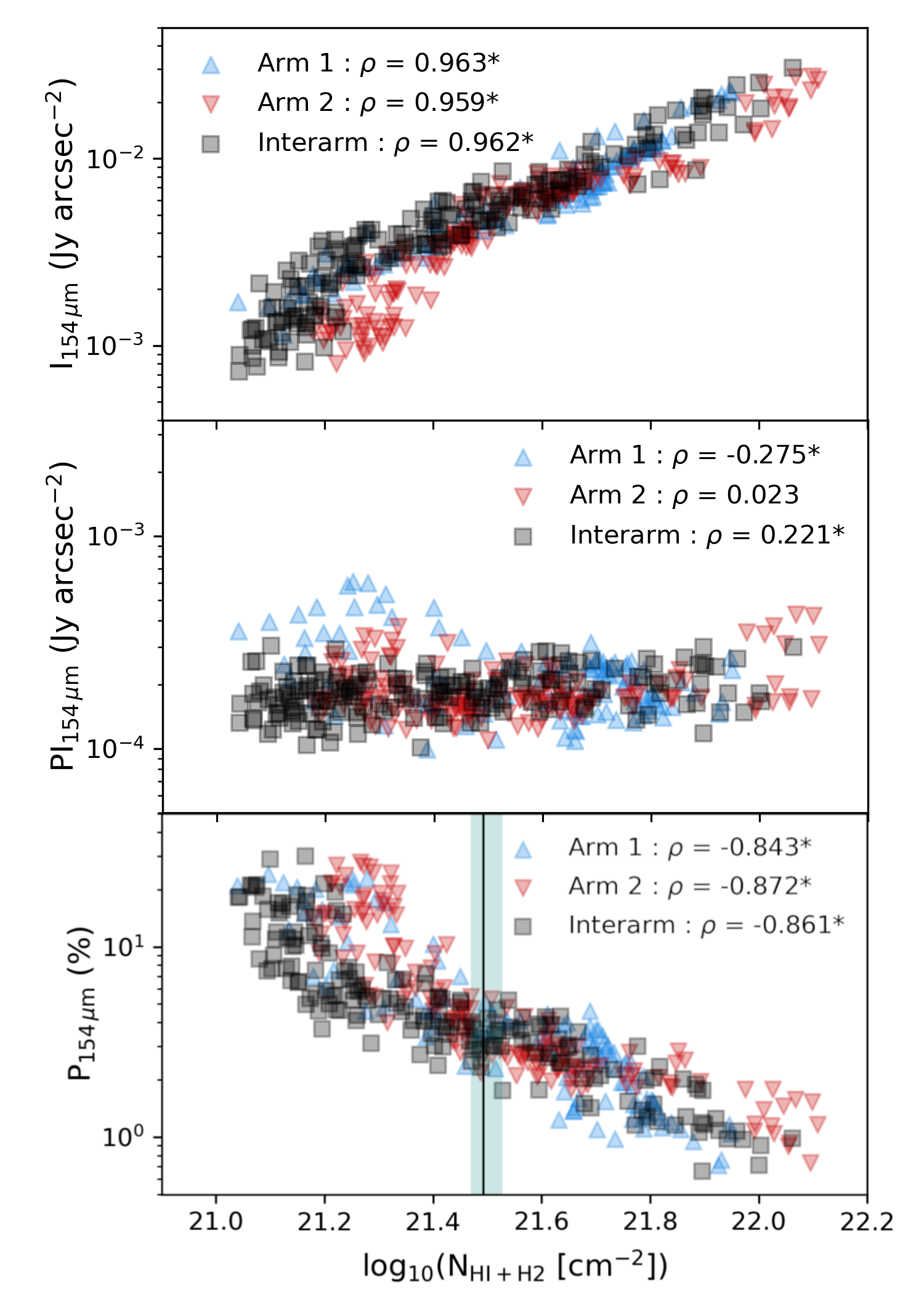} 
\includegraphics[trim={0 0 0 0}, clip, width=0.497\textwidth]{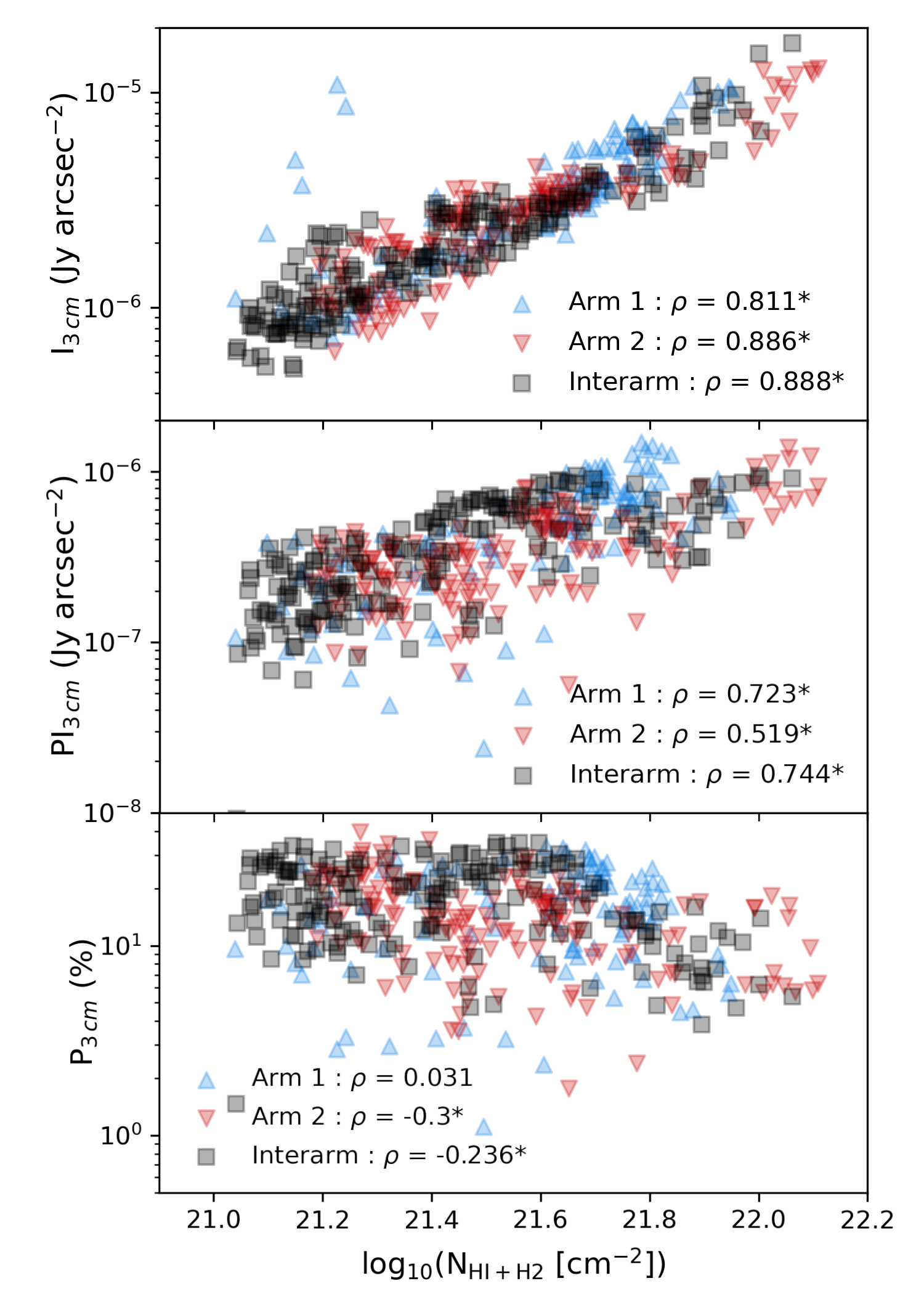}
\caption{Comparison between \mum154\ (left column) and 3\,cm (right column) of the total intensity (top row), polarized intensity (central row) and polarization fraction (bottom row) as a function of gas column density (\columndensity). Arm 1 (blue upward pointing triangle), Arm 2 (red downward pointing triangle), and interarms (black square) as defined in Figure \ref{fig:mag_pitch_arms_interarms}. See the legend on each panel for the correlation analysis. An asterisk symbol ($*$) following each $\rho$ correlation coefficient is shown if the correlation is statistically different from zero ($p<0.05$). The change in slope at $\log_{10}(\columndensity]) \sim 21.49$ is shown as a black solid line and $1-\sigma$ dashed area in the $P_{154\mu m}$--\columndensity\ plots.} 
\label{fig:m51_comparison_Nh}
\end{center}
\end{figure*}
\begin{figure*}[ht!]
 \begin{center}
\centering
 \begin{minipage}[c]{.99\textwidth}
  \centering
   \includegraphics[trim={0 0 0 0}, clip, width=0.495\textwidth]{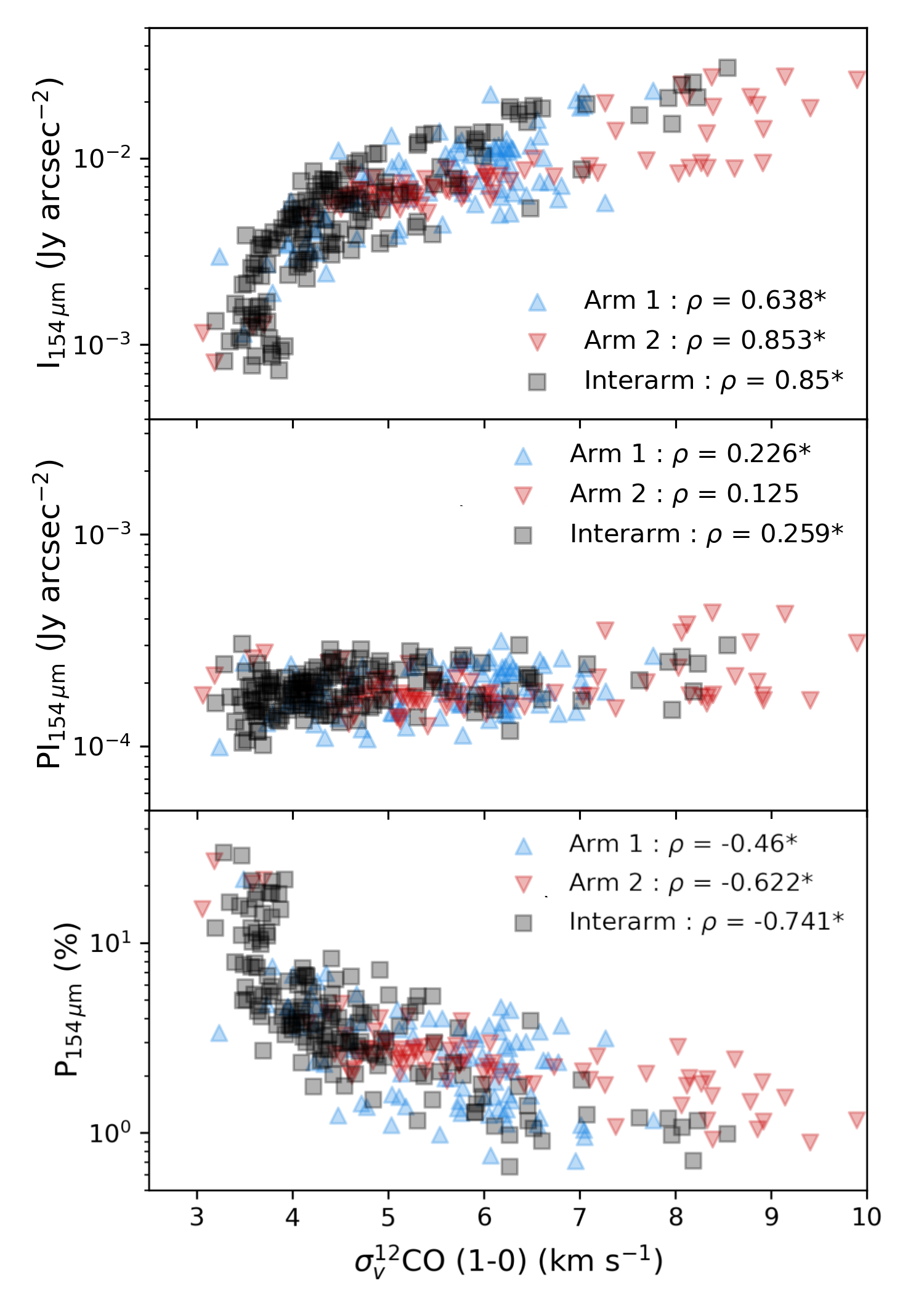} 
   \includegraphics[trim={0 0 0 0}, clip, width=0.495\textwidth]{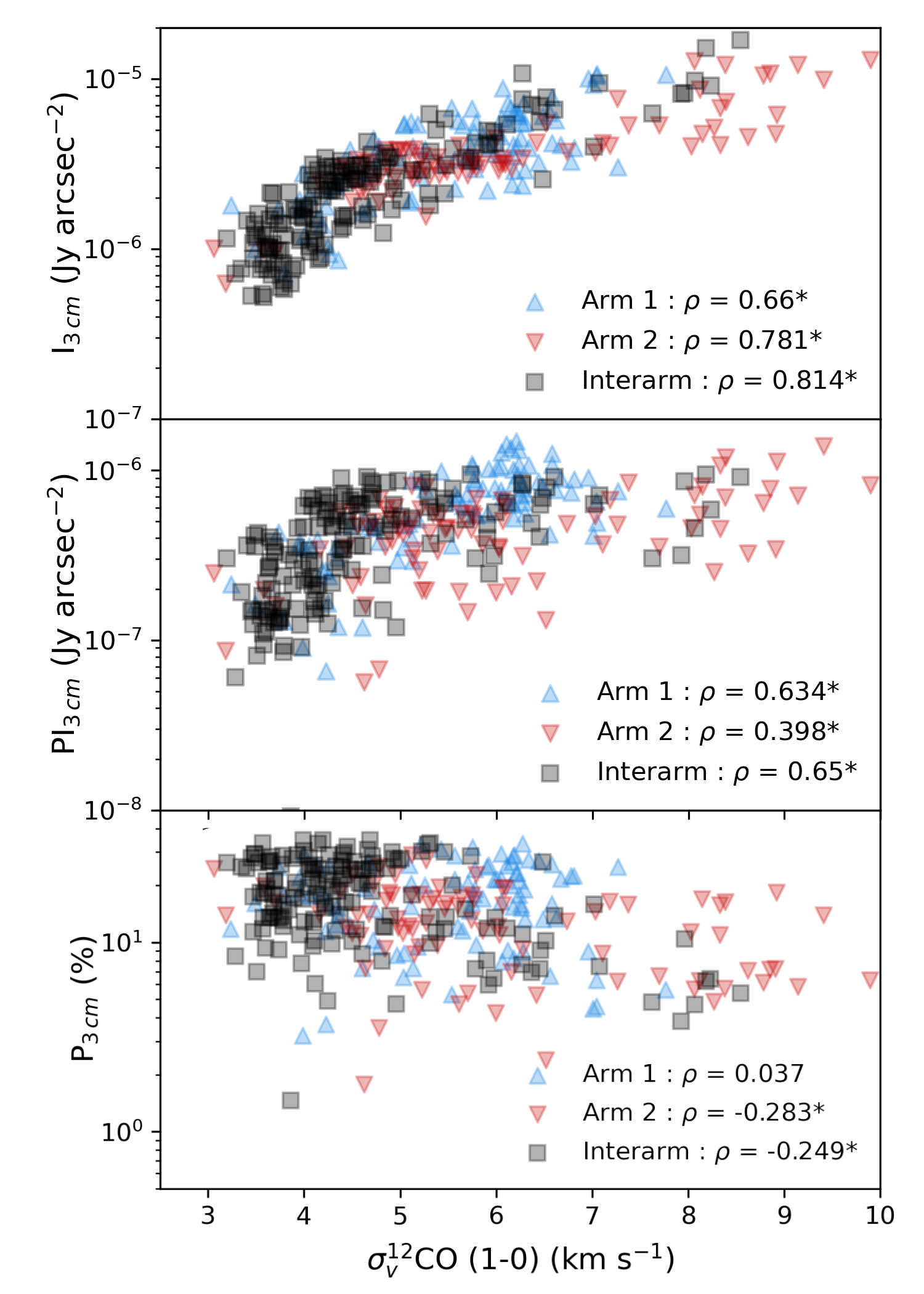}
  \end{minipage}
 \caption{Comparison between \mum154\ (left column) and 3\,cm (right column) of the total intensity (top row), polarized intensity (central row) and polarization fraction (bottom row) as a function of \lineco\ velocity dispersion ($\sigma_{v, \mathrm{^{12}CO(1-0)}}$). Arm 1 (blue upward-pointing triangle), Arm 2 (red downward-pointing triangle), and interarms (black square) as defined in Figure \ref{fig:mag_pitch_arms_interarms}. See the legend on each panel for the correlation analysis. An asterisk symbol ($*$) following each $\rho$ correlation coefficient is shown if the correlation is statistically different from zero ($p<0.05$).} 
\label{fig:m51_comparison_CO}
\end{center}
\end{figure*}

\begin{figure*}[ht!]
 \begin{center}
\centering
 \begin{minipage}[c]{.99\textwidth}
  \centering
   \includegraphics[trim={0 0 0 0}, clip, width=0.495\textwidth]{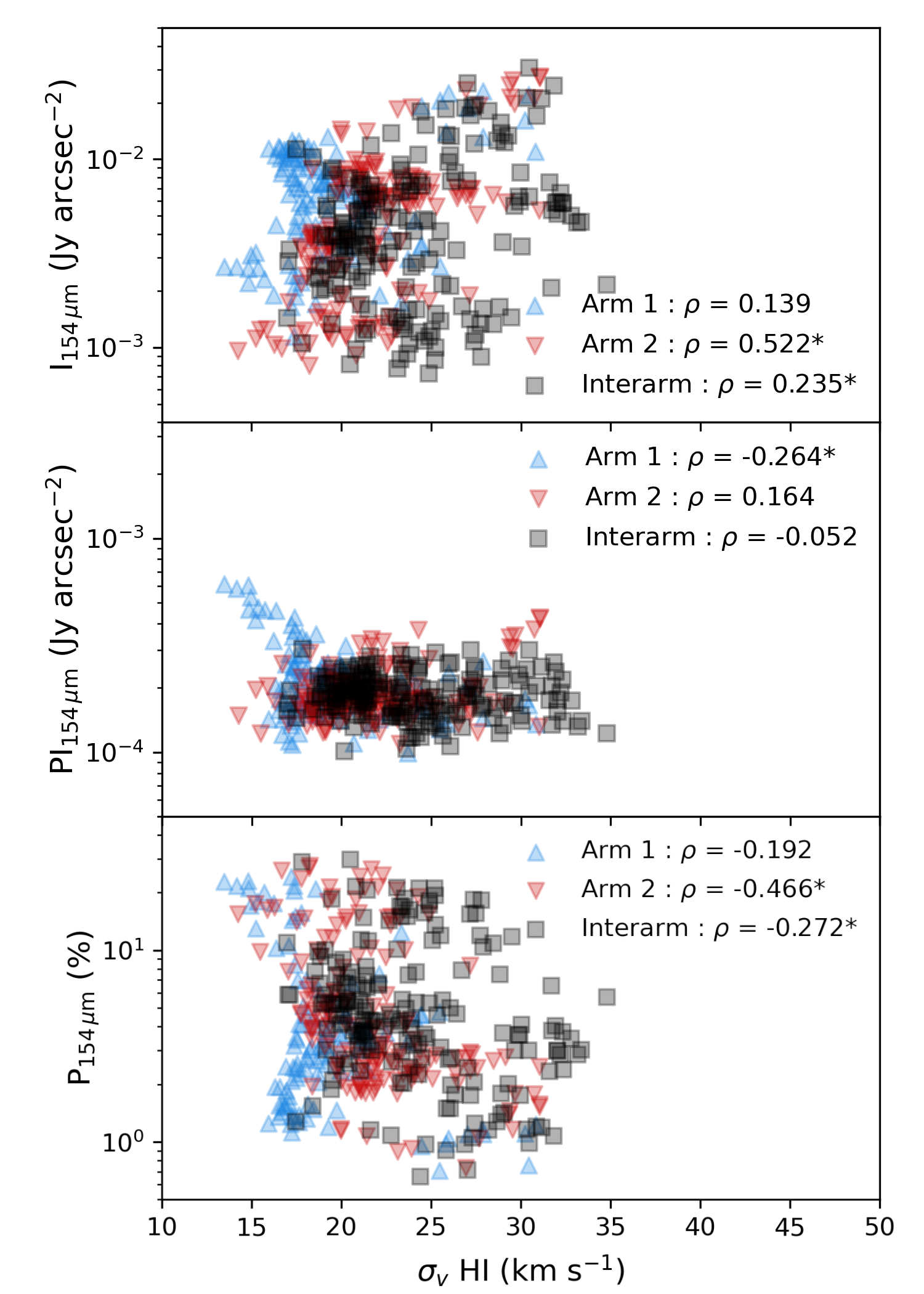}
   \includegraphics[trim={0 0 0 0}, clip, width=0.495\textwidth]{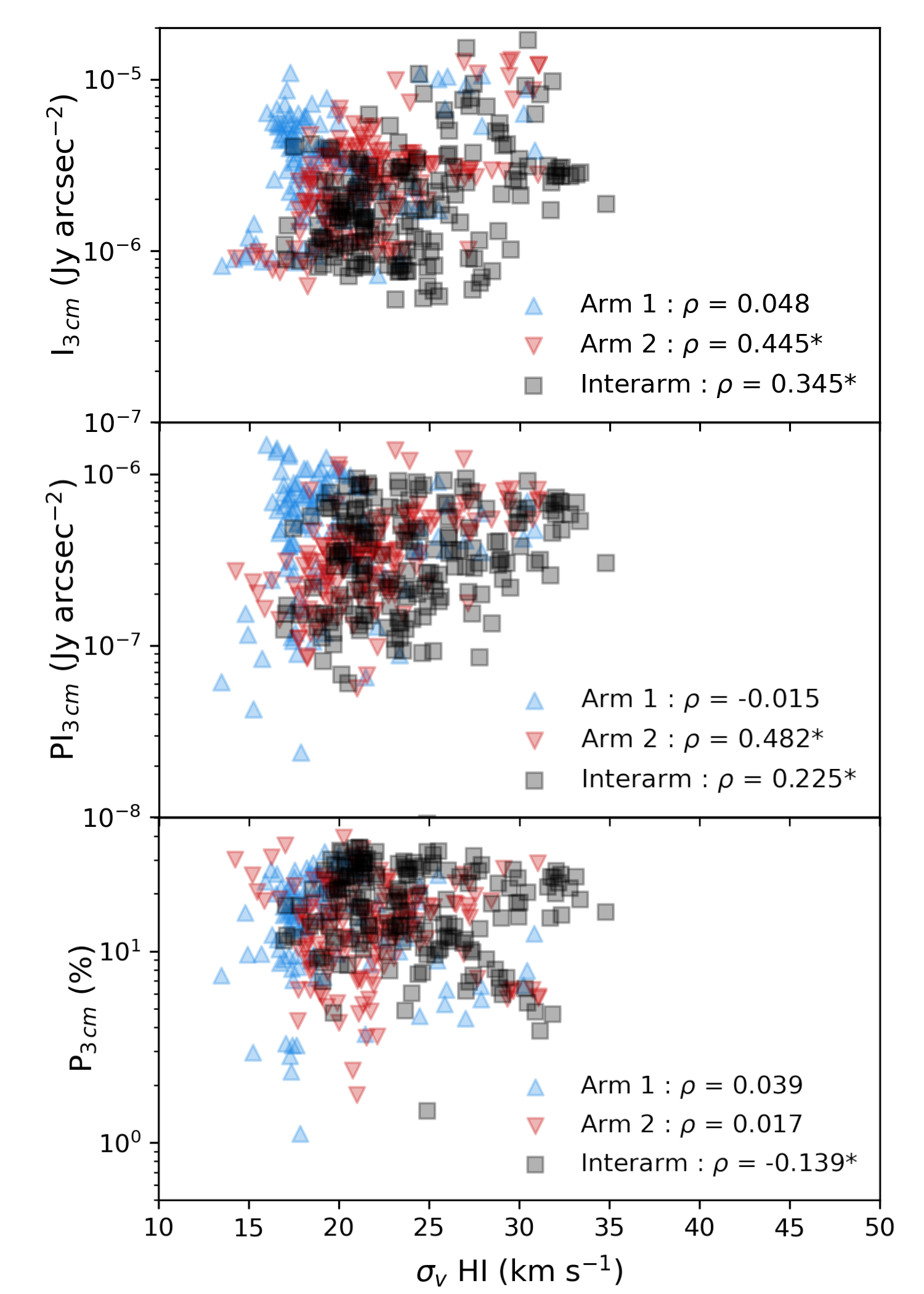}
  \end{minipage}

\caption{Comparison between \mum154\ (left column) and 3\,cm (right column) of the total intensity (top row), polarized intensity (central row) and polarization fraction (bottom row) as a function of \hi\ velocity dispersion ($\sigma_{v,{\mathrm{HI}}}$). Arm 1 (blue upward-pointing triangle), Arm 2 (red downward-pointing triangle), and interarms (black square), where each data point is a polarization measurement as shown in Figure \ref{fig:mag_pitch_arms_interarms}. See the legend on each panel for the correlation analysis. An asterisk symbol ($*$) following each $\rho$ correlation coefficient is shown if the correlation is statistically different from zero ($p<0.05$).} 
\label{fig:m51_comparison_HI}
\end{center}
\end{figure*}

To analyze how the different physical regimes of the multi-phase ISM affect the B-fields in M51, we use the velocity dispersion of the neutral and molecular gas as a proxy for the kinetic energy of the turbulence in the ISM. We also use the column density of the galactic disk to study the effect of extinction as a function of the FIR and radio polarization. 

In Fig.\,\ref{fig:m51_comparison_Nh} we analyze the variation of the total intensity (I), polarized intensity (PI), and polarization fraction (P) at \mum154\ and radio wavelengths as functions of the column density (\columndensity). All $\rho$ correlation coefficients in the figures are based on the Spearman non-parametric test. The distributions of the interarm, Arm 1, and Arm 2 regions is shown in the diagrams. We selected FIR polarization measurements with $PI/\sigma_{\mathrm{PI}} \ge 3$, $\sigma_{\mathrm{P}} \le 15$\%, $P \le 30$\%. For the selected measurements, the minimum SNR in polarization fraction equals 3. Note that we selected the cut in polarized flux such that it reduced any effects due to the positive bias of the polarization fraction. Medians of the physical parameters of Arm 1, Arm 2, and interarm zones studied in this section are shown in Table \ref{tab:multiphaseISM}. For simplicity, we only show here the diagrams in 3\,cm, but the same results are obtained in 6\,cm datasets (see Table \ref{tab:multiphaseISM}, and the 6\,cm radio polarization diagrams in Appendix \ref{appendix:PIplots_6cm}).

At 154 \um, we find a strong positive linear correlation between the total intensity and the column density \columndensity. Polarization fraction decreases with increasing column density, while the polarized intensity remains fairly constant across the full range of column densities, i.e. $\logten(\mathrm{\columndensity} [$cm$^{-2}]) = [21.0$--$22.1]$. The FIR polarization fraction is found to change in slope at $\logten(\mathrm{\columndensity} [$cm$^{-2}]) = 21.49^{+0.03}_{-0.02}$
(we follow the same method used in Sec.\,\ref{subsec:Methods_mask} to measure the \hi\ break). Using the relation between the optical extinction, $A_{V}$, and hydrogen column density, $N_{\mathrm{H}}$, relation $N_{\mathrm{H}}/A_{V} = (2.21\pm0.09)\times 10^{21}$ cm$^{-2}$ mag$^{-1}$ \citep{Guver2009}, the change in slope corresponds to an extinction of $A_{V} = 1.40^{+0.18}_{-0.12}$ mag. 



%
At radio wavelengths, the total intensity increases with the column density, with a slope of the $\logten(I)$ vs. $\logten(\mathrm{\columndensity} [$cm$^{-2}])$ relation of $1.16\pm0.03$. Radio polarization fraction is fairly constant within the full range of column densities, while the polarized intensity increases with increasing column density ($\rho>0.5$, $p<0.05$ in all components). For both FIR and radio, we find no strong, systematic differences in the trends and distribution of the Arm 1, Arm 2, and the interarm zone in any case (see $\rho$ correlation coefficients in the panels of Fig.\,\ref{fig:m51_comparison_Nh}).


In Fig.\,\ref{fig:m51_comparison_CO} we show the analysis as a function of the \lineco\ velocity dispersion ($\sigma_{v, \mathrm{^{12}CO(1-0)}}$). In the FIR, the total intensity increases with increasing the velocity dispersion of the molecular gas, the polarization fraction decreases with increasing the velocity dispersion of the molecular gas ($p<0.05$ in all components), while the polarized intensity remains fairly constant ($\rho<0.3$, not significant in Arm 2). The interarm region has lower dispersion velocity \citep[$p=4.8\cdot10^{-22}$, using the non-parametric two-sample comparison Anderson-Darling test,][]{AndersonDarling} than Arm 1 and Arm 2, dominating at $\sigma_{v, \mathrm{^{12}CO(1-0)}}\sim3$--5 km s$^{-1}$. Arm 2 present a more extended $\sigma_{v, \mathrm{^{12}CO(1-0)}}$ distribution than Arm 1, reaching values as high as 10 km s$^{-1}$. Both arms present a 1.0\% probability of having the same \lineco\ velocity dispersion. At radio wavelengths, the total intensity increases with increasing the velocity dispersion of the molecular gas, the polarization fraction is fairly constant across the full range of the velocity dispersion of the molecular gas, with $\rho\gtrsim-0.3$, and even this low trend is not statistically significant in Arm 1. We find an upward trend in the polarized intensity ($p<0.05$ in all components). As in the FIR, the radio polarization fraction is higher in the interarm with a probability of $p=4.7\cdot10^{-3}$ in FIR and $p=1.2\cdot10^{-4}$ in 3\,cm). This result is consistent with the fact that the \lineco\ velocity dispersion being lower in the interarm than in the arms. \citet{Fletcher2011} found an average polarization fraction of up to $40\%$ in the interarm regions, against a clearly reduced polarization fraction of up to $25\%$ in the spiral arms. 

\begin{deluxetable*}{lcccc}[]
\caption{Medians of the physical parameters of Arm 1, Arm 2, and interarm zones. Rows from top to bottom: 1--3) Total intensity for \mum154, 3\,cm, and 6\,cm. 4--6) Polarized intensity for \mum154, 3\,cm, and 6\,cm. 7--9) Polarization fraction for \mum154, 3\,cm, and 6\,cm. 10) \hi\ column density. 11) \lineco\ velocity dispersion. 12) \hi\ velocity dispersion.\label{tab:multiphaseISM}}
	\tablewidth{0pt}
		\tablehead{\colhead{Parameter}	& Wavelength &	\colhead{Arm 1} 	&	\colhead{Arm 2}	 &	\colhead{Interarm}\\}
		\startdata
		\multirow{3}{*}{I (Jy arcsec$^{-2}$)} & \mum154 &  $6.58^{+0.54}_{-0.36}\cdot10^{-3}$ & $5.10^{+0.56}_{-0.11}\cdot10^{-3}$ & $4.08^{+0.48}_{-0.26}\cdot10^{-3}$ \\
		& 3\,cm &  $3.48^{+0.26}_{-0.27}\cdot10^{-6}$ & $2.60^{+0.16}_{-0.16}\cdot10^{-6}$ & $1.88^{+0.17}_{-0.15}\cdot10^{-6}$ \\
		& 6\,cm &  $6.58^{+0.30}_{-0.37}\cdot10^{-6}$ & $4.68^{+0.35}_{-0.31}\cdot10^{-6}$ & $3.30^{+0.26}_{-0.19}\cdot10^{-6}$ \\
		\multirow{3}{*}{PI (Jy arcsec$^{-2}$)} & \mum154 & $2.04^{+0.15}_{-0.11}\cdot10^{-4}$ & $1.79^{+0.08}_{-0.08}\cdot10^{-4}$ & $1.86^{+0.08}_{-0.08}\cdot10^{-4}$ \\ 
		& 3\,cm & $5.41^{+0.62}_{-0.50}\cdot10^{-7}$ & $3.29^{+0.26}_{-0.26}\cdot10^{-7}$ & $3.43^{+0.18}_{-0.39}\cdot10^{-7}$ \\
		& 6\,cm & $9.80^{+0.80}_{-0.67}\cdot10^{-7}$ & $6.73^{+0.35}_{-0.34}\cdot10^{-7}$ & $7.58^{+0.63}_{-0.47}\cdot10^{-7}$\\
		\multirow{3}{*}{P (\%)} & \mum154 & $3.0^{+0.3}_{-0.3}$ & $3.5^{+0.4}_{-0.3}$ & $4.2^{+0.3}_{-0.3}$ \\
		& 3\,cm & $15.9^{+1.5}_{-1.5}$  & $13.5^{+1.0}_{-1.1}$ & $17.9^{+1.2}_{-1.1}$ \\ 
		& 6\,cm & $16.2^{+1.4}_{-1.3}$ & $14.7^{+1.0}_{-1.0}$ & $22.9^{+1.2}_{-1.3}$ \\
		$\log_{10}(\columndensity) [\mathrm{cm}^{-2}])$ &  & 	$21.66^{+0.01}_{-0.02}$ & $21.49^{+0.04}_{-0.02}$ & 	$21.40^{+0.02}_{-0.03}$        \\
		$\sigma_{^{12}\mathrm{CO}(1-0)}$ (km s$^{-1}$) &  & 	$5.77^{+0.13}_{-0.08}$ & $5.61^{+0.15}_{-0.22}$ & 	$4.29^{+0.09}_{-0.07}$ \\
		$\sigma_{\mathrm{HI}}$ (km s$^{-1}$) &  & 	$18.22^{+0.35}_{-0.20}$ & $21.34^{+0.32}_{-0.30}$ & 	$23.40^{+0.34}_{-0.21}$ \\ 
		\enddata
\end{deluxetable*}

In Fig.\,\ref{fig:m51_comparison_HI} we now present diagrams for the \hi\ velocity dispersion ($\sigma_{v,{\mathrm{HI}}}$). In general, the results show weaker correlations with \hi\ than with \lineco\ velocity dispersion in FIR and radio. The relation between the total intensity of FIR and 3\,cm with $\sigma_{v,{\mathrm{HI}}}$ presents a much lower correlation coefficient, which is only relatively mild-correlated in Arm 2 ($\rho\sim0.5$), but not well-correlated in the rest of the components. The results are similar for the polarization fraction and the polarized intensity. The FIR polarization intensity does not show any correlation with $\sigma_{v,{\mathrm{HI}}}$, and is very low in the case of 3\,cm. For the polarization fraction, we do not find any significant relation in FIR or 3\,cm with the velocity dispersion of \hi. 


\subsection{Star formation}
\label{subsec:results_sfr}
In this section, we study the relation between the star formation in the M51 disk and the magnetic fields. As described in Sec.\,\ref{sec:introduction}, one of the hypotheses that could explain potential differences between FIR and radio polarization maps is the effect of gas turbulence in star-forming regions. As supernovae explosions and winds inject the ISM with some level of turbulence, these mechanisms will generate a relationship between turbulence-driven B-fields and SFR. In addition, due to the effects of gravitational collapse, winds and star formation the magnetic field in the molecular gas clouds can present systematically different directions when compared to that of the diffuse ISM \citep[i.e.][]{Pillai2020}. 

Therefore, SFR-induced turbulence is expected to be a dominant effect. To test this hypothesis, we study the relation between polarization fraction and polarized intensity with the SFR. We obtained the SFR map from \citet[][]{Leroy2019}, which combined UV, NIR, and mid-IR photometry based on the Galaxy Evolution Explorer \citep[GALEX,][]{Martin2005} and the Wide-field Infrared Survey Explorer \citep[WISE,][]{Wright2010} and stellar population synthesis models to calibrate integrated SFR estimators. The SFR scales with the ISM density and this generally decreases with the galactocentric radius. Therefore, to compare different galactocentric radii in an equivalent way, we also normalize the SFR by the surface gas mass density to obtain the SFR efficiency in yr$^{-1}$. The gas mass density map was calculated multiplying the column density maps used in Sec.\,\ref{subsec:ISM_results} by the mean molecular weight $\mu$ and the hydrogen atomic mass ($m_\mathrm{H}$, see Sec.\,\ref{Sec:ArchivalData}). 

\begin{figure*}[t]
 \begin{center}

\includegraphics[trim={0 20 0 20}, clip, width=0.525\textwidth]{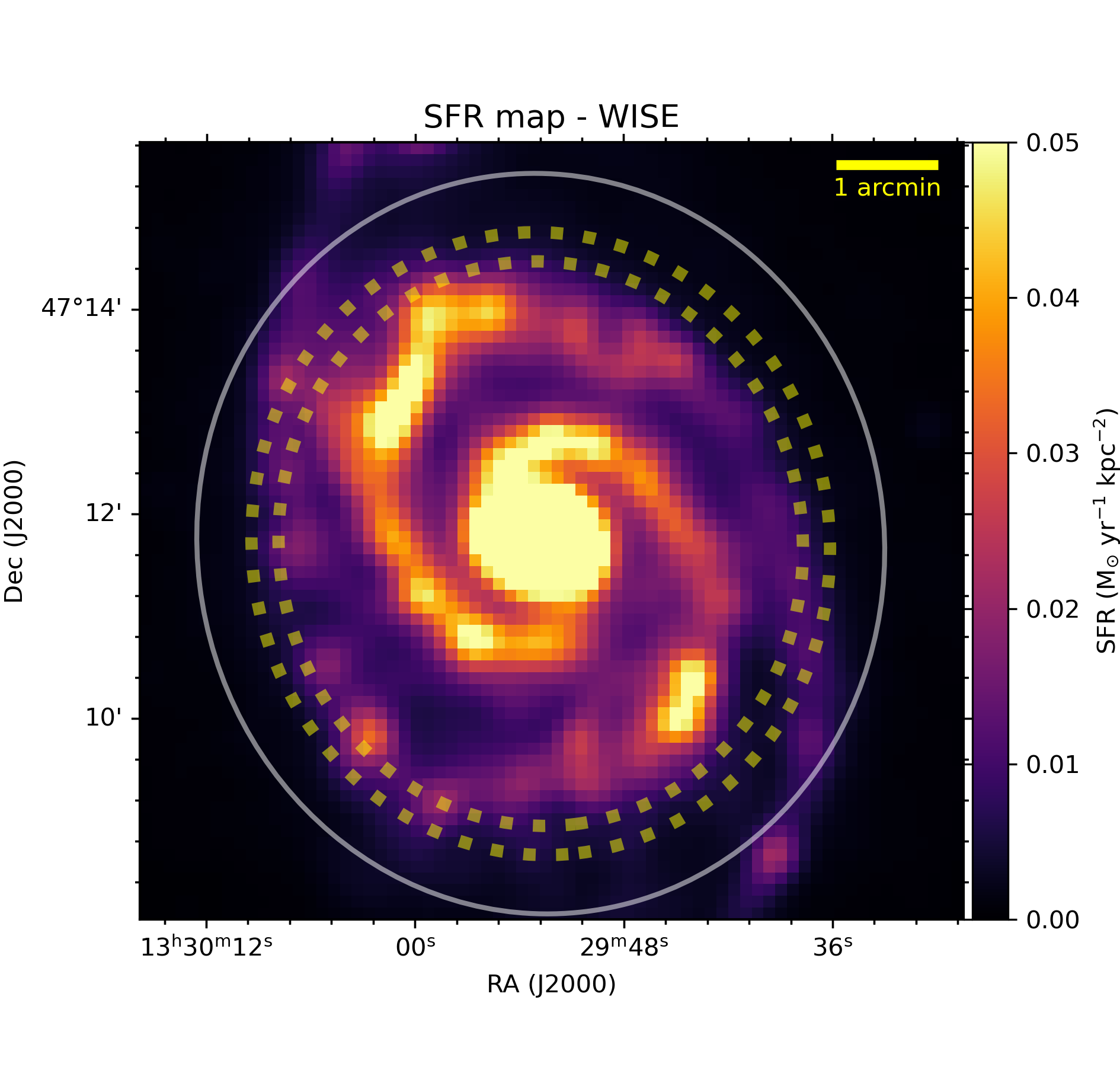}
\includegraphics[trim={50 20 0 20}, clip, width=0.468\textwidth]{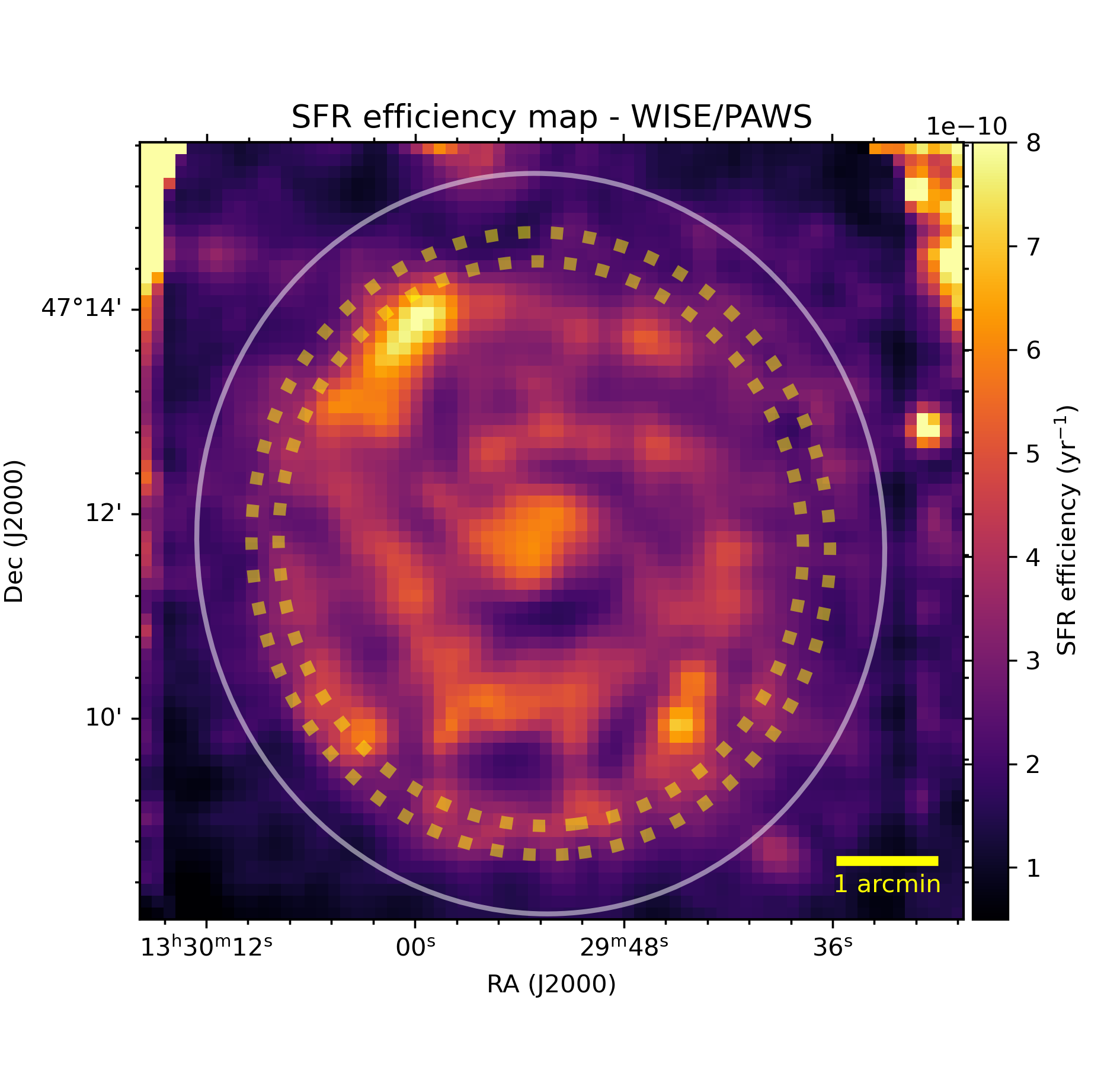}
\includegraphics[trim={60 0 75 0}, clip, width=\textwidth]{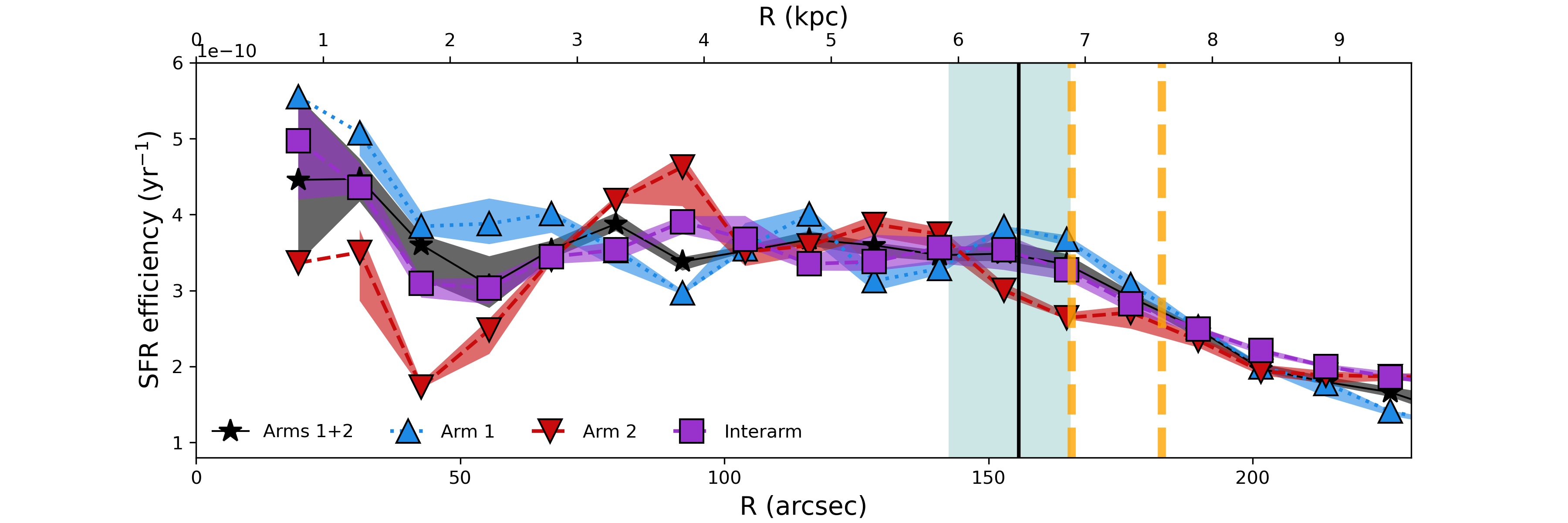}


\caption{Star formation rate efficiency analysis of \m51. \emph{Top left panel:} SFR map convolved to \hawc\ resolution, from WISE \citep{Leroy2019}. \emph{Top right panel:} SFR efficiency map, estimated from the previous SFR map and the gas mass. Yellow dashed ellipse represents the radius where the magnetic pitch angle from FIR and radio polarization observations decouple in the magnetic pitch angle profiles ($R=166\arcsec$--$183\arcsec$, see Fig.\,\ref{fig:mag_pitch_arm}). The white solid ellipse represents the maximum detection radius for \hawc\ observations. \emph{Bottom panel:} SFR efficiency radial profile, based on the two previous maps. Vertical yellow dashed lines represent the $R=166\arcsec$--$183\arcsec$ radii. Black solid vertical line and teal rectangle represent the SFR efficiency break median value and its 1$\sigma$ uncertainty interval, $R = 155.7^{+9.8}_{-13.2}$\arcsec ($6.47^{+0.41}_{-0.55}$ kpc). See the legend in the figure.} 
\label{fig:sfr_profile}
\end{center}
\end{figure*}

In Fig.\,\ref{fig:sfr_profile} we show the SFR efficiency analysis for \m51. The top panels show the SFR and SFR efficiency map for the area of M51 (reprojected to the \hawc\ resolution) where we have available FIR and radio observations. As a reference, we display two dashed ellipses at a galactocentric radius of $166\arcsec$ and $183\arcsec$ ($6.9$ and $7.6$ kpc), as an approximate limiting radius where the magnetic pitch angle profile of radio and FIR observations are compatible (Sec.\,\ref{subsubsec:results_morphological_pitch_angle}). On the one hand, the SFR map shows a smooth distribution very similar to the total intensity in FIR, with two well-defined spiral arms, and a bright inner region. On the other hand, the SFR efficiency map shows a clumpy structure, with knots of high efficiency in the outskirts of the spiral arms ($R\sim150\arcsec$, $6.2$ kpc) and lower values in the interarms. The bottom panel presents the average SFR efficiency radial profile for the spiral arms of M51. Interestingly, both spiral arms do not show similar trends in SFR efficiency. Arm 2 shows a lower value closer to the galactic center than Arm 1. Both arms present non-coincident peaks from the core to the outskirts. We find a decreasing trend in the SFR efficiency of both arms beyond $6.47^{+0.41}_{-0.55}$ kpc ($R_{\mathrm{break}}=155.7^{+9.8}_{-13.2}$ arcsec). This change in slope is significant at a $p<10^{-5}$ level. This might suggest that star formation processes might be playing a role in the same mechanism that produces the systematic differences between the FIR and radio magnetic pitch angle profiles found in Sec.\,\ref{Sec:magnetic_results}.

\begin{figure*}[ht!]
\begin{center}
\includegraphics[trim={60 0 80 20}, clip, width=\textwidth]{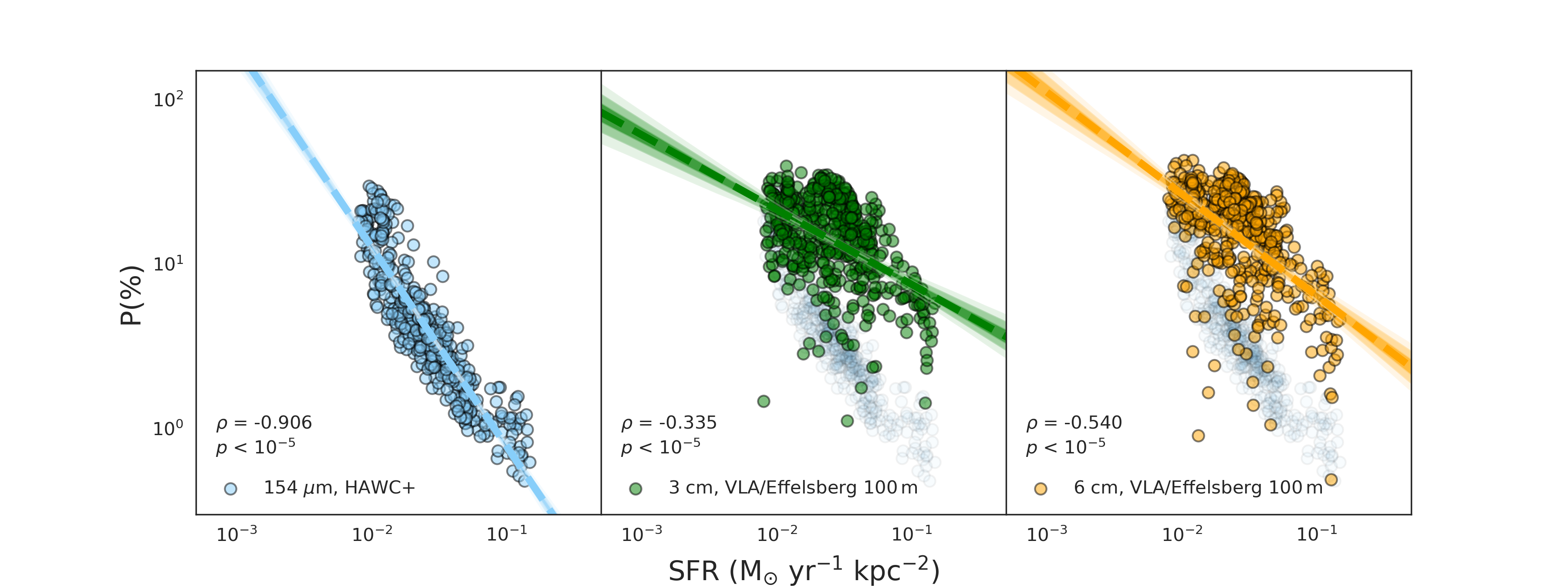}
\caption{Polarization fraction as a function of the SFR and wavelength (\mum154, 3\,cm and, 6\,cm, from left to right) for the \m51\ full disk. Each data point corresponds to an individual pixel positions in the \hawc\ and the convolved 3\,cm and 6\,cm data sets. Dashed line and contour represent the best linear fit to the diagram for each dataset. In the background of the central and right panels, we represent the \mum154\ data points, for visual reference. See the panels for the statistical correlation tests.} 
\label{fig:sfr_polarization}
\end{center}
\end{figure*}

\begin{figure*}[ht!]
\begin{center}
\includegraphics[trim={0 0 0 0}, clip, width=\textwidth]{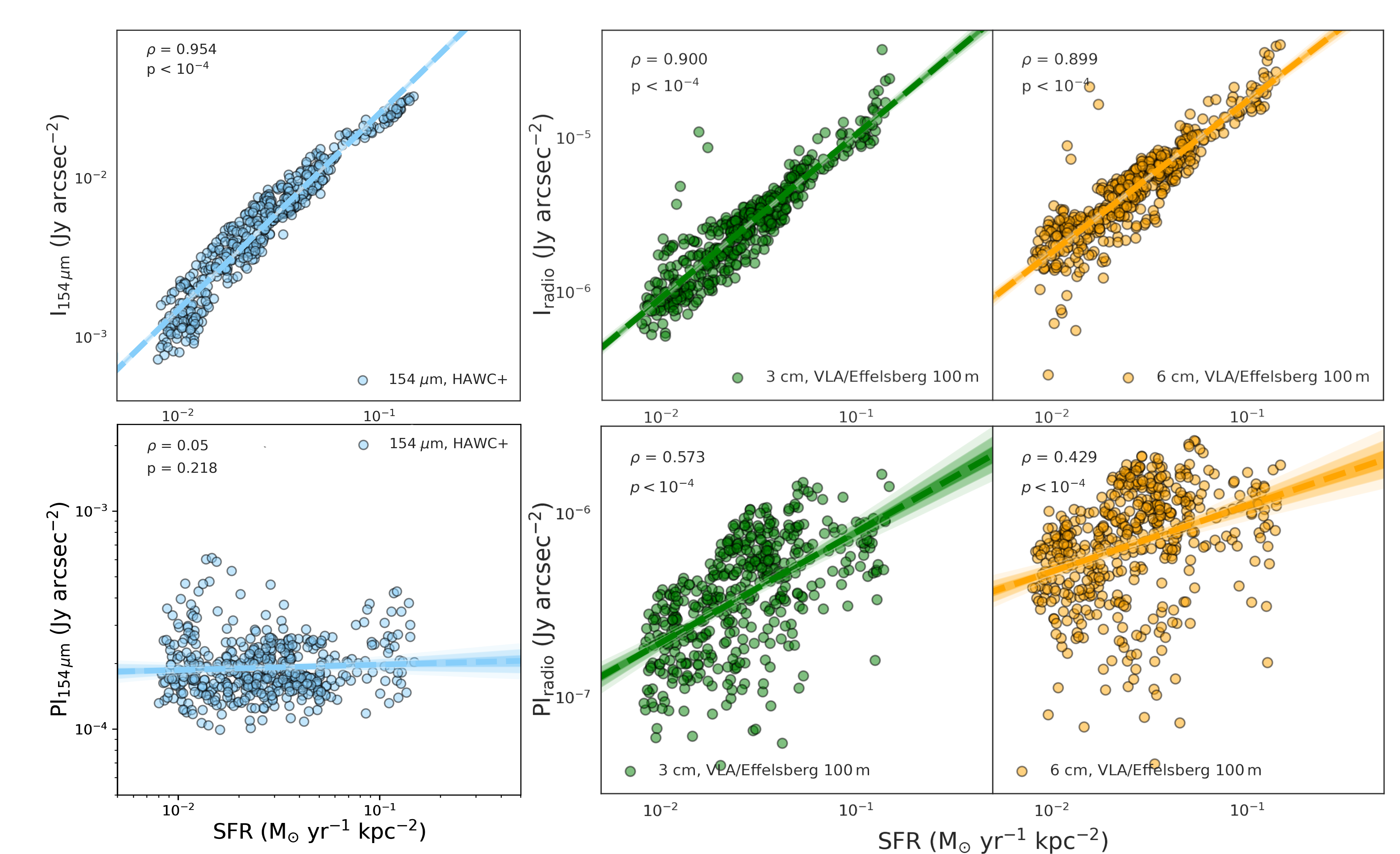}
\caption{Total (\emph{top}) and polarized intensity (\emph{bottom}) as a function of the SFR and wavelength (\mum154, 3\,cm and, 6\,cm, from left to right) for the \m51\ full disk. Linear fits are presented in Table \ref{tab:equationsSFR}. See the panels for the statistical correlation tests.} 
\label{fig:sfr_pi}
\end{center}
\end{figure*}

In Fig.\,\ref{fig:sfr_polarization} we explore the overall effect of the SFR over the polarization fraction for the \hawc, 3\,cm, and 6\,cm datasets. Interestingly, we found that there is a significant anti-correlation between the polarization fraction and the SFR in M51. This correlation is steeper and more correlated in FIR ($\rho=-0.906$) than in radio ($\rho=-0.335$ for 3\,cm, and $\rho=-0.540$ for 6\,cm). For the three wavelengths, the correlation coefficients are significant at a level of $p<10^{-5}$. Linear modelling of the log-scaled SFR and polarization fraction diagrams ($\logten(P) = a \logten{{\rm SFR}} + b$) for the different wavelengths show that the 3\,cm and 6\,cm show a variation of $P$ with the SFR shallower than that detected in the FIR data (see Table \ref{tab:equationsSFR}).

\begin{deluxetable}{crrr}[]
\tablecaption{Linear fits to the relations between total intensity (rows 1 -- 3), polarized intensity (4--6), and polarized fraction (7--9), for \mum154, 3\,cm, and 6\,cm as a function of the SFR. Row 10 shows the results for the $P_{154\,{\mu m}}$ vs. $I_{154\,{\mu m}}$ model. \label{tab:equationsSFR}}
	\tablewidth{0pt}
		\tablehead{\colhead{ID}
		 & \colhead{Equation} & Slope & Intercept}
\startdata
1 & $\logten{I_{154\,{\mu m}}} - \logten{\rm SFR}$ & $1.237^{+0.016}_{-0.016}$ & $-0.358 ^{+0.023}_{-0.023}$\\ 
2 & $\logten{I_{3\,{\rm{cm}}}} - \logten{\rm SFR}$ &  $1.061^{+0.019}_{-0.020}$ &  $-3.917^{+0.029}_{-0.030}$\\ 
3 & $\logten{I_{6\,{\rm{cm}}}} - \logten{\rm SFR}$ &  $0.981^{+0.023}_{-0.022}$ & $-3.781^{+0.035}_{-0.033}$\\
4 & $\logten{PI_{154\,{\mu m}}} - \logten{\rm SFR}$ & $0.024^{+0.021}_{-0.022}$ & $-3.68 ^{+0.033}_{-0.034}$\\
5 & $\logten{PI_{3\,{\rm{cm}}}} - \logten{\rm SFR}$ & $0.603^{+0.039}_{-0.038}$ & $-5.49^{+0.063}_{-0.061}$\\
6 & $\logten{PI_{6\,{\rm{cm}}}} - \logten{\rm SFR}$ & $0.360^{+0.038}_{-0.037}$ & $-5.59^{+0.061}_{-0.063}$\\
7 & $\logten{P_{154\,{\mu m}}} - \logten{\rm SFR}$ & $-1.212^{+0.028}_{-0.029}$ & $-1.32 ^{+0.044}_{-0.044}$\\ 
8 & $\logten{P_{3\,{\rm{cm}}}} - \logten{\rm SFR}$ &  $-0.455^{+0.036}_{-0.033}$ & $0.420^{+0.058}_{-0.054}$\\ 
9 & $\logten{P_{6\,{\rm{cm}}}} - \logten{\rm SFR}$ &  $-0.620^{+0.037}_{-0.036}$ & $0.185^{+0.061}_{-0.061}$\\
10 & $\logten{P_{154\,{\mu m}}} - \logten{I_{154\,{\mu m}}}$ &  $-0.979^{+0.016}_{-0.018}$ & $-1.670^{+0.036}_{-0.041}$\\
\enddata
\end{deluxetable}

We test the SFR correlation against the total and polarized intensity for the radio and FIR in Fig.\,\ref{fig:sfr_pi}. We find that the FIR polarized intensity does not correlate with the SFR ($p=0.218$), but we find a positive correlation in 3\,cm and 6\,cm ($\rho\sim 0.57$--$0.43$, $p<10^{-4}$). We find a positive correlation between the FIR and radio total intensity with the SFR (Table \ref{tab:equationsSFR}). This result is expected due to the FIR-radio correlation \citep{deJong1985} and the fact that the SFR is a function of total IR intensity, among other factors \citep{Leroy2019}. For radio, the total intensity increases faster than the polarized intensity with increasing of the SFR across the galaxy. 

In conclusion, we have found that there is a significant anti-correlation of the FIR polarization fraction with the SFR in M51, which does not translate into a correlation of the polarized intensity. In contrast, the radio polarized intensity does increase systematically at higher levels of SFR. The linear regression fit for the observed relation between the polarization fraction and SFR is compatible for the 3\,cm and 6\,cm observations, but not with the \mum154\ FIR dataset of M51. The observations of \hawc\ reveal that the polarization fraction in FIR is highly anti-correlated with the SFR, showing even lower values for polarization fraction at similar levels of SFR when compared to that predicted by radio observations. For the polarized intensity, we also find a different behavior in the FIR and radio: 3\,cm and 6\,cm present a positive correlation between PI and SFR, while no correlation is observed in \mum154. We discuss the relevance of these results in Sec.\,\ref{DIS:SFR}.





\section{Discussion}
\label{Sec:Discussion}

\subsection{FIR vs Radio magnetic fields}

In this work we find that the magnetic pitch angles at radio (3\,cm and 6\,cm) and FIR (\mum154) are well aligned in the inner $R<160\arcsec$ ($<6.7$ kpc) radius of \m51, i.e. $R<160\arcsec$: $\Psi_{\mathrm{FIR}} \sim \Psi_{3\,\mathrm{cm}} \sim \Psi_{3\,\mathrm{cm}}$. This result does not change when considering each one of the spiral arms independently, combined, using only the interarm region, or when analyzing the complete disk of \m51 at once. Only for the interarm region, the FIR and radio magnetic pitch angles are similar up to the largest radius ($220\arcsec$, $9.15$ kpc) of our observations, i.e. $R\le220\arcsec$: $\IAPsiFIR \sim \IAPsithreecm \sim \IAPsisixcm$. We find a significant difference between magnetic pitch angles of the arms at radio and FIR in the outer region ($R>160\arcsec$; $>6.7$ kpc) of \m51, i.e. $R>160\arcsec$: $\APsiFIR < \APsithreecm \sim \APsisixcm$. In the outskirts of \m51, the FIR magnetic spiral arms are wrapped tighter than the radio ones. The radio magnetic pitch angle seems to be more open at increasing radius from the core. Our study provides the first observational evidence of a morphological difference between the kpc-scale magnetic field structure between radio and FIR in external galaxies. 

We find that the morphological and magnetic pitch angles vary as a function of the ISM component such as $\mPsi_{\mathrm{HI}}^{\mathrm{Morph}} < \mPsi_{\mathrm{CO}}^{\mathrm{Morph}} < \MPsiFIR \sim \MPsithreecm \sim \MPsisixcm$ (see Sec.\,\ref{subsubsec:results_morphological_pitch_angle}). The spiral arms traced by the neutral gas (\hi) are wrapped tighter than those traced by the molecular gas observed in \lineco. Interestingly, the morphological pitch angles at radio and FIR are the same across the full extent ($220\arcsec$, $9.15$ kpc) of the galaxy, i.e. $\MPsiFIR \sim \MPsithreecm \sim \MPsisixcm$. However, the magnetic and morphological angles show different behavior across the galaxy disk. At low radii ($R<120\arcsec$, $R<5.0$ kpc), $\Psi^{\mathrm{Morph}}_{\mathrm{FIR,3\,cm,6\,cm}} > \Psi_{\mathrm{FIR,3\,cm,6\,cm}}$, while at larger radii $\Psi^{\mathrm{Morph}}_{\mathrm{FIR,3\,cm,6\,cm}} < \Psi_{\mathrm{FIR,3\,cm,6\,cm}}$. The exception is at FIR at radii $R>190\arcsec$ ($>7.9$ kpc), where $\Psi^{\mathrm{Morph}}_{\mathrm{FIR}} > \Psi_{\mathrm{FIR}}$. Although radio and FIR may be tracing the same morphological regions of the galaxy disk, we found that the magnetic pitch angle of the FIR differs at the outskirts of the galaxy. The FIR may be affected by a different physical mechanism in the outer regions of \m51\ (see Sec.\,\ref{DIS:multiphaseISM}).

The statistical difference found between the morphological and the magnetic pitch angles in the disk of \m51\ at the three wavelengths analyzed may be a direct hint of the independence of the $\alpha$--$\Omega$ dynamo from the spiral density waves \citep{Beck2015a}. Differences between the magnetic and morphological pitch angles have been repeatedly found by previous authors: the average magnetic pitch angle of M\,83 is about $20^{\circ}$ larger than that of the morphological spiral arms \citep{Frick2016}. In M\,101, the ordered magnetic pitch angle is found to be $\sim 8^{\circ}$ larger than those from the morphological pitch angle of the \hi\ structures \citep{Berkhuijsen2016}. \citet{VanEck2015} found that, on average, the magnetic pitch angle is $\sim5-10^{\circ}$ more open than the morphological pitch angles using a sample of $20$ nearby galaxies, a conclusion also found by \citet{Mulcahy2017} in M\,74.

In theory, spiral magnetic fields can be compressed by density waves, modifying the magnetic pitch angle. This mechanism would create a difference in the arm-interarm region across the galaxy disk. The regular magnetic field in the spiral arm should be more similar to that of the morphological pitch angle than the interarm magnetic field. The magnetic pitch angle may be first compressed and ordered in the interface between the arm-interarm region. 
There may be a temporary and spatial disconnect between the morphological spiral arm and the magnetic spiral arm due to the relative action of the large-scale dynamos and the small-scale dynamos. Detailed modeling of the \m51\ galactic system based on these observations would be required to test the interaction of the spiral density waves with the $\alpha$--$\Omega$ dynamo. 




\subsection{The magnetic fields in the multi-phase ISM}
\label{DIS:multiphaseISM}


In Sec.\,\ref{subsec:ISM_results} we found that the radio and FIR total intensity emission are both tightly correlated with the column density \columndensity\ and the \lineco\ velocity dispersion. This result and the implicit radio--FIR correlation were explained by \citet{Niklas1997}. In addition, we find that the FIR polarization decreases with increasing the velocity dispersion of the molecular gas and increasing column density. The interarm shows lower velocity dispersion and a higher degree of polarization than the arms. As the velocity dispersion is used as a proxy for the turbulent kinetic energy in the disk, a possible interpretation is that the small-scale turbulent magnetic field may be relatively more significant at higher velocity dispersion of the molecular gas and column densities than the large-scale ordered field.  

In addition, our results show that the FIR and radio total intensity, polarized intensity, and polarization fraction do not correlate with turbulence in \hi. Using magneto-hydrodynamic simulations, \citet{Dobbs2008} suggested that the small-scale turbulent component is produced by the velocity dispersion of the dust and cold gas. This turbulent component would be generated by the passage through a spiral shock. The authors found that without the cold gas component, the B-field remains well ordered apart from being compressed in the spiral shocks. Our results suggest that the small-scale turbulent field is then coupled to the molecular gas motions but not to the neutral gas of \m51. The molecular gas motions are more concentrated in the densest regions of the spiral arm and spatially coincident with the star-forming regions along the arms. 

These results suggest that the regions with higher column density and higher levels of turbulence of the molecular gas \lineco\ reduce the measured FIR polarization fraction inside each beam.  The polarized intensity is not affected by these quantities. The polarization fraction at radio wavelengths seems to be insensitive to the column density and the level of turbulence of the molecular gas, instead, the polarized radio emission is affected by these quantities. We find that both FIR and radio are insensitive to the turbulence in the neutral gas (\hi) across the galaxy disk. Interestingly, \citet{Beck2019} found no evidence of a spiral modulation of the root-mean-square turbulent speed when compared the velocity dispersion of \hi\ with the radio polarization of several spiral galaxies (\m51\ included).


\subsection{Star-formation and magnetic fields}
\label{DIS:SFR}



In Sec.\,\ref{subsec:results_sfr} we found a systematic anti-correlation between the polarization fraction and the SFR. Similar results were obtained earlier by \citet[using H$_\alpha$ emission and 6.2\,cm radio polarization]{Frick2001} and \citet{Tabatabaei2013} in NGC6946. The results of our work indicate that both FIR and radio polarization fraction are anti-correlated with the SFR. Interestingly, the polarized intensity at \mum154\ shows a negligible correlation with the SFR, \columndensity, and \lineco\ velocity dispersion, whereas PI increases at 3\,cm and 6\,cm. In the diffuse ISM, the polarization fraction will decrease due to 1) an increase of the relative contribution of unpolarized thermal emission from SFR, 2) Faraday depolarization, and 3) variations of the B-field orientation within the beam and along the LOS. Processes related to star formation (small-scale dynamo) would induce the formation of an anisotropic B-field component from the isotropic turbulent field, hence increasing the polarized intensity in 3\,cm and 6\,cm. The polarized intensity may increase if the relative contribution of anisotropic turbulent fields increases within the beam. However, the PI distributions in FIR show no correlation with SFR, \columndensity, or turbulence. Two different scenarios may explain this result:

\begin{enumerate}
    \item Different magnetic field directions in the same line of sight or within the same beam decrease the polarization intensity in FIR \citep{Fissel2016}

    \item Effects on the dust grain alignment efficiency as a function of the total intensity towards regions of high column density \citep{Hoang2021}.
\end{enumerate}


In the first scenario, the turbulence and morphological complexity of the B-field in and around the molecular clouds may cause beam depolarization at FIR wavelengths. Considering this hypothesis, the relative physical size of the HAWC+ beam at 154$\mu$m is 13.6\arcsec, approximately 565\,pc at a distance of 8.58\,Mpc. If we compare this with the size distribution of the giant molecular clouds of \m51, which ranges from 9 to 190\,pc in radius, with an average of $\sim$50\,pc \citep{Hughes2013}, we find that the vast majority of these clouds and their structure are unresolved with our spatial resolution. Thus, the complex B-field in the plane of the sky within our beam and/or tangled B-field along the LOS towards the cores of these structures causes a drop of polarization in our observations (i.e. depolarization).

%

The second proposed mechanism is based on a loss of dust grain alignment efficiency towards regions of high column density and gas turbulence. According to the Radiative Alignment Torques theory \citep[RAT,][]{Dolginov1976, Lazarian2007}, dust grain alignment efficiency decreases for grains smaller than a certain size ($a_{\mathrm{crit}}$) with column density due to collision dumping effect \citep{Hoang2021}. Specifically, higher gas density causes a stronger loss of alignment by gas-collision (which affects more efficiently smaller grains). This effect changes the population of aligned grain sizes to larger dust grains, i.e. the grain-size distribution of aligned grain is narrower, which makes P decrease with increasing intensity and \columndensity. In addition, P decreases with increasing gas turbulence (velocity dispersion of gas) because the gas turbulence randomizes and/or changes the position angle of polarization along the LOS. This effect results in a decrease of P as $\sigma_{v, \mathrm{^{12}CO(1-0)}}$ increases, consistent with the first proposed scenario considering RATs.

To quantify this effect, the polarization fraction has been found to depend on a certain power of the total intensity \citep[P $\propto$ I$^{\xi}$,][]{Hoang2021}. The power depends on the dust grains alignment efficiency, where $\xi=0$ corresponds to full alignment (perfectly polarized dust grain population), $\xi=-1$ to pure random alignment, and $\xi=-0.5$ alignment dominated by gas turbulence. As PI$ = $P$\cdot$I, PI becomes constant as $\xi$ decreases. In the case of \m51, we measure $\xi = -0.979^{+0.016}_{-0.018}$ (see Table \ref{tab:equationsSFR}), which implies a pure random alignment regime. In this regime, PI is constant with I, \columndensity, and $\sigma_{v, \mathrm{^{12}CO(1-0)}}$.

These hypotheses for the variation of the polarization fraction of intensity, as well as the lower anti-correlation found in radio observations when compared to FIR, require further investigation, which is beyond the scope of this manuscript.

We cannot connect directly the variation of the polarization fraction with the inner structure of the magnetic field in radio. In order to do that, we would need to take into account the added factor of Faraday depolarization \citep{Sokoloff1998} and the increase in unpolarized thermal emission, which can be significant at 3\,cm and 6\,cm. Regions with higher SFR present higher molecular gas densities, cold gas velocity dispersion, and higher neutral gas column densities. This can be associated with a decrease in the polarization fraction in FIR, but also with an increase of the total FIR intensity.

The SFR efficiency profile does show a significant decrease at $R = 155.7^{+9.8}_{-13.2}$\arcsec ($6.47^{+0.41}_{-0.55}$ kpc) of the galactic disk. The different SFR efficiency profiles between both arms suggest that an asymmetric structure, possibly triggered by the interaction of the galactic disk with the companion galaxy, M51b. In addition, we do not observe the misalignment of the FIR magnetic field outside the spiral arms. This distortion is found in the outermost radius of \m51, close to the radii where Arm\,2 is closer to M51b. While there is an agreement of the general structure between the magnetic field of the spiral arms in the molecular gas and the diffuse ISM for the inner region, the magnetic pitch angle break is only found in FIR and not in the radio polarization observations. These results may suggest that the molecular disk might be more affected by the interaction with M51b than the diffuse gas. This result is expected since the molecular gas is a kinematically colder component of the galactic disk than the diffuse, more dispersion-supported gas. \citet{Iono2005} found significant differences ($\Delta v > 50$ km s$^{-1}$) between the diffuse gas and molecular disk kinematics in the rotation curves of a sample of galaxy interacting pairs observed in \hi\ and \co. This suggests that the distortion of the magnetic pitch angle profile found in the outskirts could be produced by the interaction of M51b with the cold dense molecular disk, visible on both sides of the galaxy due to the effect of gravitational tidal forces \citep{Duc2013}. Galaxy interactions could affect the diffuse gas differently from the molecular gas, which is kinematically colder, with a highly rotation supported distribution \citep{Drzazga2011}. In that case the location of the molecular clouds preferentially in the spiral arms of \m51\ could explain that the find a misalignment between the two components of the magnetic field. Indeed, large angular dispersion in the measured magnetic field due to the interaction of galaxies has been recently found using $89$ \um\ polarization data of Centaurus\,A by \citet{ELR2021b}. This author found that the small-scale turbulent fields have a larger contribution than large-scale ordered fields in the molecular gas of the remnant warped disk. The fact that we find the distortion on both spiral arms would require a detailed MHD study of the effects of tidal forces on galactic disks, and the previous history of the \m51\ interaction. The most drastic feature is the down-bending break of the magnetic pitch angle profile in FIR.

\citet{VanEck2015} and \citet{Chyzy2017} found a tight relationship between the specific SFR and the total magnetic field strength, which implies that the process of amplifying magnetic fields in galaxies is mainly driven by small-scale dynamo mechanisms from local SFR \citep{Gressel2008b,SB2013}. The results from \citet{Chyzy2017} show that the total magnetic field is correlated with the density of the cold molecular gas (H$_2$) but not with the warm diffuse \hi\ interstellar medium, a result that is compatible with our findings in Sec.\,\ref{subsec:ISM_results}. This shows that the amplification of the B-fields may be taking place in the star-forming regions of \m51. This amplification may be driven by small-scale turbulent dynamos, where small-scale refers to scales smaller than our beam size and spatially correlated with the star-forming regions along the spiral arms.

\section{Conclusions}
\label{Sec:Conclusions}

One of the most important and unexplored questions in galaxy evolution is \emph{Can magnetic fields shape galaxies?} \citep{Battaner2007, RuizGranados2010, Tsiklauri2011, RuizGranados2012, Jalocha2012a, Jalocha2012b, Elstner2014}. Previous analysis on this topic based their conclusions on the structure of the radio polarization magnetic field, corresponding to the diffuse ISM. In this paper, we present quantitative evidence that the kpc-scale structure of the magnetic field in the molecular gas and the diffuse ISM of the grand design face-on spiral galaxy M51 shows significant differences in the structure:

\begin{enumerate}
    \item Within the inner $150\arcsec$ ($6.24$ kpc) of \m51\, we found a general agreement of the magnetic field orientation (measured as the magnetic pitch angle)  between the \mum154, 3\,cm and 6\,cm bands. At $R>150\arcsec$ ($>6.24$ kpc), the magnetic pitch angle profile at \mum154\ shows a significant break towards lower pitch angles, which is not detectable in 3\,cm or 6\,cm. 
    
    \item When the two individual spiral arms are compared, they show significantly different magnetic pitch angle profiles, consistently at all three wavelengths studied. The exception is found at the outer region ($R>150\arcsec$, $>6.24$ kpc) in 154\,\um. 
    
    \item Longer wavelengths have higher magnetic pitch angles in the arms, i.e. $\Psi_{\mathrm{FIR}}^{\mathrm{Arms}} < \Psi_{\mathrm{3\,cm}}^{\mathrm{Arms}} \sim \Psi_{\mathrm{6\,cm}}^{\mathrm{Arms}}$. 
    
    \item We do not find significant differences in the magnetic pitch angles of the interarm regions, i.e. $\Psi_{\mathrm{FIR}}^{\mathrm{IA}} \sim \Psi_{\mathrm{3\,cm}}^{\mathrm{IA}} \sim \Psi_{\mathrm{6\,cm}}^{\mathrm{IA}}$. 
    
    \item The morphological pitch angles at FIR and radio wavelengths are similar across the full disk of M 51. However, we found that morphological pitch angles change as a function of the multi-phase ISM, such as  $\Psi_{\mathrm{HI}}^{\mathrm{Morph}} < \Psi_{\mathrm{CO}}^{\mathrm{Morph}} < \Psi_{\mathrm{FIR}}^{\mathrm{Morph}} \sim \Psi_{3\,\mathrm{cm}}^{\mathrm{Morph}} \sim \Psi_{6\,\mathrm{cm}}^{\mathrm{Morph}}$.
    
    \item At FIR and radio and at radius $<100\arcsec$ ($<4.16$ kpc), the magnetic pitch angles are wrapped tighter than the morphological pitch angles.
    
    \item At FIR and radio and at radius $>100\arcsec$ ($>4.16$ kpc), the magnetic pitch angles of the spiral arms are larger than those from the morphological structure. The exception is the FIR, whose magnetic pitch angle becomes tighter than the morphological pitch angle at radius $>200\arcsec$ ($>8.32$ kpc).
    
\end{enumerate}

We also compared the FIR and radio polarization with the properties of the multi-phase ISM using the column density, velocity dispersion of the neutral (\hi) and molecular (\lineco) gas, and the SFR. Our results are:

\begin{enumerate}
    \item The FIR and radio total intensity are positively correlated with the hydrogen column density, and \lineco\ velocity dispersion.

    \item The FIR polarization fraction is negatively correlated with the total hydrogen column density (\columndensity) and the \lineco\ velocity dispersion. At radio, the polarization fraction is flat with these quantities.
    
    \item The FIR polarized intensity is flat with the column density and \lineco\ velocity dispersion. At radio, the polarization intensity increases with these quantities. Two different mechanisms (beam depolarization and dust grain alignment efficiency) are proposed in Sec.\,\ref{DIS:SFR} to explain the different trends observed in FIR.
    
    \item We found no correlation between the FIR and radio with the \hi\ velocity dispersion.
    
    \item The polarization intensity presents a significant correlation with the SFR in 3\,cm and 6\,cm, but none in \mum154\ observations. We found a tight anti-correlation between the polarization fraction and SFR in \mum154, 3\,cm and 6\,cm. 
    
    \item The two spiral arms show different trends as a function of SFR efficiency. Arm 2 shows a lower value closer to the galactic center than Arm 1. Both arms present non-coincident peaks from the core to the outskirts.
    
    \item We found a decreasing trend in the SFR efficiency of both arms beyond $6.47^{+0.41}_{-0.55}$ kpc ($R_{\mathrm{break}}=155.7^{+9.8}_{-13.2}$ \arcsec).

\end{enumerate}

The results detailed above point to an important observation: the multi-phase of the ISM affect the B-field structure in the galaxy. This effect can be disentangled by performing a multi-wavelength approach using the FIR and radio polarization observations. Our observations support the presence of a clear interlinked scenario between the SFR and the magnetic field in different phases of the ISM. Lower polarization fractions may be due to the presence of magnetized but complex structures in the regions with denser molecular clouds. The location of the arm, interarm and core components used to produce the diagrams of polarization fraction and intensity support this interpretation. 

The diffuse ISM presents a much more regular magnetic field than the cold dense molecular gas, and this is revealed in the structure of the magnetic pitch angle profiles. It is interesting that these magnetic fields show differences from the pitch angle structure of the morphological arms, supporting the separation of the $\alpha$--$\Omega$ dynamo from the density waves. The observed differences between the radio parameters and those of the FIR might be produced by kinematic decoupling between the diffuse and dense ISM through the tidal forces with the companion galaxy M51b. However other effects, such as internal kinematic phenomena associated with density wave resonances cannot be ruled out. These effects are beyond the scope of this paper and will be studied in a forthcoming publication. 

It remains unknown if the magnetic fields can systematically influence the global kinematics of the star-forming regions inside the molecular clouds, enhancing stellar migration. Observational testing of such a hypothesis can only be obtained through a revision of our analysis based on the magnetic field structure of molecular clouds in galaxies. High-resolution, FIR polarization observations of galaxies such as those provided by HAWC+/SOFIA are vital to understanding the role of magnetic fields in the evolution of the Universe. Ongoing efforts like the SOFIA Legacy Program (PIs: Lopez-Rodriguez \& Mao) will provide deeper FIR polarimetric observations of a sample of nearby galaxies, where, combining them with observations of radio and other tracers, we should be able to disentangle the relation between the SFR and the magnetic structure of the molecular clouds within the galactic disks. 

\begin{acknowledgments}
We thank Joan Schmelz, Mahboubeh Asgari-Targhi, Kassandra Bell, Rick Fienberg for their support and advice during this work. A.B. was supported by an appointment to the NASA Postdoctoral Program at the NASA Ames Research Center, administered by Universities Space Research Association under contract with NASA. K.T. has received funding from the European Research Council (ERC) under the European Unions Horizon 2020 research and innovation programme under grant agreement No. 771282. J.E.B was supported by project P/308603 of the Instituto de Astrofísica de Canarias. S.E.C. acknowledges support by the Friends of the Institute for Advanced Study Membership. Based on observations made with the NASA/DLR Stratospheric Observatory for Infrared Astronomy (SOFIA) under the 70\_0509, 76\_0003, and 08\_0260 Programs. SOFIA is jointly operated by the Universities Space Research Association, Inc. (USRA), under NASA contract NNA17BF53C, and the Deutsches SOFIA Institut (DSI) under DLR contract 50 OK 0901 to the University of Stuttgart. This work made use of THINGS, 'The \hi\ Nearby Galaxy Survey' \citep{Walter2008}.
\end{acknowledgments}

\vspace{5mm}
\facilities{SOFIA (\hawc)}


\software{ \textsc{Python} \citep{van1995python}, \textsc{R} \citep{R}, \textsc{Astropy} \citep{Astropy}, 
\textsc{APLpy} \citep{Robitaille2012},
\textsc{Matplotlib} \citep{Matplotlib}, \textsc{Pandas} \citep{pandas}, , \textsc{Anaconda} \citep{anaconda}, \textsc{Seaborn} \citep{waskom2020seaborn}, 
          }

\clearpage

\appendix




\section{Mock magnetic field test}
\label{appendix:mock_pitch}

\begin{figure*}[t]
 \begin{center}
\includegraphics[trim={0 0 0 0}, clip, width=0.49\textwidth]{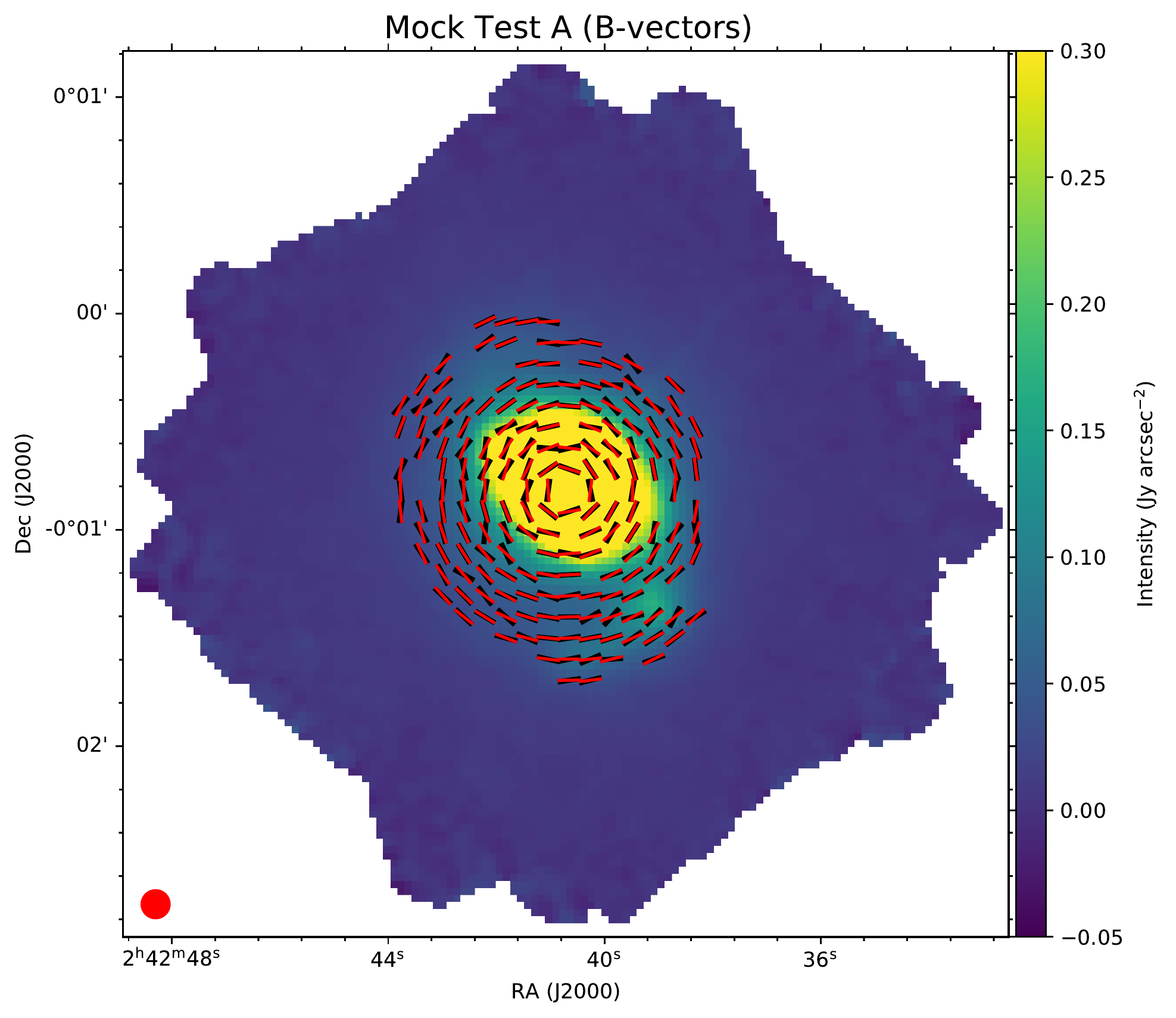}
\includegraphics[trim={0 0 0 0}, clip, width=0.49\textwidth]{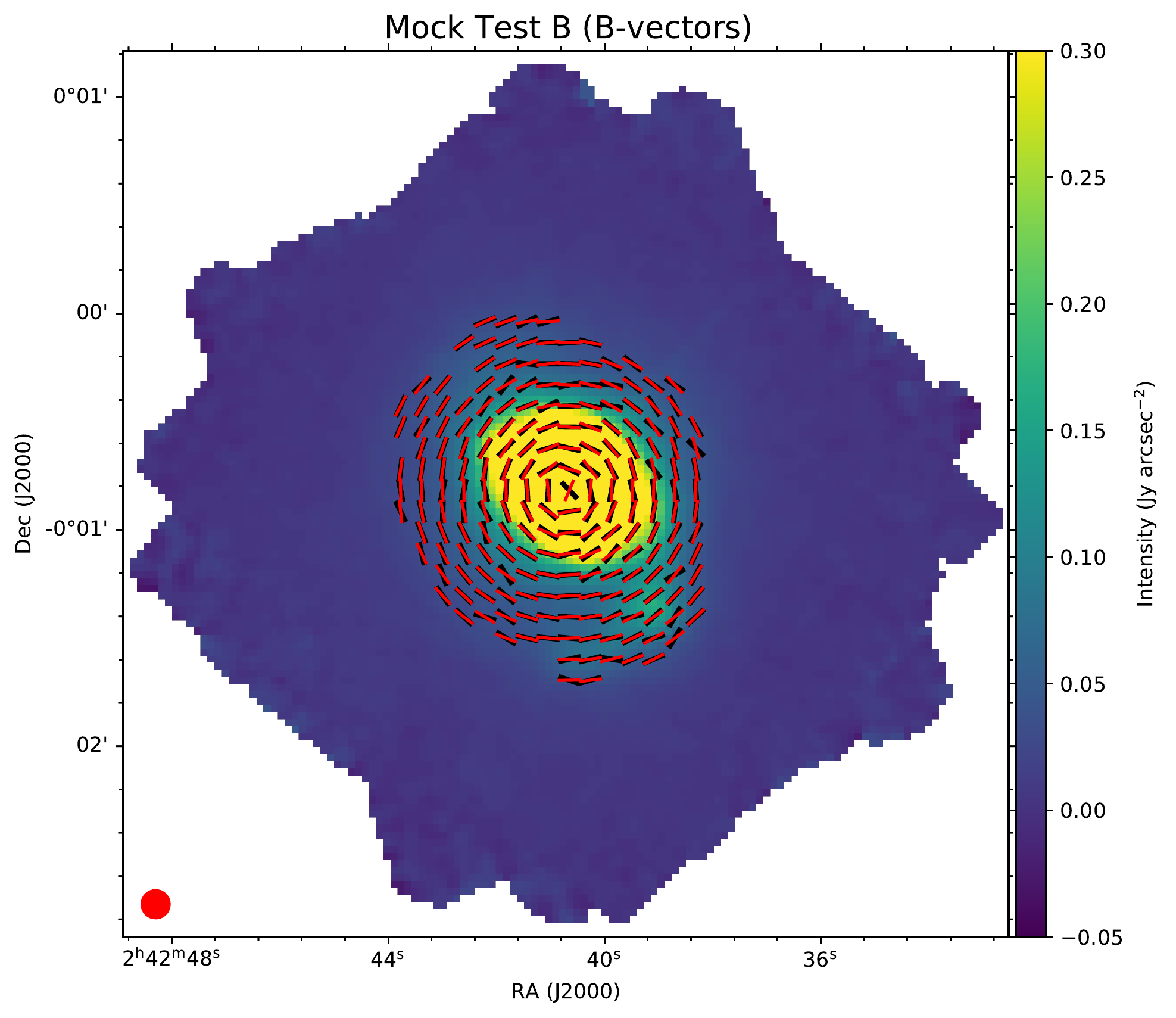}
\includegraphics[trim={0 0 0 0}, clip, width=0.49\textwidth]{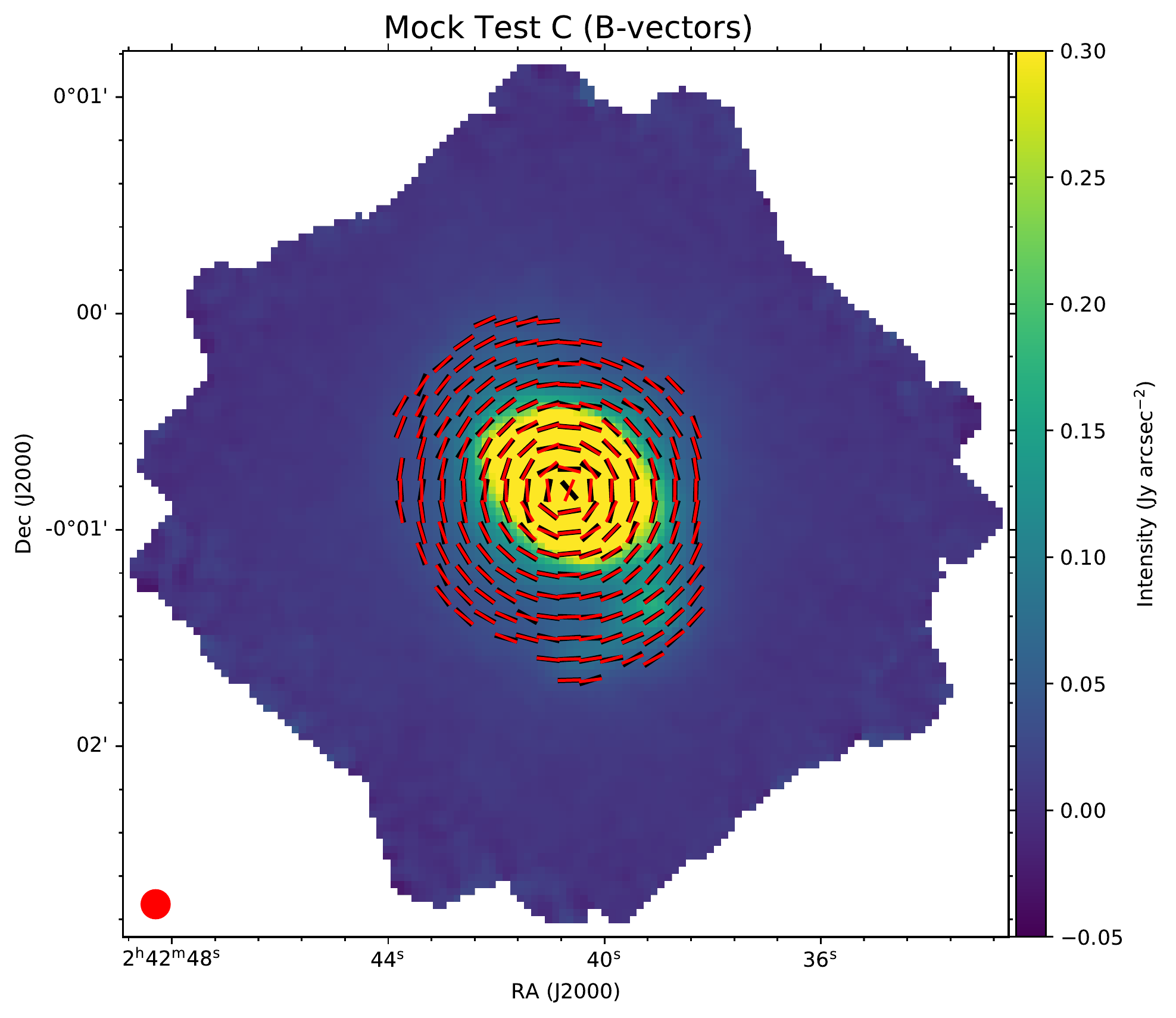}
\includegraphics[trim={0 0 0 0}, clip, width=0.49\textwidth]{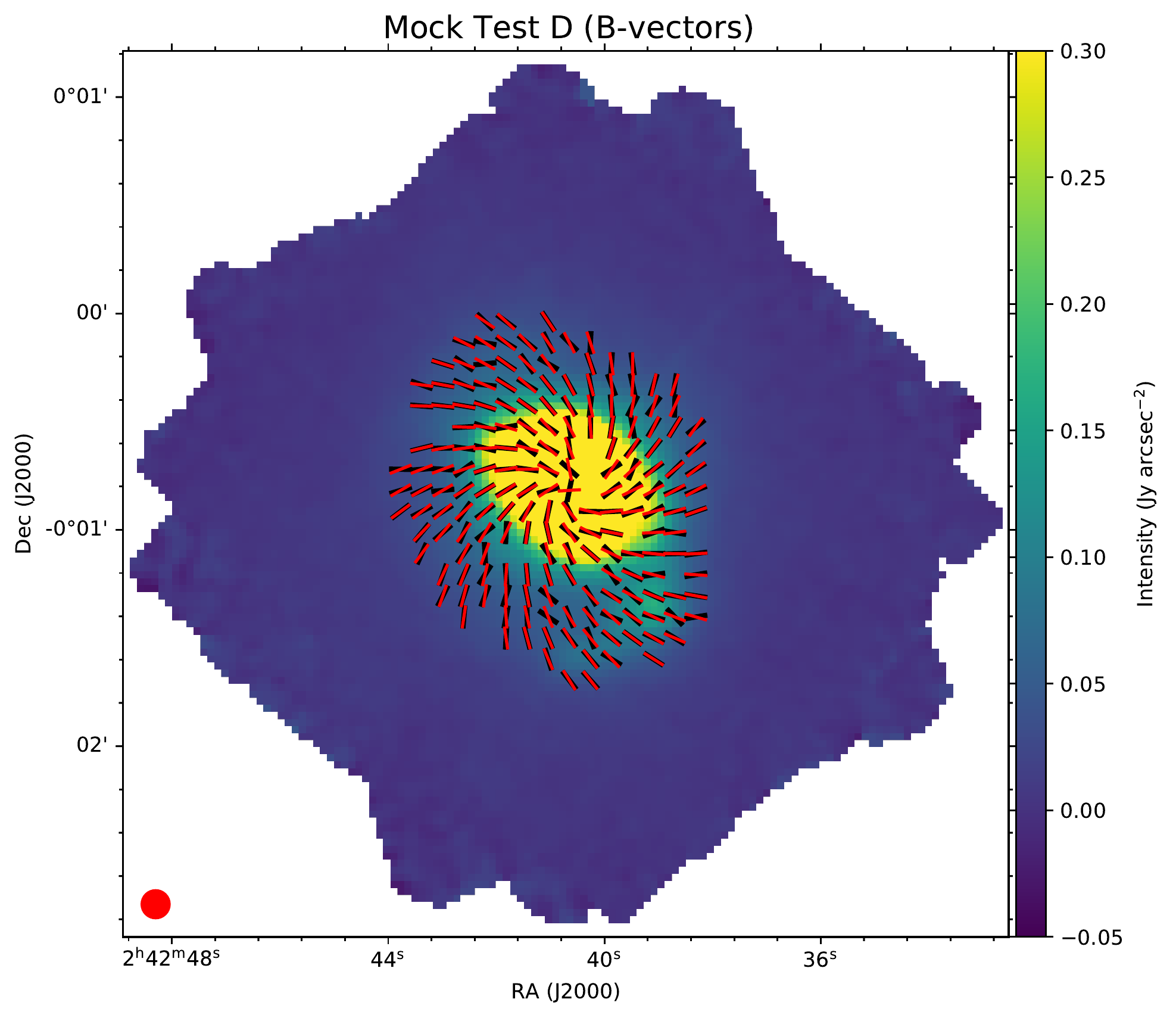}

\caption{Mock observations of the spiral B-field. Total intensity (colorscale) maps show the $89$ \um~HAWC+ observations of NGC1068 by \citet{LopezRodriguez2020}. Mock B-field orientations (black) and model (red) are shown for tests ABCD with the parameters shown in Table \ref{tab:app_mocktests}.}
\label{fig:mock_pitch_angle1}
\end{center}
\end{figure*}

\begin{figure*}[t]
 \begin{center}
\includegraphics[trim={0 0 0 0}, clip, width=0.49\textwidth]{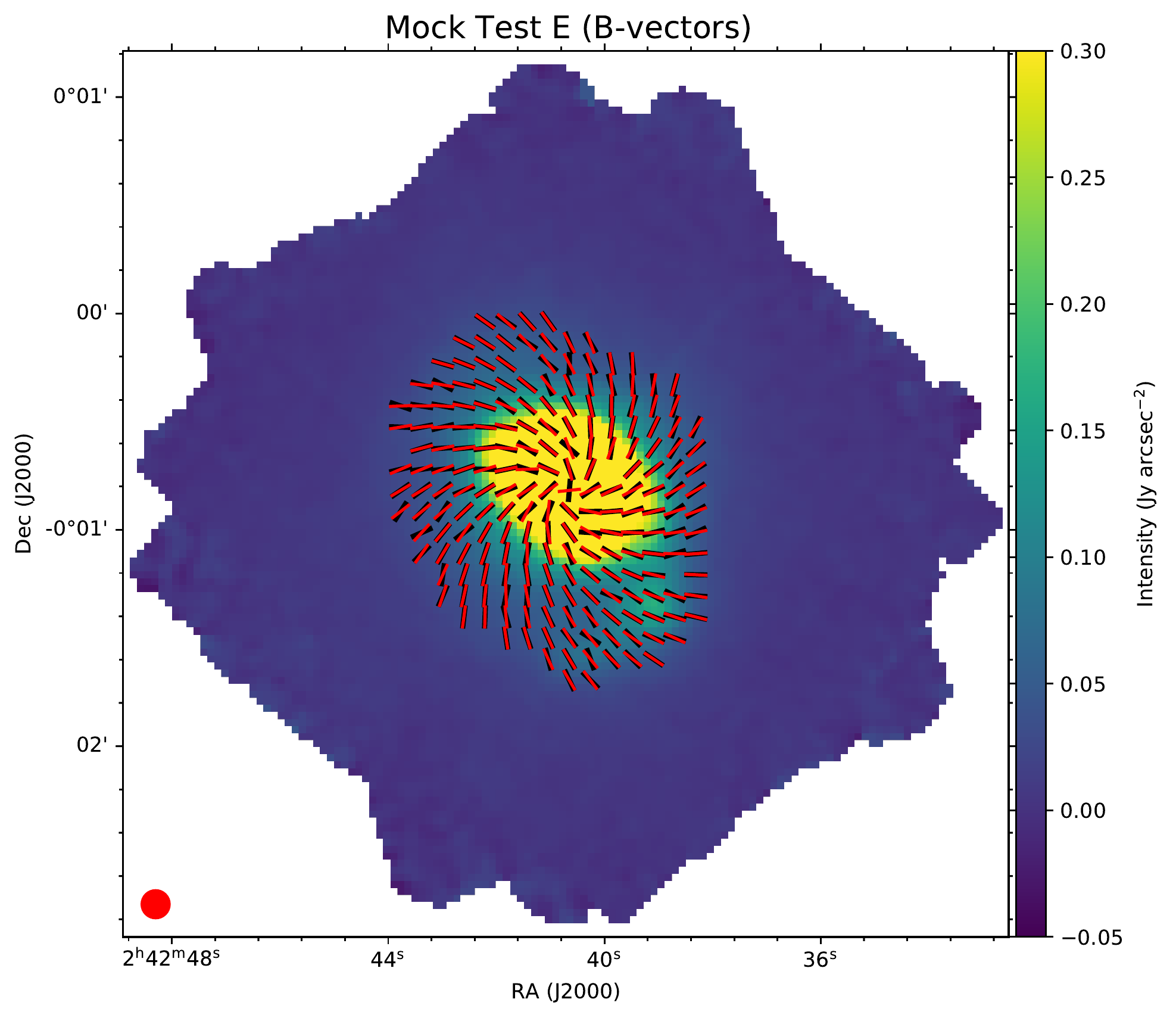}
\includegraphics[trim={0 0 0 0}, clip, width=0.49\textwidth]{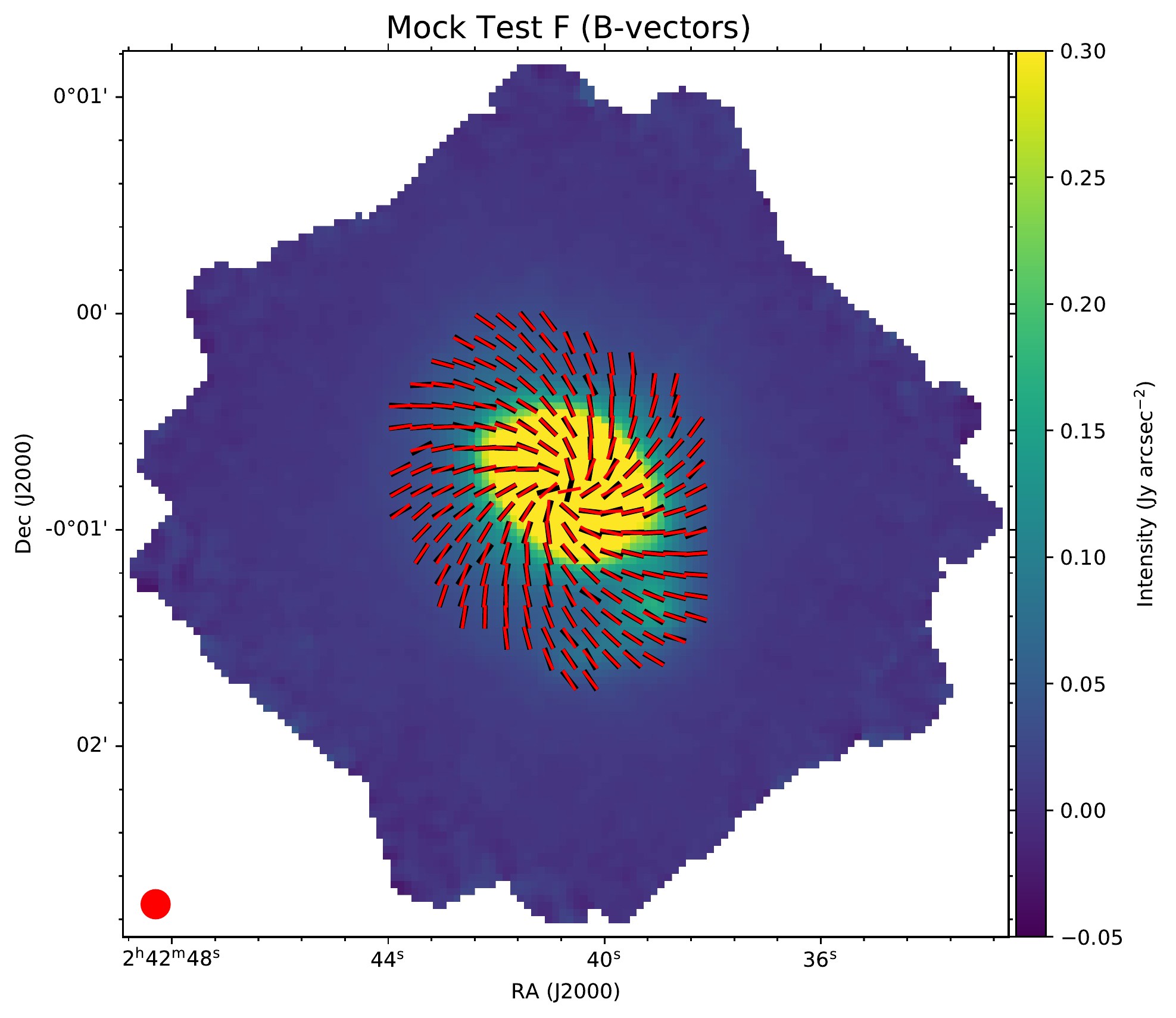}
\includegraphics[trim={0 0 0 0}, clip, width=0.49\textwidth]{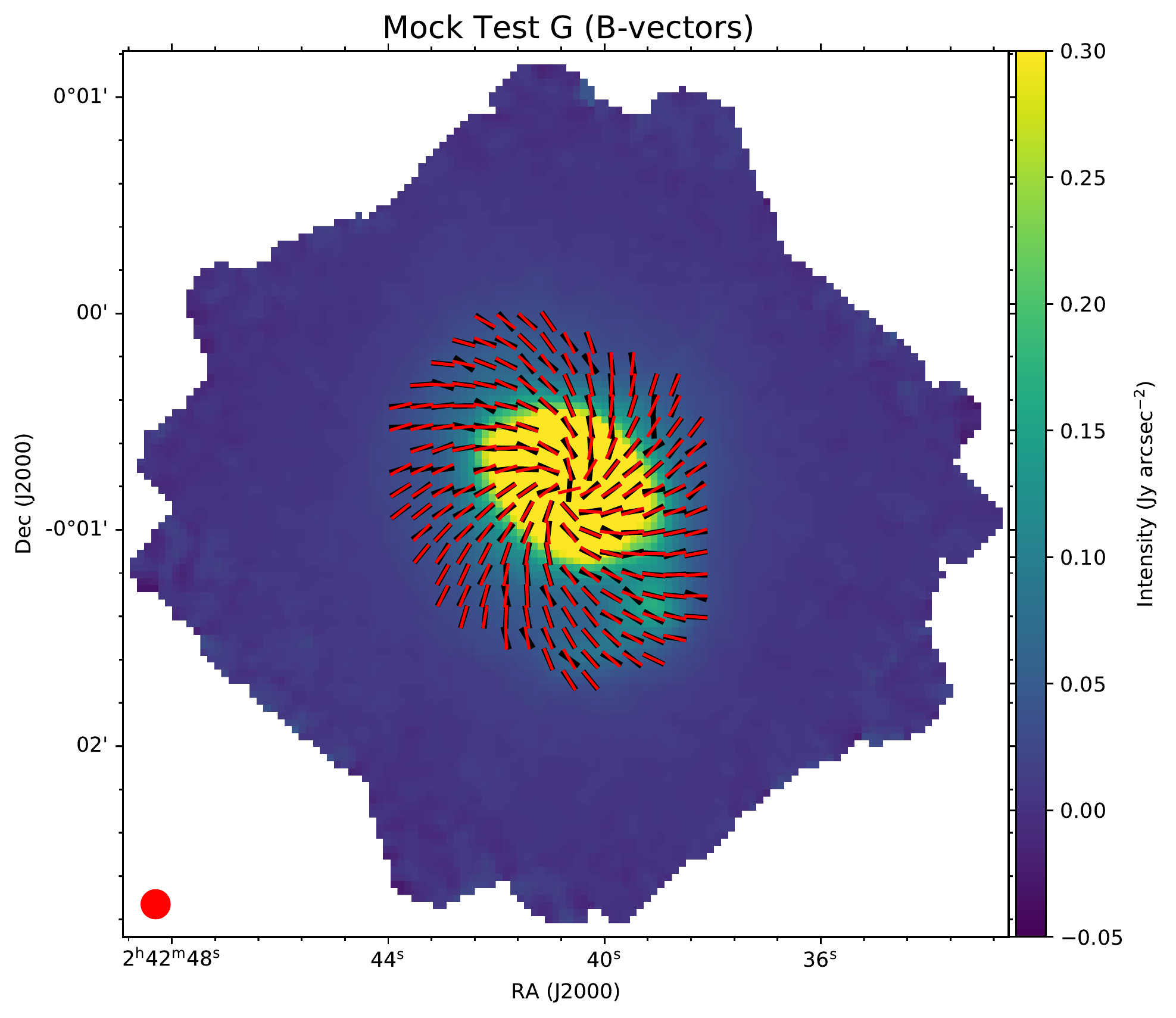}
\includegraphics[trim={0 0 0 0}, clip, width=0.49\textwidth]{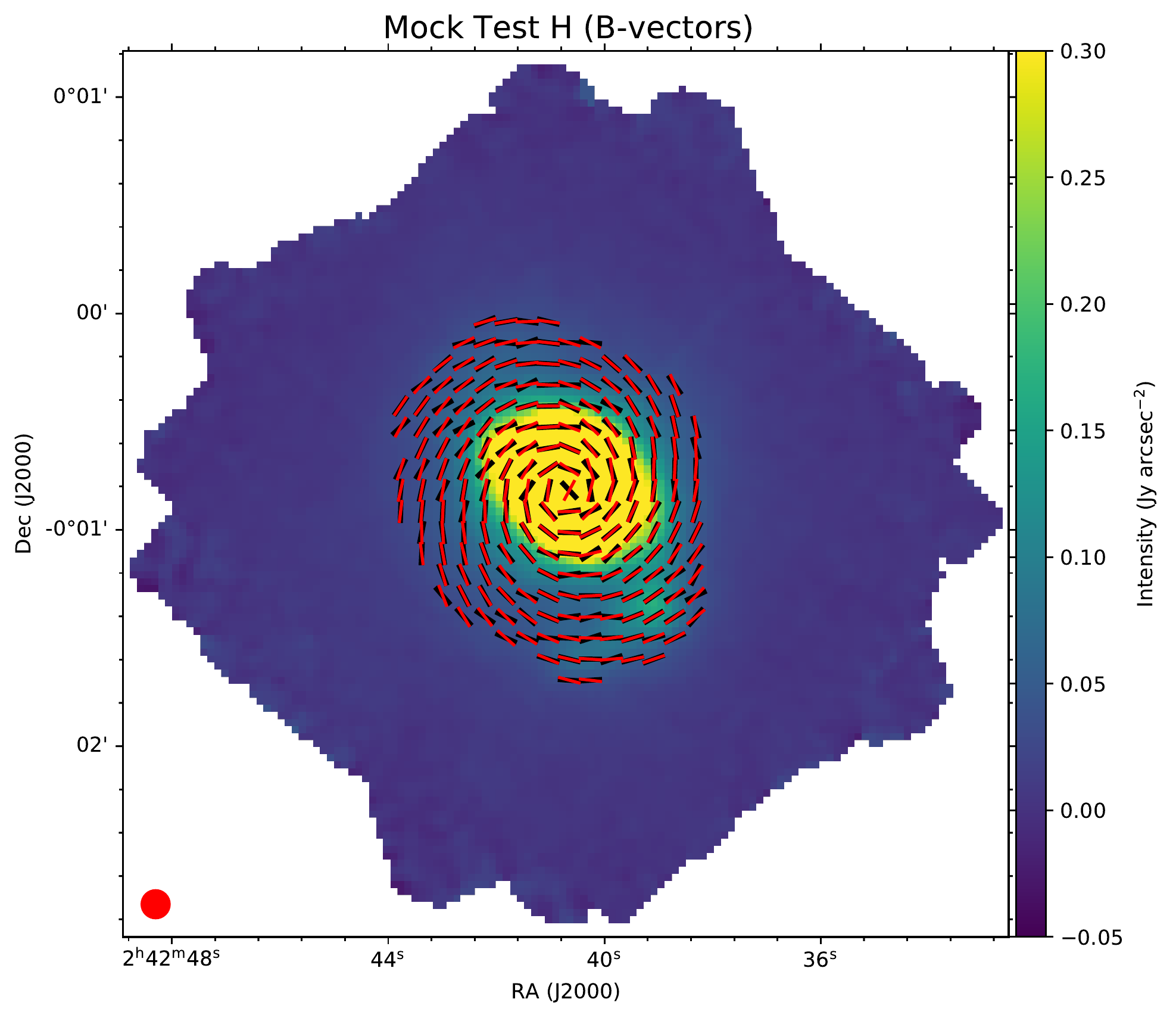}
\caption{Same as Fig. \ref{fig:mock_pitch_angle1} for tests EFGH.}
\label{fig:mock_pitch_angle2}
\end{center}
\end{figure*}

\begin{figure*}[t]
 \begin{center}
\includegraphics[trim={50 0 70 0}, clip, width=\textwidth]{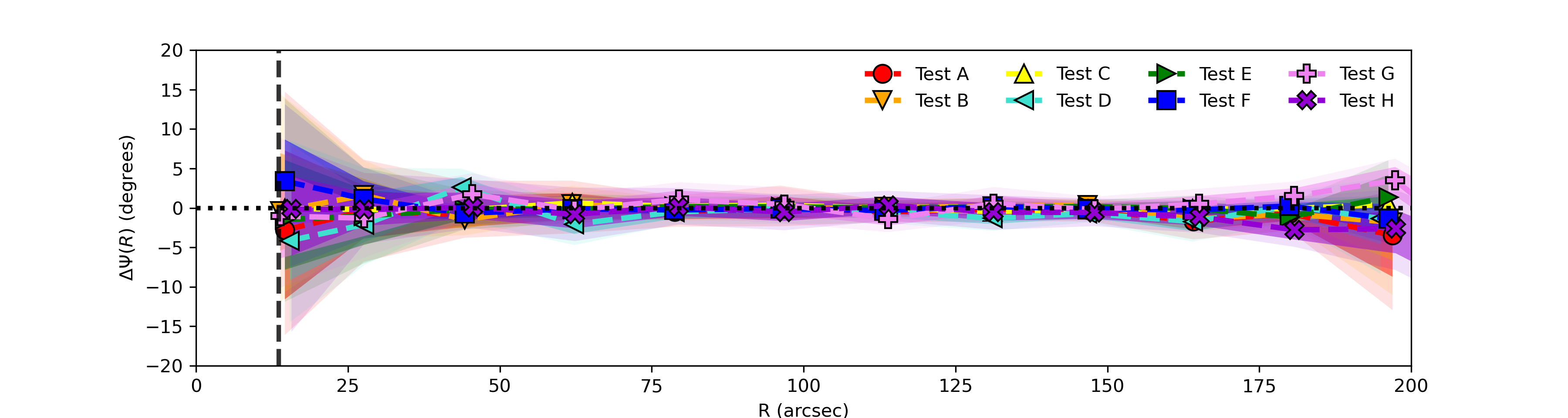}
\caption{Magnetic pitch angle mock dataset analysis. On the vertical axis we represent the magnetic pitch angle profile $\Psi(R)$ minus the simulated angle $\Psi_{Mock}$ as a function of radius for the mock observations shown in Table \ref{tab:app_mocktests}.}
\label{fig:mock_pitch_angle3}
\end{center}
\end{figure*}

In this section, we detail the tests performed to ensure the quality of the magnetic pitch angle profiles. We use a set of $8$ mock observations with different configurations in terms of magnetic pitch angle ($\Psi$), position angle (PA), inclination ($i$), and SNR. These tests were performed following a single-blind setup, where a member of the team produced the mock observations and another member of the team performed the data analysis without knowing the parameters of the models. This approach ensures the unbiased quality of the results. We use the $89$ \um\ HAWC+ observations of NGC\,1068 presented by \citet{LopezRodriguez2020} to setup the HAWC+ array configuration and the total intensity. Figures \ref{fig:mock_pitch_angle1} and \ref{fig:mock_pitch_angle2} show the total intensity of NGC1068 at 89 \um. Stokes $QU$ were replaced by the mock observations with the parameters shown in Table \ref{tab:app_mocktests}. A logarithmic spiral function with a single pitch angle, $\Psi$, across the image was used. This B-field model was then inclined and tilted to produce the projected B-field orientation in the plane of the sky. Noise was added using a Gaussian profile with mean $\mu=0$ and standard deviation $\sigma=$max$(IQU)/$SNR, that is, the noise level is specified by the desired SNR from the peak pixel. Figure \ref{fig:mock_pitch_angle3} shows the difference between the fixed parameter in the model with the estimated pitch angle following the approach in Section \ref{subsec:Methods_magnetic_pitch_angle}. An accuracy $\le5^{\circ}$ is achieved for polarization measurements with $P/\sigma_{P}\ge 2$. The large uncertainties at the inner and outer radii are due to the small amount of polarization measurements to produce enough statistical analysis. At these radii, a maximum angular uncertainty of $\sim15^{\circ}$ is expected.

\begin{deluxetable*}{cccccl}[ht!]
\tablecaption{Parameters of mock observations for the pitch angle estimations. \emph{Columns, left to right}: 1) ID. 2) Inclination of the model. 3) Position angle. 4) B-field pitch angle. 5) Signal to noise ratio (SNR). 6) Brief description on the individual models.\label{tab:app_mocktests}}
	\tablewidth{0pt}
		\tablehead{\colhead{Test}	&	\colhead{Inclination} 	&	\colhead{PA} & \colhead{Pitch}	 &	\colhead{SNR} &	\colhead{Comments} \\
			&	\colhead{($^{\circ}$)} 	&	\colhead{($^{\circ}$)} & \colhead{($^{\circ}$)}}
		\startdata
		A   &   0   &   0   &   0   &   1   &   Face-on, Azimuthal Profile, Low SNR \\
		B   &   0   &   0   &   0   &   3   &   Face-on, Azimuthal Profile \\
		C   &   0   &   0   &   0   &   10  &   Face-on, Azimuthal Profile, High SNR \\
		D   &   0   &   0   &   60   &   1   &  Face-on, Large pitch angle, Low SNR \\
		E   &   0   &   0   &   60   &   3   &  Face-on, Large pitch angle \\
		F   &   0   &   0   &   60   &  10   &  Face-on, Large pitch angle, High SNR \\
		G   &   30   &   0   &   60   &   1   &  Inclined, Large pitch angle, Low SNR \\
		H   &   30   &   47   &   17   &   1   &  Inclined, Tilted, Small pitch angle, Low SNR \\
		 \enddata
\end{deluxetable*}

\section{Polarization position angle diagrams}
\label{appendix:PA}

\begin{figure*}[ht!]
 \begin{center}
\centering
 \begin{minipage}[c]{.99\textwidth}
  \centering
   \includegraphics[trim={0 0 0 0}, clip, width=0.495\textwidth]{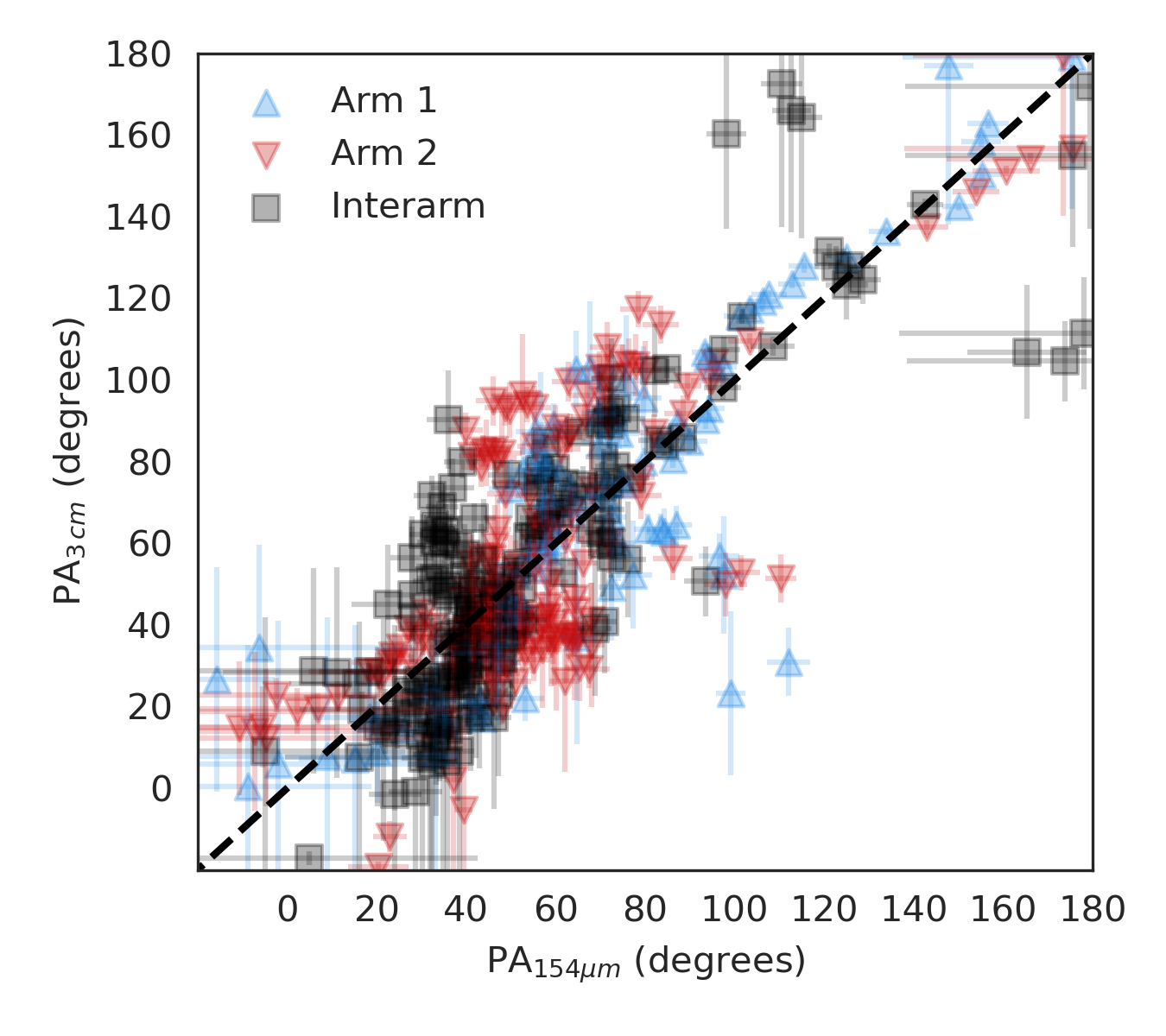} 
   \includegraphics[trim={0 0 0 0}, clip, width=0.495\textwidth]{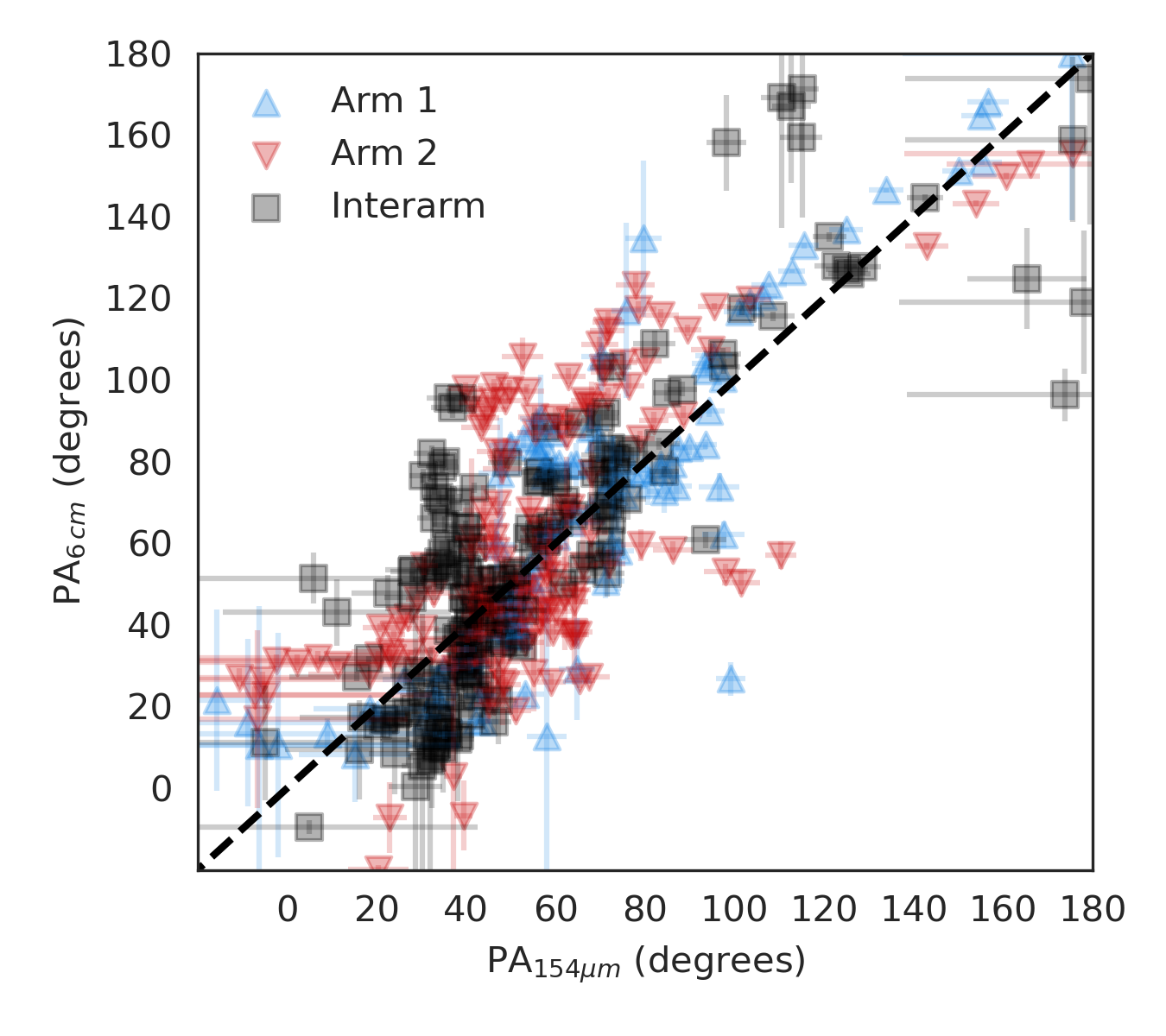}
  \end{minipage}
  
  
\caption{Distribution of the position angles of the 90$^{\circ}$-rotated polarization orientations in 3\,cm and 6\,cm, as a function of those obtained in 154\,$\mu$m. The dashed diagonal represents the 1:1 relation. See the legend for the symbols identifying the different morphological components of \m51.} 
\label{fig:PA}
\end{center}
\end{figure*}

In Fig.\,\ref{fig:PA} we represent the position angle of the 90$^{\circ}$-rotated polarization orientations of the 3\,cm and 6\,cm radio datasets as a function of those of SOFIA/HAWC+ in 154\,$\mu$m, for the different morphological components of \m51. We refer to Fig.\,11 of \citet{Jones2020} for a version of this figure with a subset of the HAWC+ observations presented in this work. The observed variation from a 1:1 relation are expected in these diagrams, suggesting that FIR and radio polarization observations do not trace the same magnetic field structure, agreeing with the main results of the present work (see Sec.\,\ref{Sec:Conclusions}).

\section{Polarization diagrams at 6\,cm}
\label{appendix:PIplots_6cm}

In this appendix we show the plots for the total intensity, polarized intensity, and polarization fraction at $6$ cm as a function of the column density, and velocity dispersion of the neutral gas, \hi, and molecular gas, \lineco. Section \ref{subsec:ISM_results} presents the analysis.

\begin{figure*}[ht!]
 \begin{center}
\centering
 \begin{minipage}[c]{.99\textwidth}
  \centering
   \includegraphics[trim={0 0 0 0}, clip, width=0.495\textwidth]{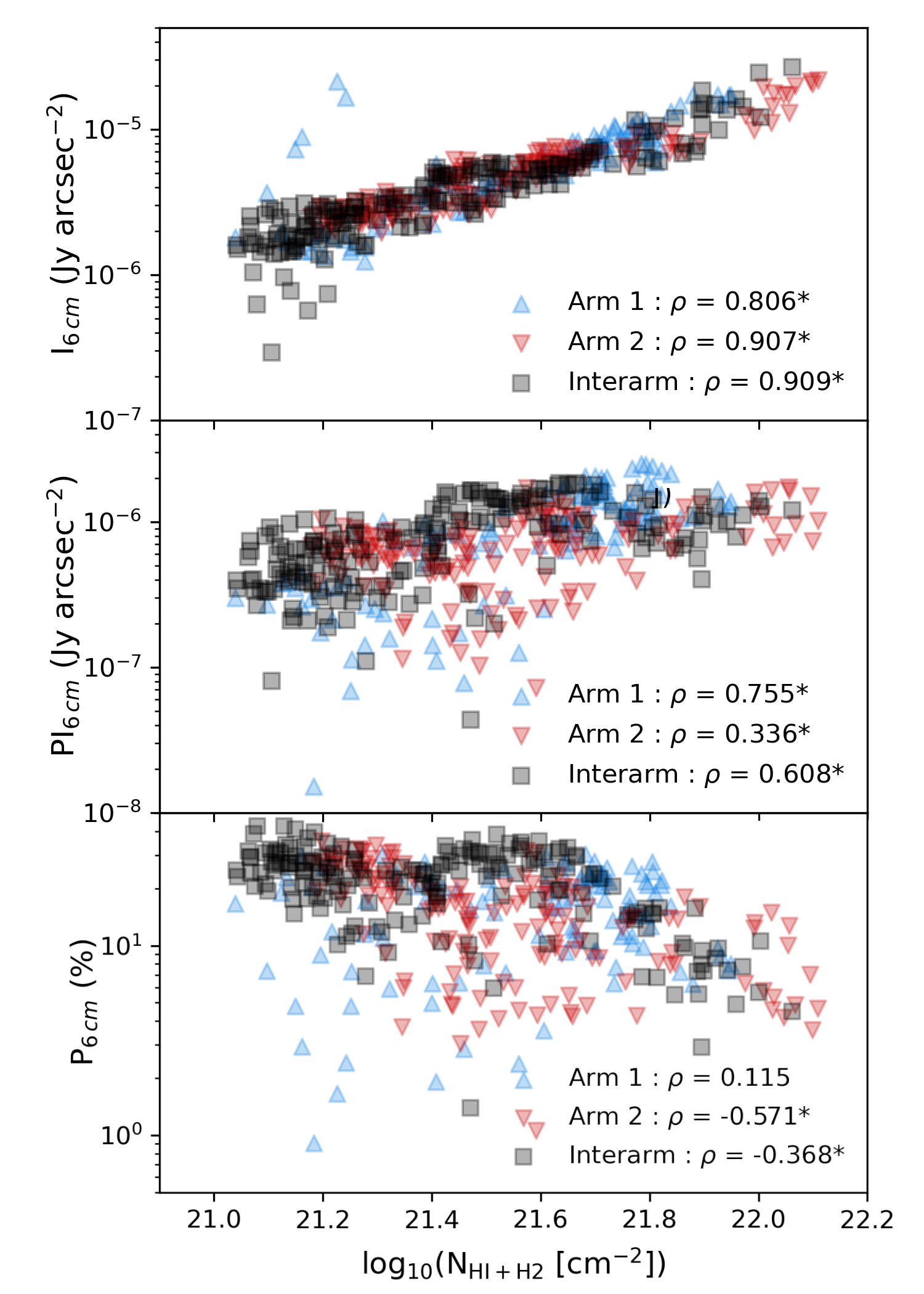} 
   \includegraphics[trim={0 0 0 0}, clip, width=0.495\textwidth]{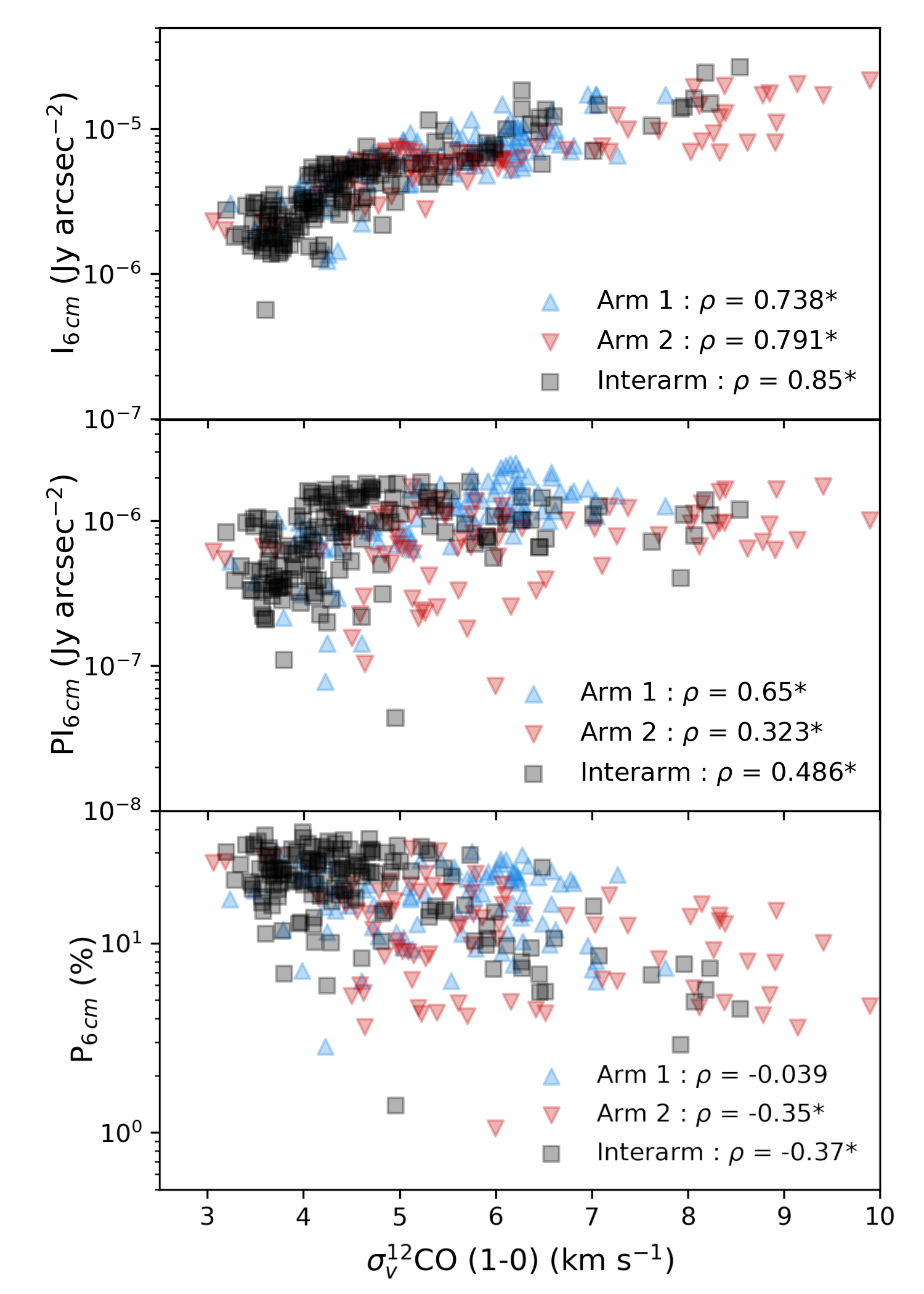}
  \end{minipage}
  
  
\caption{Distribution of 6\,cm total intensity (top row), polarized intensity (central row) and polarization fraction (bottom row) as a function of gas column density (\columndensity, left column) and \lineco\ velocity dispersion ($\sigma_{v, \mathrm{^{12}CO(1-0)}}$, right column). Symbols represent Arm 1 (blue upward pointing triangle), Arm 2 (red downward pointing triangle), and interarms (black square), where each data point is a polarization measurement as shown in Figure \ref{fig:mag_pitch_arms_interarms}. See the legend on each panel for the correlation analysis. An asterisk symbol ($*$) following each $\rho$ correlation coefficient is shown if the correlation is statistically different from zero ($p<0.05$).} 
\label{fig:m51_comparison_NHCO_6cm}
\end{center}
\end{figure*}

\begin{figure}[ht!]
\begin{center}
\includegraphics[trim={0 0 0 0}, clip, width=0.495\textwidth]{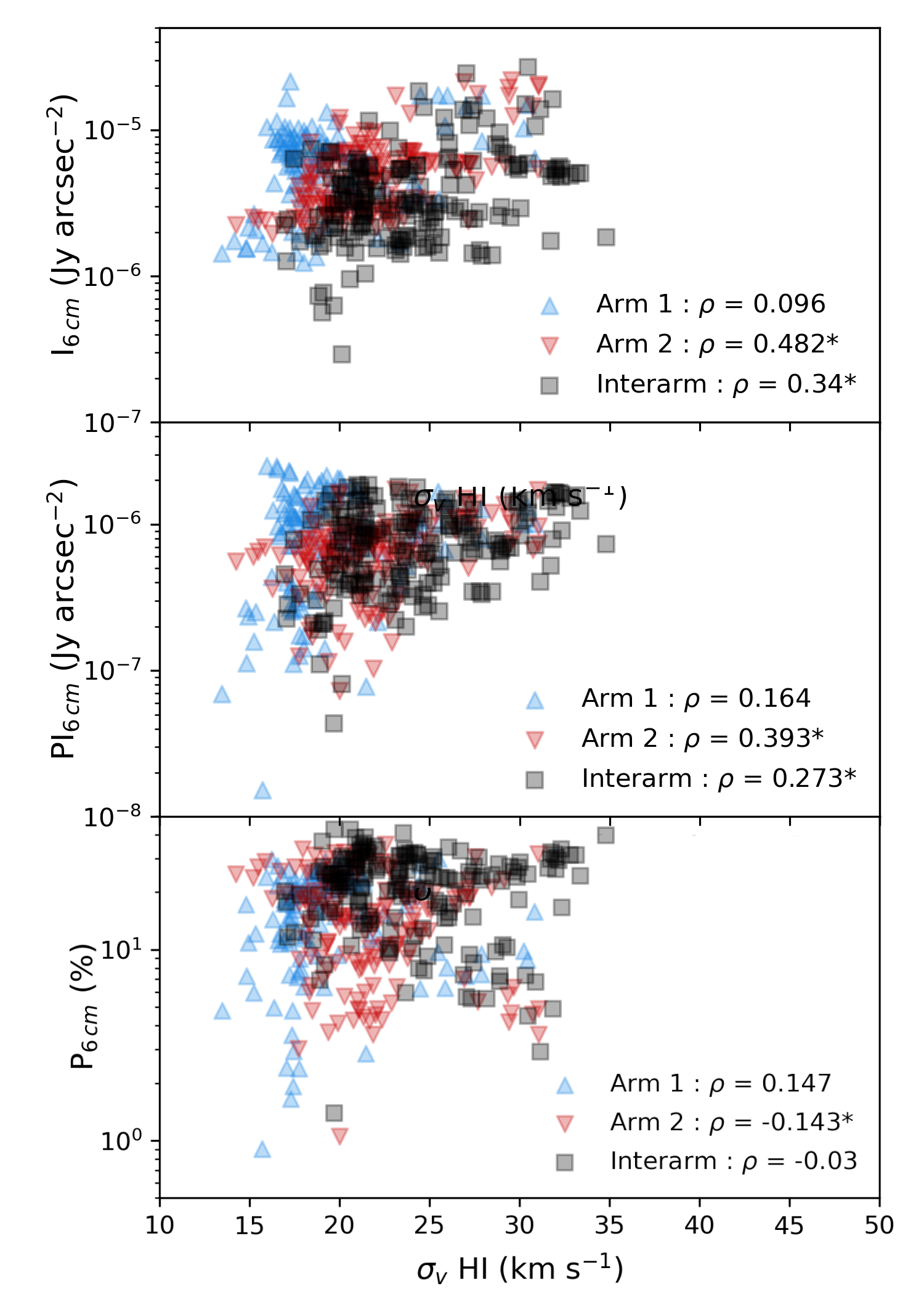} 
\caption{Distribution of 6\,cm total intensity (top row), polarized intensity (central row) and polarization fraction (bottom row) as a function of the \hi\ velocity dispersion ($\sigma_{v,{\mathrm{HI}}}$). Symbols represent Arm 1 (blue upward pointing triangle), Arm 2 (red downward pointing triangle), and the interarm region (black square), where each data point is a polarization measurement as shown in Figure \ref{fig:mag_pitch_arms_interarms}. See the legend on each panel for the correlation analysis. An asterisk symbol ($*$) following each $\rho$ correlation coefficient is shown if the correlation is statistically different from zero ($p<0.05$).}
\label{fig:m51_comparison_HI_6cm}
\end{center}
\end{figure}



%
\bibliography{M51_references}
\bibliographystyle{aasjournal}






\end{document}